\documentclass[11pt,a4paper]{article}

\usepackage[utf8]{inputenc}
\usepackage[T1]{fontenc}
\usepackage{lmodern}
\usepackage{amsmath,amssymb}
\usepackage{graphicx}
\usepackage{booktabs}
\usepackage{multirow}
\usepackage{array}
\usepackage{float}
\usepackage{caption}
\usepackage{subcaption}
\usepackage{xcolor}
\usepackage{url}
\usepackage{hyperref}
\usepackage{geometry}
\usepackage{enumitem}
\usepackage{amsmath}

\usepackage{booktabs}
\usepackage{makecell}
\usepackage{adjustbox}

\hypersetup{
    colorlinks=true,
    linkcolor=blue,
    citecolor=blue,
    urlcolor=blue
}

\title{
\textbf{Benchmarking Quantum Feature Encoding Strategies for Binary Classification with QSVM}
}

\author{
Murat Kurt\\
\texttt{kmurat.physics@gmail.com}
}

\date{}

\begin{document}

\maketitle

\begin{abstract}

\setlength{\emergencystretch}{3em}

The way in which classical data are encoded into quantum states plays a significant role in both classification performance and quantum circuit complexity in Quantum Machine Learning. In this study, the effects of different quantum feature encoding strategies on Quantum Support Vector Machine performance were investigated using five binary classification datasets. In particular, the statistical relationships between features were incorporated into quantum circuits through \(RY(\theta)\) and controlled-\(RY(\theta)\) gates, and this approach was compared with conventional quantum feature maps. The results demonstrate that incorporating statistical relationships into the encoding process can influence classification performance. However, more complex and densely entangled circuits do not necessarily yield higher performance. In addition, a composite evaluation metric was employed to jointly assess predictive performance, generalization, and circuit cost. The findings across the five datasets indicate that the choice of quantum feature encoding strategy should account for the underlying structure of the data and that predictive performance should be evaluated together with quantum circuit complexity.

\end{abstract}


\section{Introduction}
\setlength{\emergencystretch}{3em}
\label{sec:introduction}

Quantum computing (QC) is a rapidly developing research field that enables 
computational problems to be addressed within a framework fundamentally 
different from classical approaches by exploiting quantum mechanical 
properties such as superposition and entanglement. In particular, with the 
development of Noisy Intermediate Scale Quantum (NISQ) devices, algorithms 
aimed at the efficient utilization of limited quantum resources have gained 
increasing importance 
\cite{benioff1980computer,deutsch1985quantum,simon1997power,
shor1997polynomial,grover1997quantum}. In this context, Quantum Machine 
Learning (QML) has emerged at the intersection of classical machine learning 
and QC as an important research area that aims to address 
problems such as classification, regression, pattern recognition, and 
optimization using quantum computational methods 
\cite{schuld2015introduction,biamonte2017quantum,allcock2019quantum,
zeguendry2023quantum,peralgarcia2024systematic,lu2024quantum,
devadas2025quantum,rodriguezdiaz2026survey,nadeem2026quantum}.

One of the fundamental stages of QML is the 
representation of classical data in quantum systems. For a classical dataset 
to be processed by a quantum algorithm, the data must first be encoded into 
quantum states. This process is performed through different data encoding 
methods or quantum feature maps 
\cite{rath2024quantum,goto2021universal,albrecht2023quantum,
nguyen2024quantum,hayashi2025effective,jha2026comparative,
jager2026quantum,kurt2026correlation}. Quantum feature maps transform 
classical feature vectors into quantum states in a Hilbert space, thereby 
enabling data samples to be represented within a quantum system. Therefore, 
the structure of the encoding method is one of the fundamental factors that 
determine the representation of classical data in quantum space and the 
similarities formed between quantum states.

QSVMs represent an important application 
of quantum feature maps to classification problems. In the QSVM approach, 
classical data samples are transformed into quantum states through quantum 
feature maps, and a quantum kernel matrix is constructed based on the 
similarities between these states. The resulting kernel matrix is then used 
in the classification process of a support vector machine. Consequently, the 
performance of a QSVM depends not only on the classification algorithm but 
also on how classical data are mapped into the quantum state space and how 
the selected quantum feature map represents the underlying structure of the 
data 
\cite{rebentrost2014quantum,gentinetta2024complexity,
suzuki2024quantum,jager2023universal,
zhang2023quantum,nohara2022pairwise,xu2026qsvm}.

Various encoding and feature mapping approaches have been developed to map 
classical data into quantum states. Angle encoding, which maps classical 
features onto the angles of parameterized quantum rotation gates, is one of 
the commonly used approaches due to its relatively simple circuit structure. 
In this context, $R_Y(\theta)$ rotations allow each classical feature to be 
represented through the rotation angle of the corresponding qubit. In 
addition, Pauli-based quantum feature maps such as Z feature Map and 
ZZ feature Map are widely used, particularly in quantum kernel methods. 
Z feature Map focuses on encoding individual features into quantum states, 
whereas ZZ feature Map contains two-qubit operations that enable interactions 
between pairs of features to be represented within the quantum circuit. 
These interactions can be constructed using different entanglement 
topologies, such as linear, circular or fully connected structures 
\cite{vasques2023application,farooq2024enhanced,
ajibosin2024implementation,alvarez2025benchmarking,
toufah2025investigating,bansal2025enhancing,
schnabel2025quantum}.

In conventional feature maps, interactions between qubits are generally 
constructed according to predefined entanglement structures, such as linear, 
circular or fully connected topologies. However, these connectivity patterns 
are not directly determined based on the statistical relationships between 
features in the dataset. This raises the question of how the dependency 
structure inherent in classical data can be incorporated into the design of 
quantum feature maps. Therefore using statistical relationships between 
features to determine interactions within a quantum circuit represents a 
promising direction for developing data-driven quantum feature maps.

In classical data analysis, various statistical measures are used to 
identify relationships between variables. The Pearson correlation 
coefficient evaluates the direction and strength of linear relationships 
between variables, whereas Spearman correlation and Kendall's Tau are used 
to identify rank-based monotonic relationships. Mutual Information provides 
a means of evaluating more general dependency structures between variables 
within an information-theoretic framework while Distance Correlation can 
capture nonlinear dependencies 
\cite{george2014eeg,battiti1994mutual,li2012distance,
tan2022distance,ratnasingam2023distance,das2024feature}. Although these 
methods are based on different mathematical principles they can provide 
complementary information for identifying relationships among features in 
the data.

Incorporating these statistical relationships between features into the 
quantum data encoding process provides an alternative approach for directly 
transferring the structure of classical data into a quantum circuit. In 
standard angle encoding, each feature can be represented independently by a 
rotation angle. By contrast, incorporating relationships between feature 
pairs through controlled quantum gates enables dependency information 
between features, in addition to individual feature information, to be 
embedded into the quantum state. In this way, interactions between qubits 
can be constructed not only according to a predefined connectivity pattern 
but also by taking into account the statistical characteristics of the 
analyzed dataset.

In this study, the effect of incorporating statistical relationships between 
classical features into quantum feature maps on QSVM classification is 
systematically investigated. For this purpose, classical features are 
encoded into quantum states using $R_Y(\theta)$ rotations, while the 
relationships between feature pairs are incorporated into the quantum 
circuit using controlled-$R_Y(\theta)$ (CRY) gates. Five different statistical 
dependency measures---Pearson correlation, Spearman correlation, Kendall's 
Tau, Mutual Information, and Distance Correlation---are considered to 
determine feature interactions implemented through CRY gates. This framework 
enables the data structures identified by different dependency measures to 
be incorporated into the quantum encoding process and their effects on QSVM 
performance to be examined within a common experimental setting.

The proposed approach is evaluated on five different datasets. For 
comparison, standard $R_Y(\theta)$ angle encoding, Z Feature Map, and ZZ Feature Map with 
linear, circular, and full entanglement structures are employed. The models 
are evaluated using classification metrics including accuracy, F1 score, 
ROC-AUC, and log-loss. In addition, the differences between training and test 
performance are examined to assess the generalization behavior of the 
models. Circuit depth and gate counts are also included in the analysis to 
evaluate the computational requirements of the quantum feature maps. Thus, 
the different data encoding approaches are compared not only in terms of 
classification performance but also with respect to their quantum circuit 
structures.

The main contribution of this study is the direct incorporation of 
statistical dependency information among classical features into the 
construction of quantum feature maps and the systematic comparison of the 
effects of different dependency measures on this process. Through experiments 
conducted on five different datasets, feature interactions based on Pearson 
correlation, Spearman correlation, Kendall's Tau, Mutual Information, and 
Distance Correlation are evaluated within a common framework alongside 
standard quantum feature maps. In this context, the study provides an 
experimental framework for investigating the relationship between the 
statistical structure of classical data and quantum data encoding, and for 
evaluating data-aware quantum feature maps for QSVM based classification.

\section{Materials and Methods}
\label{sec:materials_methods}

In this study, a state vector based computational framework was developed to
investigate the effect of incorporating statistical relationships among
features into the quantum data encoding process on QSVM performance. The
experimental workflow consisted of dataset preparation, data preprocessing,
estimation of statistical relationships among features, construction of
quantum feature maps, computation of quantum kernel matrices, and evaluation
of classification performance.

\subsection{Datasets}
\label{subsec:datasets}

Five datasets representing different binary classification problems were
used in this study. All datasets were obtained from the Kaggle data-sharing
platform. The use of datasets with different characteristics was intended to
evaluate the proposed approach under different feature distributions and
inter-feature dependency structures rather than under conditions specific
to a single dataset.

For each dataset, a target variable was defined and the problem was treated
as a binary classification task. Identifier-like variables that did not
provide meaningful information for classification were excluded from the
analysis. Categorical variables were converted into numerical form so that
they could be processed by the QSVM without unnecessarily increasing the
input dimensionality. Each dataset was divided into training and test sets,
and scaling parameters were estimated exclusively from the training data to
prevent data leakage. Before quantum encoding, feature values were scaled to
an appropriate numerical range compatible with quantum rotation angles.

\subsection{Determination of Statistical Relationships Among Features}
\label{subsec:statistical_relations}

Five statistical dependency measures were employed to characterize
relationships between pairs of features: Pearson correlation, Spearman
correlation, Kendall's Tau, Mutual Information (MI), and Distance
Correlation. These measures were selected because they capture
different forms of statistical dependence within the data.

The Pearson correlation coefficient was used to quantify the direction and
strength of the linear relationship between two variables. For two variables
$X$ and $Y$, the Pearson correlation coefficient is calculated as

\begin{equation}
r_{XY} =
\frac{\sum_{i=1}^{N}(x_i-\bar{x})(y_i-\bar{y})}
{\sqrt{\sum_{i=1}^{N}(x_i-\bar{x})^2}
\sqrt{\sum_{i=1}^{N}(y_i-\bar{y})^2}}.
\end{equation}

Spearman correlation was used to evaluate monotonic relationships based on
the ranks of the variables. When rank differences are used, the Spearman
coefficient is expressed as

\begin{equation}
\rho =
1-\frac{6\sum_{i=1}^{N}d_i^2}{N(N^2-1)},
\end{equation}

where $d_i$ denotes the difference between the ranks of the corresponding
observations.

Kendall's Tau evaluates the ordinal association between two variables based
on concordant and discordant observation pairs. The coefficient is expressed
as

\begin{equation}
\tau =
\frac{N_c-N_d}
{\frac{1}{2}N(N-1)},
\end{equation}

where $N_c$ and $N_d$ denote the numbers of concordant and discordant pairs,
respectively.

MI was used to quantify statistical dependence between two variables within
an information-theoretic framework and is defined as

\begin{equation}
I(X;Y) =
\sum_{x,y} p(x,y)
\log
\left(
\frac{p(x,y)}
{p(x)p(y)}
\right).
\end{equation}

DC was employed as a measure capable of identifying nonlinear as well as
linear dependencies. Thus, the five dependency measures, which are based on
different mathematical principles, were compared within the same
experimental framework to investigate their effects on quantum data
encoding.

\subsection{Quantum Data Encoding}
\label{subsec:quantum_encoding}

The proposed encoding approach consists of two main stages. In the first
stage, each classical feature is encoded into a quantum state using a
single qubit $R_Y$ rotation. An $R_Y$ rotation is defined as

\begin{equation}
R_Y(\theta)=
\begin{pmatrix}
\cos(\theta/2) & -\sin(\theta/2) \\
\sin(\theta/2) & \cos(\theta/2)
\end{pmatrix}.
\end{equation}

Accordingly, for a data sample
$\mathbf{x}=(x_1,x_2,\ldots,x_n)$ containing $n$ features, each feature
$x_i$ determines the rotation angle of the $R_Y$ gate applied to the
corresponding qubit.

In the second stage, the statistical relationships calculated between
feature pairs are incorporated into the quantum circuit using controlled
$R_Y$ (CRY) gates. If the dependency measure between features $i$ and $j$
is denoted by $D_{ij}$, the corresponding controlled rotation is generally
represented as

\begin{equation}
CRY_{ij}(\theta_{ij}),
\end{equation}

where $\theta_{ij}$ denotes the rotation angle derived from the statistical
relationship calculated for the corresponding feature pair. Thus, individual
feature values are encoded using $R_Y$ gates while the statistical
dependency structure among features is incorporated into the quantum feature
map through CRY gates.

The same underlying $R_Y$ encoding structure was retained for Pearson,
Spearman, Kendall's Tau, MI, and Distance Coleration . The principal difference among these
experiments was the statistical dependency measure used to determine the
CRY interactions. This design enabled the effects of different dependency
measures on the QSVM kernel and classification performance to be compared
under equivalent experimental conditions.

\subsection{Baseline Quantum Feature Maps}
\label{subsec:baseline_featuremaps}

To evaluate the proposed approach, standard $R_Y$ angle encoding,
Z Feature Map, and ZZ Feature Map were employed as baseline encoding strategies.

Z Feature Map maps classical features into quantum states through single qubit
phase operations, whereas ZZ Feature Map additionally incorporates interactions
between feature pairs into the quantum feature map. For ZZ Feature Map, three
entanglement topologies were evaluated: linear, circular, and full
entanglement. These structures allowed the proposed statistical
dependency-based interaction strategy to be compared with predefined
quantum feature-map connectivity patterns.

\subsection{state vector Based Quantum Kernel Computation}
\label{subsec:state vector}

The experiments in this study were not executed on a physical Quantum
Processing Unit (QPU). Instead, the quantum circuits and the resulting
quantum states were evaluated on a classical computer using state vector based
numerical computation. Therefore, the reported results are not experimental
measurements obtained from physical quantum hardware but are based on the
state vector representation of ideal quantum states.

For each data sample $\mathbf{x}$, let $U_{\phi}(\mathbf{x})$ denote the
corresponding quantum feature map. The resulting quantum state is given by

\begin{equation}
|\psi(\mathbf{x})\rangle =
U_{\phi}(\mathbf{x})|0\rangle^{\otimes n}.
\end{equation}

The quantum kernel value between two data samples $\mathbf{x}_i$ and
$\mathbf{x}_j$ was calculated as the squared magnitude of the inner product
between their corresponding state vectors:

\begin{equation}
K(\mathbf{x}_i,\mathbf{x}_j)
=
\left|
\langle
\psi(\mathbf{x}_i)
|
\psi(\mathbf{x}_j)
\rangle
\right|^2.
\end{equation}

This calculation was performed for all pairs of samples in the training set
to construct the training kernel matrix $K_{\mathrm{train}}$. During
testing, quantum kernel values between the test and training samples were
calculated to obtain $K_{\mathrm{test}}$. The resulting kernel matrices
were subsequently supplied to the support vector machine as precomputed
kernels.

The state vector approach eliminates sampling uncertainty associated with a
finite number of measurement shots and excludes noise effects specific to
physical quantum hardware from the experimental comparison. The primary
objective was therefore to compare the behavior of different quantum data
encoding and feature-map designs under ideal conditions. Consequently, the
reported results should not be interpreted as direct indicators of
performance on physical quantum hardware, but rather as a state vector-based
comparison of the theoretical behavior of the investigated quantum feature
maps.

\subsection{QSVM Classification}
\label{subsec:qsvm}

The training and test kernel matrices computed for each quantum feature map
were used for QSVM classification. During training, the support vector
machine was fitted using $K_{\mathrm{train}}$ and the corresponding training
labels. During testing, class predictions were generated using
$K_{\mathrm{test}}$.

The same training--test split and preprocessing procedure were used for all
methods to ensure comparisons under equivalent conditions. This design was
intended to ensure that observed performance differences could primarily be
associated with the selected quantum feature map and statistical dependency
measure rather than variations in data splitting or preprocessing.

\subsection{Evaluation of Classification Performance and Circuit Complexity}
\label{subsec:evaluation}

Classification performance was evaluated using accuracy, F1 score, ROC-AUC,
and log-loss. Training and test performance were also jointly examined to
evaluate the generalization behavior of the models.

Circuit depth and gate count were calculated to evaluate the quantum feature
maps in terms of circuit structure in addition to classification
performance. This enabled the structural complexity of the quantum circuits
generated by different encoding strategies to be considered together with
their predictive performance.

Throughout all experiments, data preprocessing, training--test splitting,
and evaluation procedures were kept fixed to systematically isolate and
compare the effects of the quantum data encoding strategies.

\subsection{Datasets and Data Preprocessing}

Five datasets were used to evaluate the effects of different data structures
on quantum feature encoding strategies: the EEG Eye State
Classification Dataset, Heart Failure Prediction Dataset,
Credit Risk Dataset, Student Performance Dataset, and
\textit{Alzheimer's Disease Dataset}. All datasets were obtained from
Kaggle. Each dataset was formulated as a binary classification problem, and
the same training--test split and preprocessing steps were retained across
all quantum feature encoding strategies compared within each dataset.

\paragraph{EEG Eye State Classification Dataset.}
The EEG Eye State Classification Dataset is a binary classification dataset
designed to determine whether the eyes are open or closed from
electroencephalography (EEG) signals \cite{eeg_kaggle}. The dataset is based
on continuous EEG measurements acquired using an Emotive EEG Neuro headset.
To reduce the influence of class imbalance on the experimental results, a
balanced subset containing an equal number of samples from the two classes
was constructed, resulting in a total of 2000 samples. After separating the
target variable, all 14 input features corresponding to the 14 EEG channels
were retained. Consequently, the quantum feature maps were constructed with
an input dimension of $n=14$.

\paragraph{Heart Failure Prediction Dataset.}
The Heart Failure Prediction Dataset was obtained from Kaggle
\cite{heart_kaggle}. The original dataset contains 918 observations and 11
clinical input features associated with the prediction of heart disease.
The target variable was defined as Heart Disease. In this study,
FastingBS was excluded from the analysis, while Age,
Sex, ChestPainType, RestingBP,
Cholesterol, RestingECG, MaxHR,
ExerciseAngina, Oldpeak, and ST\_Slope were
retained, resulting in a total of 10 input features. Categorical variables
were converted into numerical values without increasing the input
dimensionality, and the experiments were therefore conducted with $n=10$.

\paragraph{Credit Risk Dataset.}
The Credit Risk Dataset was obtained from Kaggle
\cite{credit_kaggle}. The dataset contains demographic, financial, and
credit-related variables used to classify credit default risk. To control
the computational cost of the state vector-based experiments, 1000
observations were used in this study. The target variable was separated from
the input features, and categorical variables were converted into numerical
values without increasing the number of columns. Following preprocessing, a
total of nine input features were retained. Thus, all quantum feature
encoding strategies were evaluated in the same $n=9$ dimensional feature
space.

\paragraph{Student Performance Dataset.}
The Student Performance Dataset was obtained from Kaggle
\cite{student_kaggle}. It contains variables describing student
demographics, study habits, absences, tutoring support, parental support,
extracurricular activities, and Grade Point Average (GPA). The
StudentID variable was excluded from the analysis. A binary target
label was derived from GPA: students with $GPA \geq 2.5$ were classified as
successful ($y=1$), whereas those with $GPA<2.5$ were classified as
unsuccessful ($y=0$). Because GPA was used to construct the target label, it
was removed from the input features. The remaining 13 variables were used
in the QSVM experiments, resulting in an input dimension of $n=13$.

\paragraph{Alzheimer's Disease Dataset.}
The Alzheimer's Disease Dataset was obtained from Kaggle
\cite{alzheimer_kaggle}. The original dataset contains 2149 patient records
with demographic information, lifestyle factors, medical history, clinical
measurements, cognitive and functional assessments, and symptom-related
variables. The target variable was defined as Diagnosis, while
PatientID and other identifier-like variables without meaningful
classification information were excluded from the analysis. To reduce the
memory requirements associated with the high-dimensional feature space in
state vector-based computation, Principal Component Analysis (PCA) was
applied and the first 14 principal components (PCs) were retained.
Consequently, all quantum feature encoding strategies were evaluated within
the same reduced $n=14$ dimensional feature space.

\section{Results and Discussion}
\label{sec:results_discussion}

\subsection{EEG Eye State Classification Dataset}
\label{subsec:eeg_results}

The experiments conducted on the EEG Eye State Classification Dataset
compared Angle-RY, H+RY, RY+CRY encodings based on different statistical
dependency measures, Z Feature Map, and ZZ Feature Map with different
entanglement structures. The evaluation was performed in terms of
classification performance, generalization behavior, quantum circuit
complexity, and computational cost.

\subsubsection{Classification Performance}

\begin{table*}[!htbp]
\setlength{\tabcolsep}{4.5pt}
\centering
\caption{Classification performance of quantum feature encoding strategies
on the EEG Eye State Classification Dataset.}
\label{tab:eeg_performance}

\footnotesize
\setlength{\tabcolsep}{3pt}
\renewcommand{\arraystretch}{1.20}

\begin{tabular}{lccccccc}
\toprule
\textbf{Encoding} &
\makecell{\textbf{Train}\\\textbf{Acc.}} &
\makecell{\textbf{Test}\\\textbf{Acc.}} &
\makecell{\textbf{Acc.}\\\textbf{Gap}} &
\makecell{\textbf{Test}\\\textbf{Prec.}} &
\makecell{\textbf{Test}\\\textbf{Rec.}} &
\makecell{\textbf{Test}\\\textbf{F1}} &
\makecell{\textbf{Test}\\\textbf{AUC}} \\
\midrule

Angle-RY        & 0.749 & 0.748 & 0.001 & 0.824 & 0.560 & 0.667 & 0.823 \\
H+RY            & 0.749 & 0.748 & 0.001 & 0.824 & 0.560 & 0.667 & 0.823 \\
RY+CRY-Pearson  & 0.883 & 0.806 & 0.077 & 0.811 & 0.742 & 0.775 & 0.889 \\
RY+CRY-Spearman & 0.865 & 0.802 & 0.063 & 0.818 & 0.720 & 0.766 & 0.883 \\
RY+CRY-Kendall  & 0.842 & 0.786 & 0.056 & 0.814 & 0.680 & 0.741 & 0.873 \\
RY+CRY-MI       & 0.829 & 0.778 & 0.051 & 0.800 & 0.676 & 0.733 & 0.868 \\
RY+CRY-Distance & 0.876 & 0.796 & 0.080 & 0.806 & 0.720 & 0.761 & 0.886 \\
Z               & 0.894 & 0.834 & 0.060 & 0.866 & 0.747 & 0.802 & 0.910 \\
ZZ linear       & 0.994 & 0.842 & 0.152 & 0.888 & 0.742 & 0.809 & 0.918 \\
ZZ circular     & 0.995 & 0.846 & 0.149 & 0.898 & 0.742 & 0.813 & 0.927 \\
ZZ full         & 1.000 & 0.850 & 0.150 & 0.908 & 0.742 & 0.817 & 0.928 \\

\bottomrule
\end{tabular}

\end{table*}

\begin{figure}[!htbp]
\centering
\includegraphics[width=0.90\columnwidth]{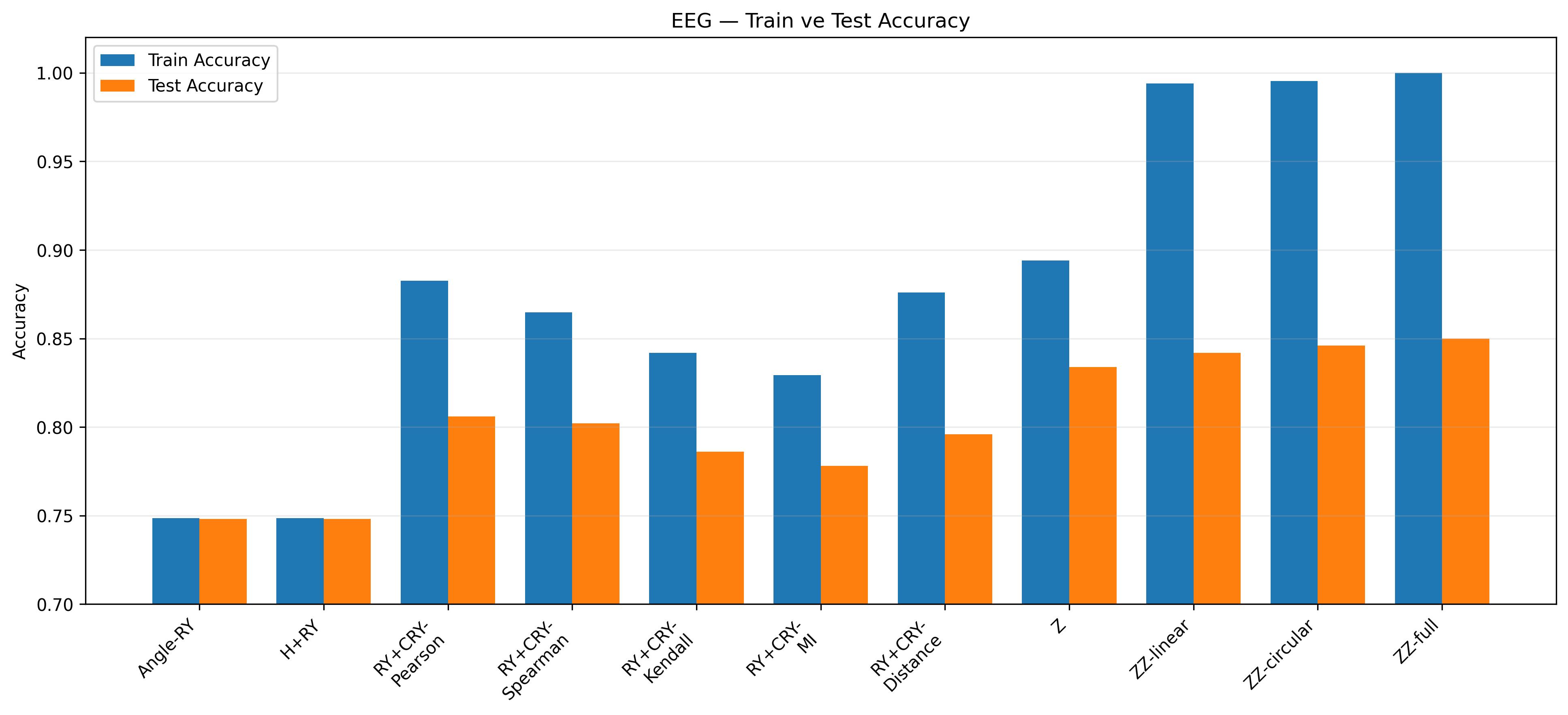}
\caption{Comparison of training and test accuracy across encoding strategies
for the EEG Eye State Classification Dataset.}
\label{fig:eeg_accuracy}
\end{figure}

Table~\ref{tab:eeg_performance} and Fig.~\ref{fig:eeg_accuracy} show that
the choice of quantum feature encoding strategy has a clear effect on QSVM
performance for the EEG dataset. The lowest test accuracies were obtained
with Angle-RY and H+RY, both achieving a test accuracy of 0.748. The
identical classification results obtained with H+RY and Angle-RY indicate
that, under the present experimental conditions, applying Hadamard gates
before the RY rotations did not provide an additional performance benefit.

Incorporating statistical relationships between features into the encoding
through CRY gates generally resulted in higher test performance than
Angle-RY. Among the RY+CRY methods, the highest test accuracy was obtained
with Pearson-based encoding at 0.806, followed by Spearman (0.802),
Distance Correlation (0.796), Kendall's Tau (0.786), and MI (0.778).

The absolute increase in test accuracy achieved by Pearson-based RY+CRY
relative to Angle-RY was

\begin{equation}
0.806-0.748=0.058,
\end{equation}

corresponding to an improvement of approximately 5.8 percentage points.
Thus, transferring inter-feature relationships into controlled rotation
angles can produce a more discriminative quantum feature space than
independent RY angle encoding.

Nevertheless, the highest test accuracies were observed for Z feature Map
and, in particular, the ZZ based feature maps. Z feature Map achieved a test
accuracy of 0.834, whereas ZZ linear, ZZ circular, and ZZ full achieved
0.842, 0.846, and 0.850, respectively. Therefore, the highest test accuracy
for the EEG dataset was obtained with ZZ full.


\begin{figure}[!htbp]
\centering
\includegraphics[width=0.95\columnwidth]
{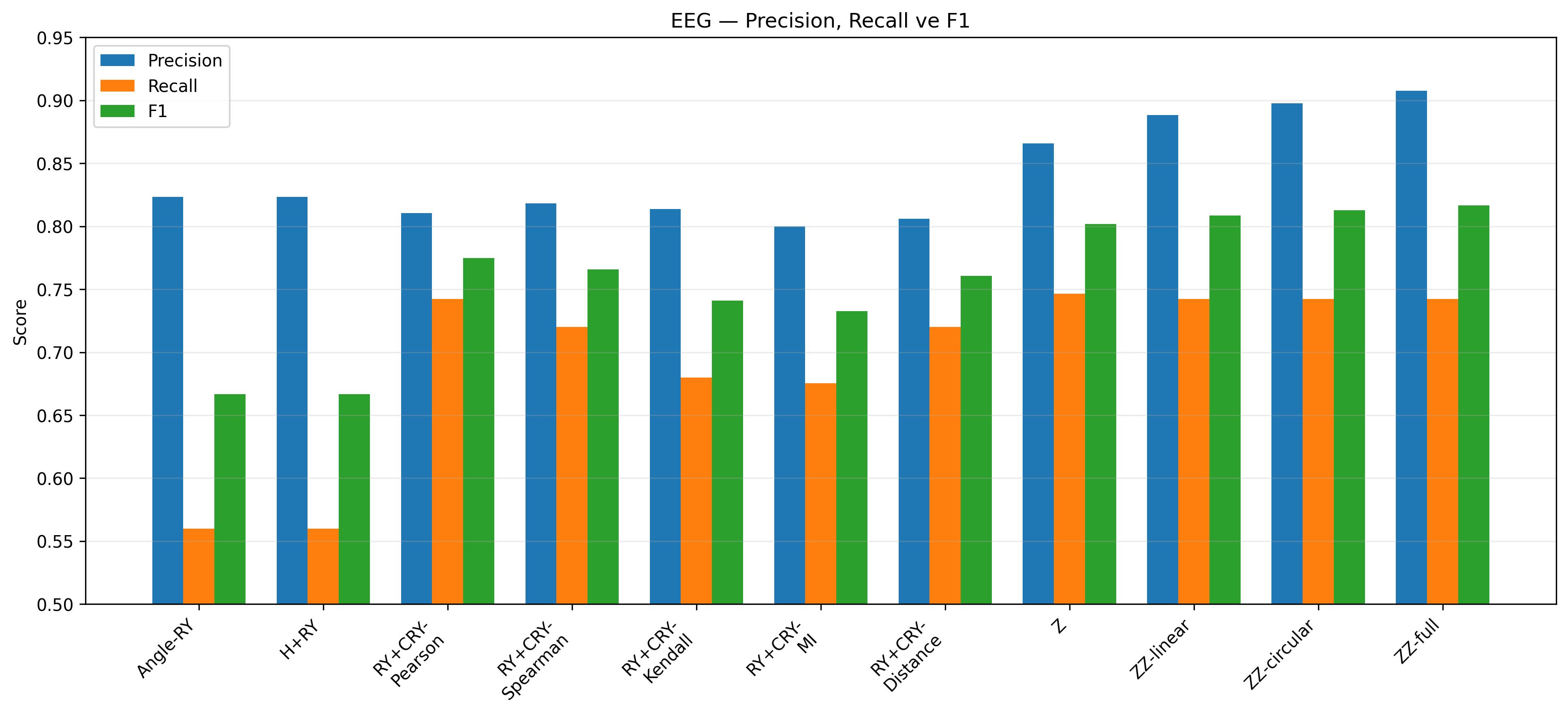}
\caption{Comparison of test Precision, Recall, and F1 scores for the EEG
Eye State Classification Dataset.}
\label{fig:eeg_prf}
\end{figure}

Figure~\ref{fig:eeg_prf} provides a more detailed view of the accuracy
results in terms of Precision, Recall, and F1. The F1 score increased from
0.667 for Angle-RY to 0.775 for Pearson-based RY+CRY. Z feature Map achieved
an F1 score of 0.802, whereas ZZ linear, ZZ circular, and ZZ full achieved
0.809, 0.813, and 0.817, respectively.

The highest Precision value, 0.908, was obtained with ZZ full.
ZZ circular and ZZ linear achieved Precision values of 0.898 and 0.888,
respectively. In contrast, the Recall values of all three ZZ encodings
remained at approximately 0.742. This result indicates that the increases
in F1 and Accuracy observed from ZZ linear to ZZ full were primarily
associated with improvements in Precision rather than Recall.

The Recall value of Pearson-based RY+CRY, 0.742, is also noteworthy, as it
is very close to those obtained with the ZZ based methods. However, its
lower Precision resulted in a lower F1 score than those of the ZZ methods.


\begin{figure}[!htbp]
\centering
\includegraphics[width=0.90\columnwidth]{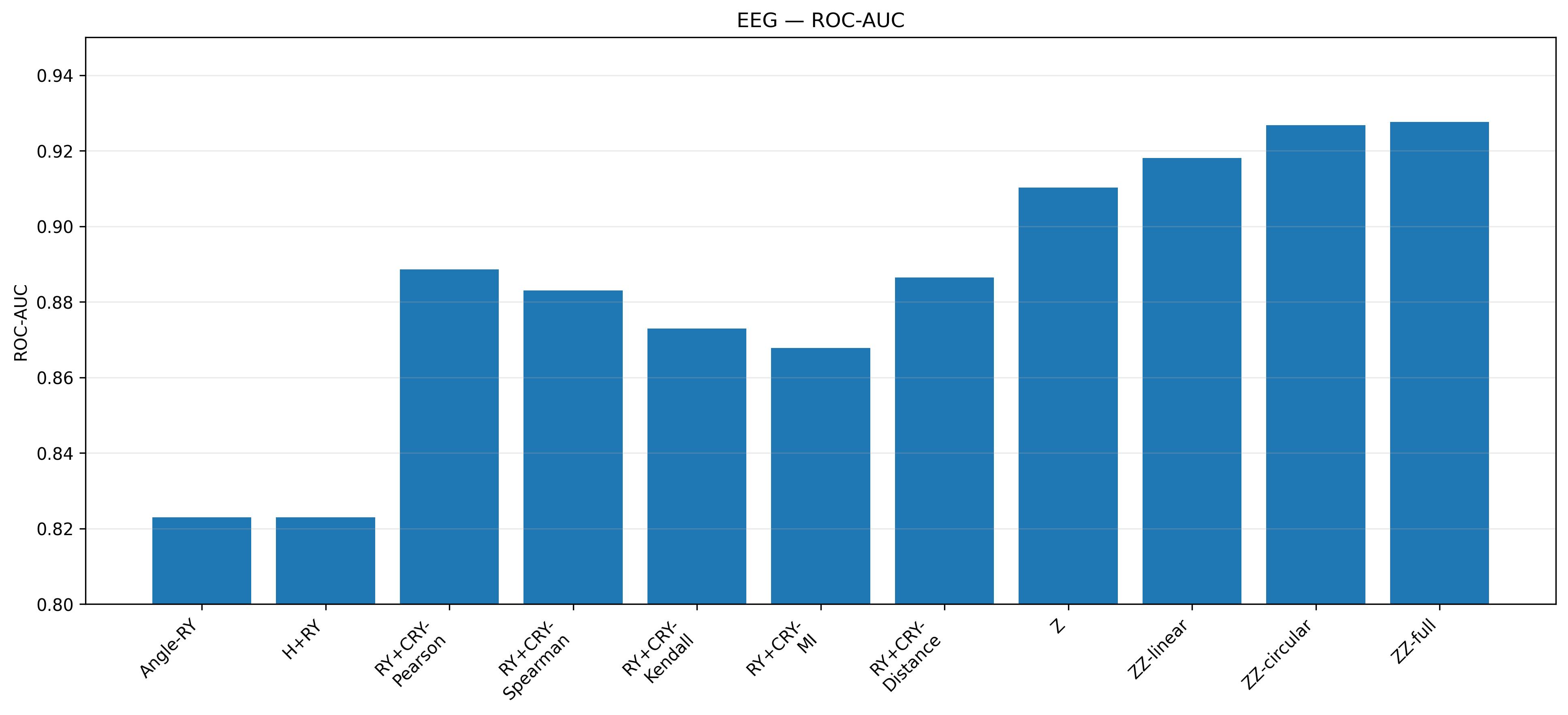}
\caption{Comparison of test ROC-AUC values across encoding strategies for
the EEG Eye State Classification Dataset.}
\label{fig:eeg_auc}
\end{figure}

The ROC-AUC results exhibited a similar trend
(Fig.~\ref{fig:eeg_auc}). The ROC-AUC value of 0.823 obtained with
Angle-RY and H+RY increased to 0.889 with Pearson-based RY+CRY and to
0.910 with Z feature Map. ZZ linear, ZZ circular, and ZZ full achieved
ROC-AUC values of 0.918, 0.927, and 0.928, respectively.

The difference in ROC-AUC between ZZ circular and ZZ full was only 0.001.
Similarly, their test accuracies differed by only 0.004. Therefore, the
additional contribution of the full connectivity structure over the
circular structure to class discrimination was limited. The significance
of this finding becomes more apparent when considered together with the
circuit complexity results.

\subsubsection{Generalization Behavior and Log-Loss}

\begin{table*}[!htbp]
\centering
\caption{Training--test performance gaps and Log-Loss values for the EEG
Eye State Classification Dataset.}
\label{tab:eeg_generalization}

\footnotesize
\begin{adjustbox}{max width=\textwidth}
\begin{tabular}{lcccccccc}
\toprule
\textbf{Encoding} &
\textbf{Prec. Gap} &
\textbf{Recall Gap} &
\textbf{F1 Gap} &
\textbf{AUC Gap} &
\textbf{Train F1} &
\textbf{Test F1} &
\textbf{Train Log-Loss} &
\textbf{Test Log-Loss} \\
\midrule
Angle-RY          & 0.022 & -0.022 & -0.009 & 0.029 & 0.658 & 0.667 & 0.479 & 0.513 \\
H+RY              & 0.022 & -0.022 & -0.009 & 0.029 & 0.658 & 0.667 & 0.479 & 0.513 \\
RY+CRY-Pearson    & 0.101 & 0.075 & 0.087 & 0.059 & 0.862 & 0.775 & 0.297 & 0.416 \\
RY+CRY-Spearman   & 0.090 & 0.057 & 0.072 & 0.054 & 0.837 & 0.766 & 0.323 & 0.424 \\
RY+CRY-Kendall    & 0.083 & 0.053 & 0.065 & 0.051 & 0.806 & 0.741 & 0.354 & 0.440 \\
RY+CRY-MI         & 0.091 & 0.030 & 0.055 & 0.047 & 0.788 & 0.733 & 0.375 & 0.449 \\
RY+CRY-Distance   & 0.102 & 0.085 & 0.093 & 0.057 & 0.854 & 0.761 & 0.307 & 0.420 \\
Z                 & 0.068 & 0.075 & 0.072 & 0.048 & 0.874 & 0.802 & 0.262 & 0.373 \\
ZZ linear         & 0.112 & 0.244 & 0.185 & 0.082 & 0.993 & 0.809 & 0.030 & 0.361 \\
ZZ circular       & 0.101 & 0.249 & 0.182 & 0.073 & 0.995 & 0.813 & 0.021 & 0.345 \\
ZZ full           & 0.092 & 0.258 & 0.183 & 0.072 & 1.000 & 0.817 & 0.001 & 0.339 \\
\bottomrule
\end{tabular}
\end{adjustbox}
\end{table*}


\begin{figure}[!htbp]
\centering
\includegraphics[width=0.90\columnwidth]{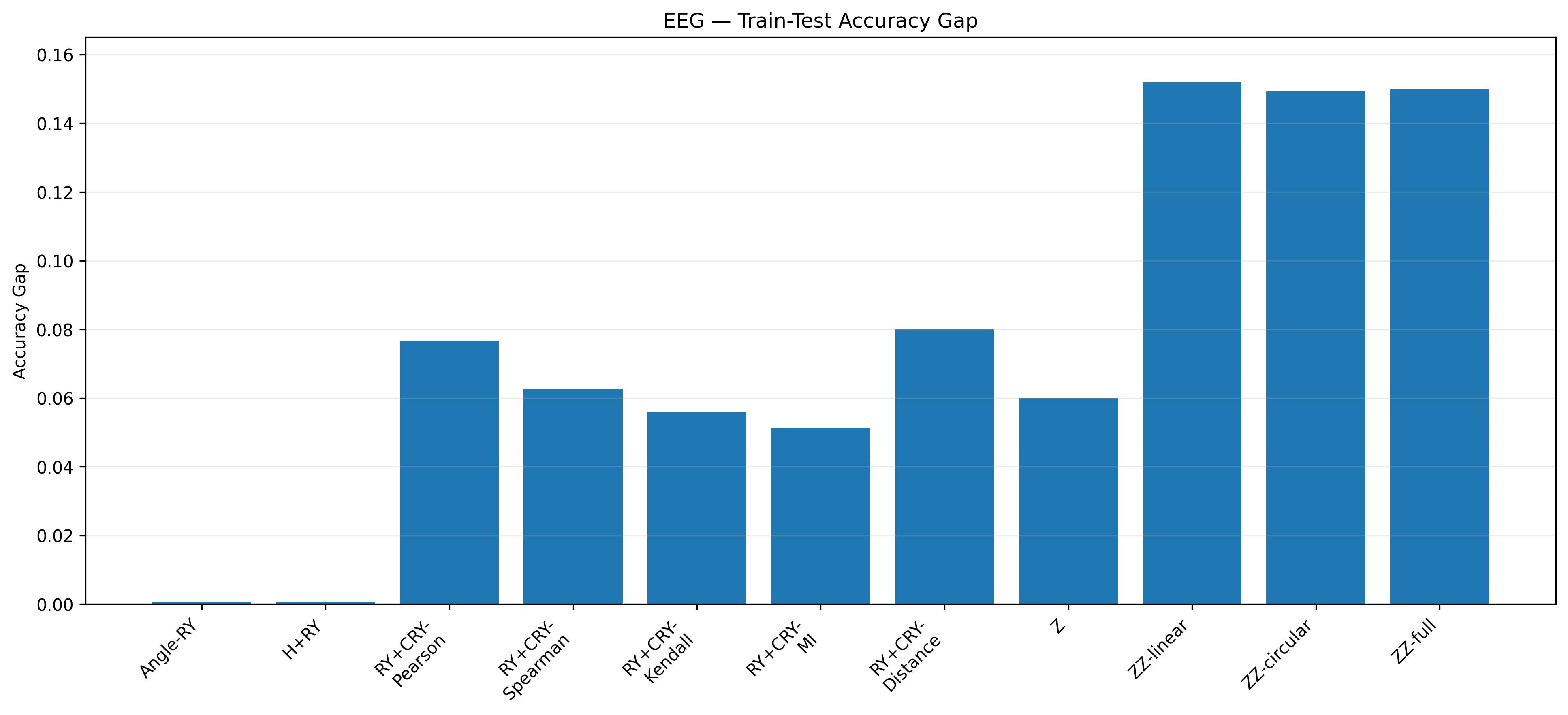}
\caption{Accuracy Gap between training and test accuracy for the EEG Eye
State Classification Dataset.}
\label{fig:eeg_accuracy_gap}
\end{figure}

Figure~\ref{fig:eeg_accuracy_gap} illustrates how the improvement in
classification performance is reflected in model generalization. The
Accuracy Gap for Angle-RY and H+RY was approximately 0.001. However, since
both methods achieved a test accuracy of only 0.748, the small gap should
not by itself be interpreted as evidence of superior generalization.
Instead, it indicates that training and test performance remained similar
but at a relatively low level.

For the RY+CRY methods, the Accuracy Gap ranged from 0.051 to 0.080.
Pearson-based RY+CRY achieved the highest test accuracy within this group,
0.806, with an Accuracy Gap of 0.077. Spearman-based RY+CRY achieved a
similar test accuracy of 0.802 with a smaller gap of 0.063. Therefore,
Spearman-based RY+CRY provides a noteworthy balance between performance and
generalization within this group.

Z feature Map achieved a test accuracy of 0.834 with an Accuracy Gap of
0.060. This indicates that the method maintained relatively high test
performance with a smaller training--test difference than the ZZ based
methods.

In contrast, the Accuracy Gap values for ZZ linear, ZZ circular, and
ZZ full were 0.152, 0.149, and 0.150, respectively. Although the training
accuracy of ZZ full reached 1.000, its test accuracy was 0.850, indicating
a very strong fit to the training data. Thus, the high test performance of
the ZZ-based feature maps was accompanied by a more pronounced
training--test gap.


\begin{figure}[!htbp]
\centering
\includegraphics[width=0.90\columnwidth]{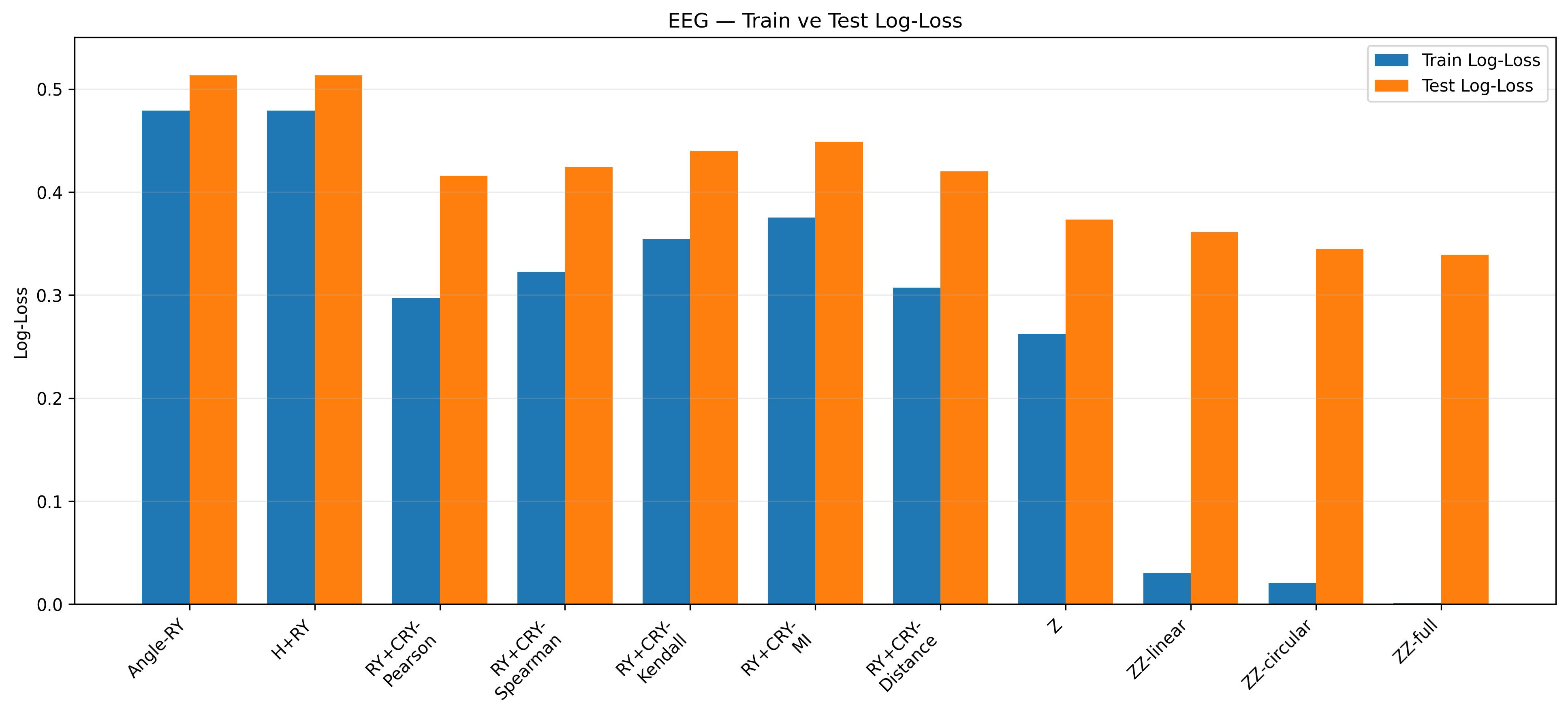}
\caption{Comparison of training and test Log-Loss values for the EEG Eye
State Classification Dataset.}
\label{fig:eeg_logloss}
\end{figure}

The Log-Loss results support these observations
(Fig.~\ref{fig:eeg_logloss}). For Angle-RY, the Train and Test Log-Loss
values were 0.479 and 0.513, respectively, whereas for Pearson-based
RY+CRY they decreased to 0.297 and 0.416. Z feature Map achieved a Train
Log-Loss of 0.262 and a Test Log-Loss of 0.373.

For the ZZ-based feature maps, the training loss decreased much more
sharply. The Train Log-Loss values for ZZ linear, ZZ circular, and ZZ full
were 0.030, 0.021, and 0.001, respectively, whereas the corresponding Test
Log-Loss values remained at 0.361, 0.345, and 0.339.

In particular, the training loss of 0.001 and training accuracy of 1.000
obtained with ZZ full indicate an extremely strong fit to the training
data. However, it is also important that the same method achieved the
lowest Test Log-Loss among all methods, at 0.339. Therefore, the result of
ZZ full should not be interpreted as poor test performance, but rather as
\textbf{strong fitting to the training data and a pronounced training--test
gap accompanied by high test performance}.

\subsubsection{Quantum Circuit Complexity}

\begin{table*}[!htbp]
\centering
\caption{Logical and decomposed circuit complexities of quantum feature
encoding strategies for the EEG Eye State Classification Dataset.}
\label{tab:eeg_circuit}

\footnotesize
\begin{adjustbox}{max width=\textwidth}
\begin{tabular}{lrrrrrrrrr}
\toprule
\textbf{Encoding} &
\textbf{H} &
\textbf{RY} &
\textbf{CRY} &
\textbf{CX} &
\textbf{Total} &
\textbf{Depth} &
\textbf{Dec. CX} &
\textbf{Dec. Gates} &
\textbf{Dec. Depth} \\
\midrule
Angle-RY          & 0  & 14 & 0  & 0   & 14  & 1  & 0   & 14  & 1 \\
H+RY              & 14 & 14 & 0  & 0   & 28  & 2  & 0   & 28  & 2 \\
RY+CRY-Pearson    & 0  & 14 & 91 & 0   & 105 & 26 & 182 & 378 & 89 \\
RY+CRY-Spearman   & 0  & 14 & 91 & 0   & 105 & 26 & 182 & 378 & 89 \\
RY+CRY-Kendall    & 0  & 14 & 91 & 0   & 105 & 26 & 182 & 378 & 89 \\
RY+CRY-MI         & 0  & 14 & 91 & 0   & 105 & 26 & 182 & 378 & 89 \\
RY+CRY-Distance   & 0  & 14 & 91 & 0   & 105 & 26 & 182 & 378 & 89 \\
Z                 & 14 & 0  & 0  & 0   & 28  & 2  & 0   & 28  & 2 \\
ZZ linear         & 14 & 0  & 0  & 26  & 67  & 41 & 26  & 67  & 41 \\
ZZ circular       & 14 & 0  & 0  & 28  & 70  & 44 & 28  & 70  & 44 \\
ZZ full           & 14 & 0  & 0  & 182 & 301 & 77 & 182 & 301 & 77 \\
\bottomrule
\end{tabular}
\end{adjustbox}
\end{table*}


\begin{figure*}[!htbp]
\centering
\begin{subfigure}{0.48\textwidth}
\centering
\includegraphics[width=\linewidth]{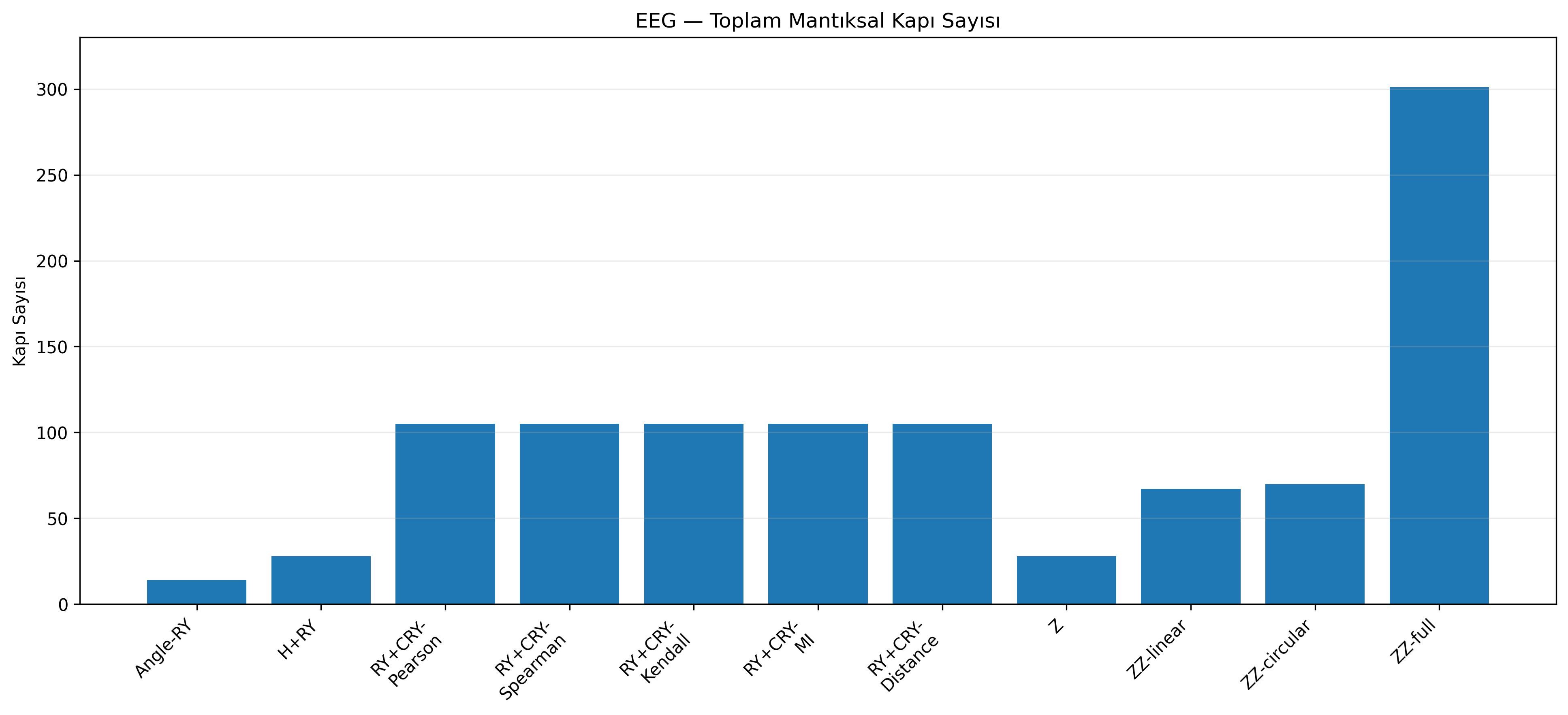}
\caption{Total number of logical gates.}
\label{fig:eeg_total_gates}
\end{subfigure}
\hfill
\begin{subfigure}{0.48\textwidth}
\centering
\includegraphics[width=\linewidth]{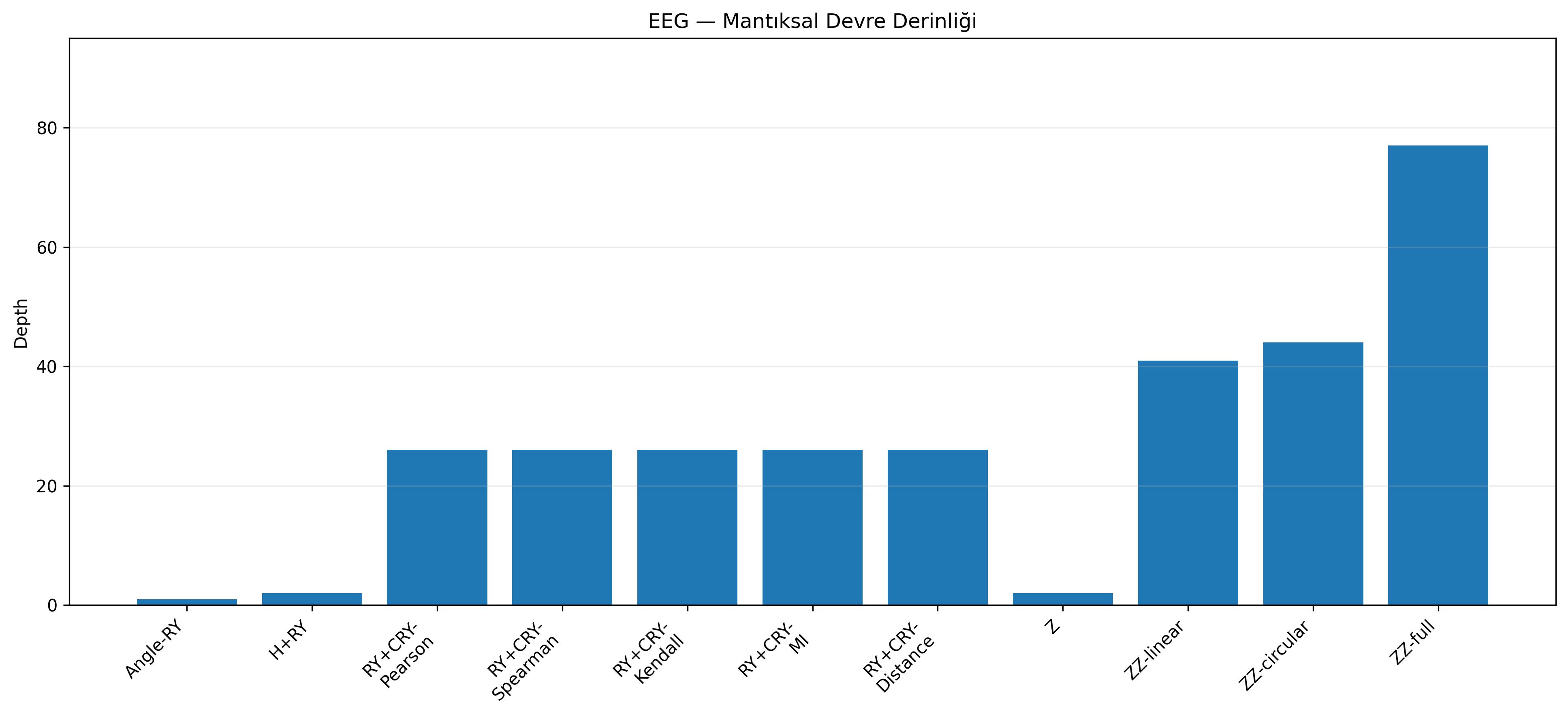}
\caption{Logical circuit depth.}
\label{fig:eeg_depth}
\end{subfigure}
\caption{Comparison of total gate count and circuit depth for the EEG Eye
State Classification Dataset.}
\label{fig:eeg_gate_depth}
\end{figure*}

Table~\ref{tab:eeg_circuit} and Fig.~\ref{fig:eeg_gate_depth} reveal
substantial differences in circuit complexity among the encoding
strategies. Angle-RY produced the simplest circuit structure, with only
14 RY gates and a circuit depth of 1. Although H+RY contained 28 logical
gates and had a depth of 2, it produced exactly the same classification
performance as Angle-RY. Thus, adding the Hadamard layer increased circuit
cost without providing a performance gain in the present experiment.

The RY+CRY methods used 91 CRY operations in addition to 14 RY gates.
For 14 features, the number of unique feature pairs is

\begin{equation}
\binom{14}{2}=91.
\end{equation}

Therefore, the employed structure connects all feature pairs through
controlled rotations. The fact that the five dependency-based methods have
identical gate counts and circuit depths indicates that their performance
differences arise not from circuit size but from the statistical dependency
measure used to determine the CRY rotation angles.


\begin{figure}[!htbp]
\centering
\includegraphics[width=0.95\columnwidth]
{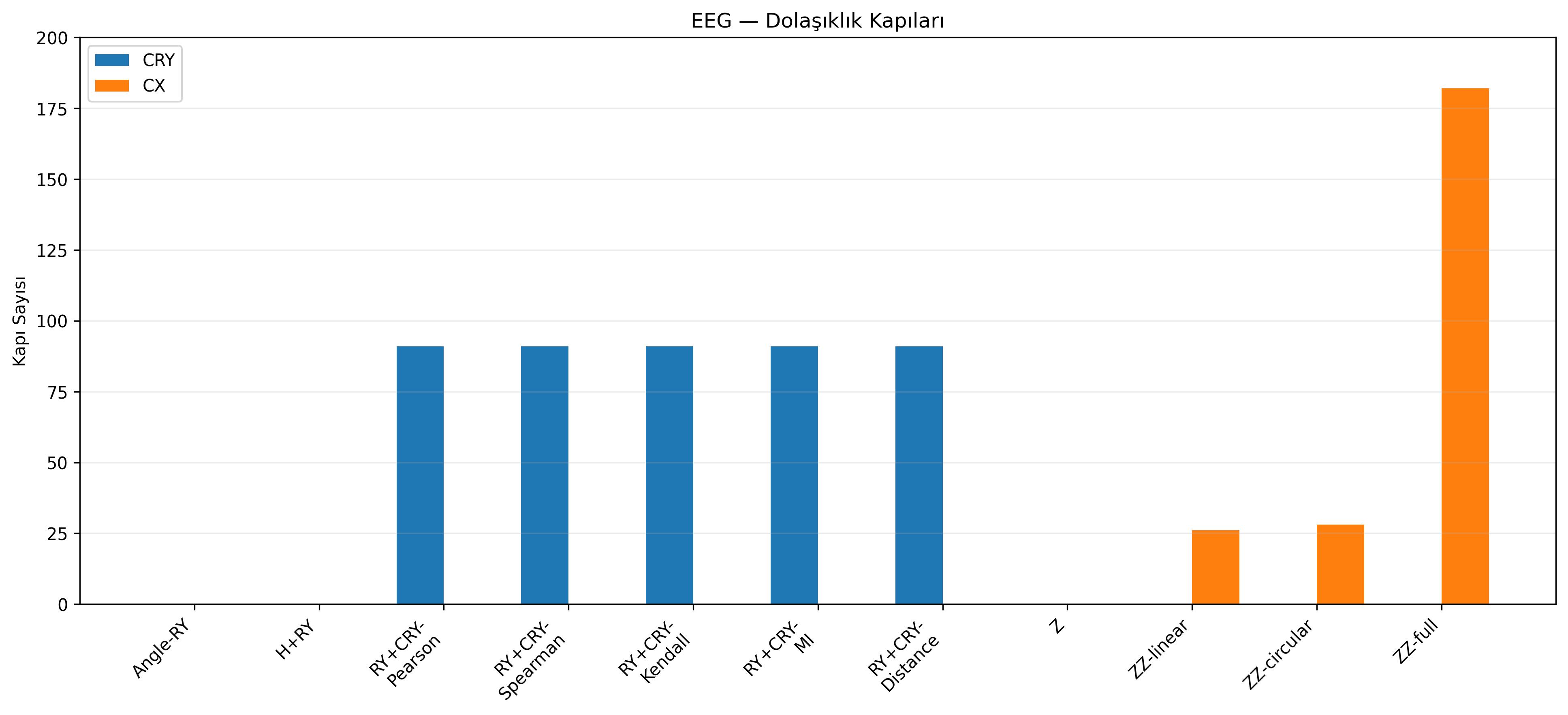}
\caption{Comparison of CRY and CX entangling gates for the EEG Eye State
Classification Dataset.}
\label{fig:eeg_entangling}
\end{figure}

As shown in Fig.~\ref{fig:eeg_entangling}, the RY+CRY structures contain
91 CRY gates at the logical level. In contrast, ZZ full contains 182 CX
gates, resulting in a substantially denser connectivity structure than
ZZ linear and ZZ circular.

This result indicates that the higher classification performance of
ZZ full was obtained together with substantially denser feature
interactions. However, the magnitude of the performance improvement was
not proportional to the increase in circuit cost.


\begin{figure}[!htbp]
\centering
\includegraphics[width=0.95\columnwidth]
{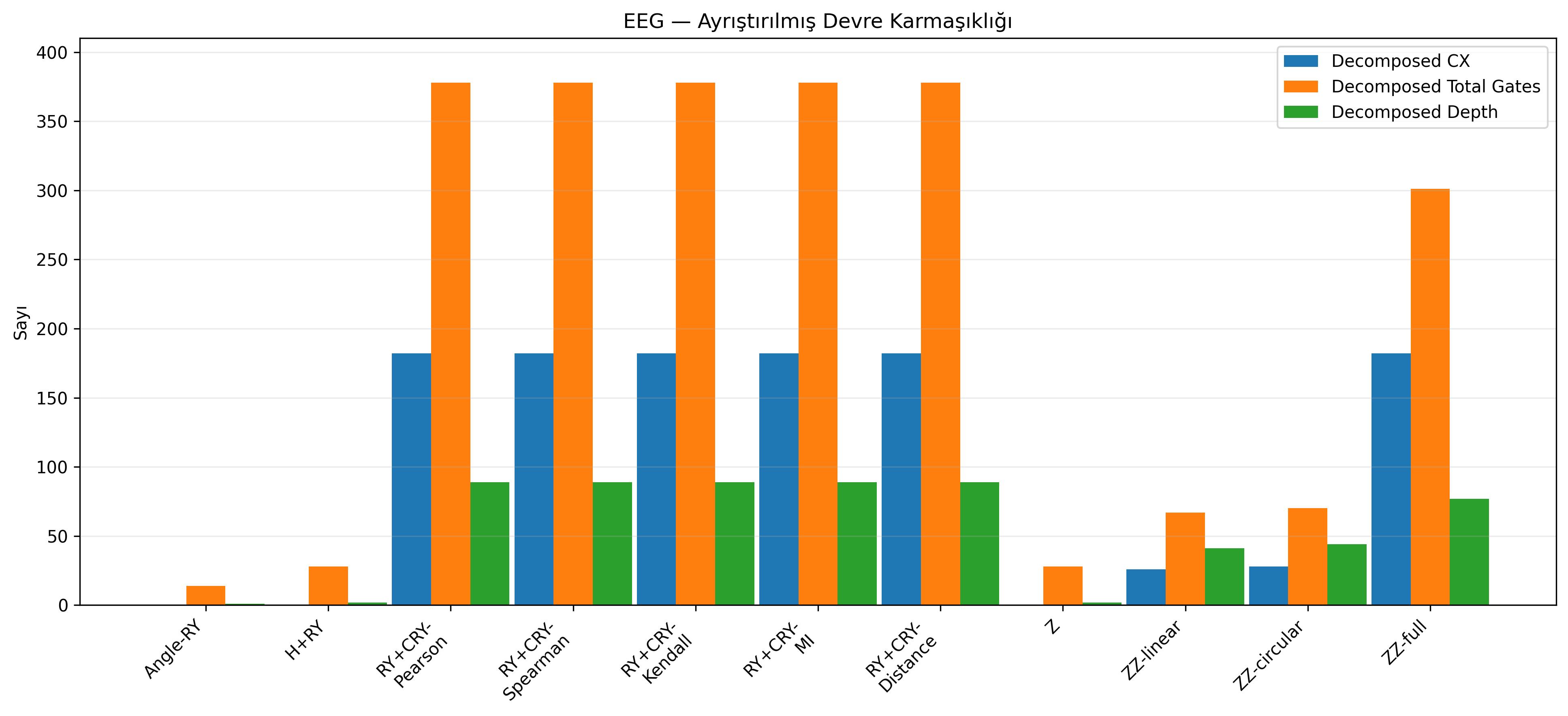}
\caption{CX gate count, total gate count, and circuit depth of the
decomposed circuits for the EEG Eye State Classification Dataset.}
\label{fig:eeg_decomposed}
\end{figure}

From a hardware implementation perspective, evaluating the decomposed
circuit structure is more informative than considering only the number of
logical gates. As shown in Fig.~\ref{fig:eeg_decomposed}, after
decomposition, the RY+CRY circuits contain 378 total gates, 182 CX gates,
and a circuit depth of 89.

Therefore, the performance improvement achieved by RY+CRY over Angle-RY
comes at the cost of a substantial increase in entangling gates and circuit
depth. In particular, encoding all 91 feature pairs is the primary source
of circuit complexity in the current structure.

Within the ZZ family, ZZ linear, ZZ circular, and ZZ full contain 67, 70,
and 301 total logical gates, respectively. Moving from ZZ circular to
ZZ full increases the total gate count by a factor of

\begin{equation}
\frac{301}{70}\approx4.30.
\end{equation}

In contrast, the increase in test accuracy is only

\begin{equation}
0.850-0.846=0.004,
\end{equation}

corresponding to 0.4 percentage points. Similarly, ROC-AUC increases only
from 0.927 to 0.928. Therefore, for the EEG dataset, the additional
classification gain provided by full ZZ connectivity over the circular
structure is relatively limited compared with the corresponding increase
in circuit complexity.

\subsubsection{Computational Cost}

\begin{table*}[!htbp]
\centering
\caption{Computation times for the EEG Eye State Classification Dataset
(seconds).}
\label{tab:eeg_time}

\footnotesize
\begin{adjustbox}{max width=\textwidth}
\begin{tabular}{lrrr}
\toprule
\textbf{Encoding} &
\textbf{state vector Time} &
\textbf{Kernel Time} &
\textbf{Training Time} \\
\midrule
Angle-RY          & 5.578   & 5.037 & 0.066 \\
H+RY              & 10.348  & 4.967 & 0.078 \\
RY+CRY-Pearson    & 56.071  & 4.949 & 0.087 \\
RY+CRY-Spearman   & 56.015  & 5.431 & 0.113 \\
RY+CRY-Kendall    & 56.845  & 5.042 & 0.083 \\
RY+CRY-MI         & 57.408  & 5.023 & 0.079 \\
RY+CRY-Distance   & 56.771  & 5.043 & 0.084 \\
Z                 & 9.964   & 5.001 & 0.087 \\
ZZ linear         & 25.799  & 4.972 & 0.168 \\
ZZ circular       & 27.975  & 5.013 & 0.182 \\
ZZ full           & 126.616 & 5.029 & 0.148 \\
\bottomrule
\end{tabular}
\end{adjustbox}
\end{table*}


\begin{figure}[!htbp]
\centering
\includegraphics[width=0.90\columnwidth]
{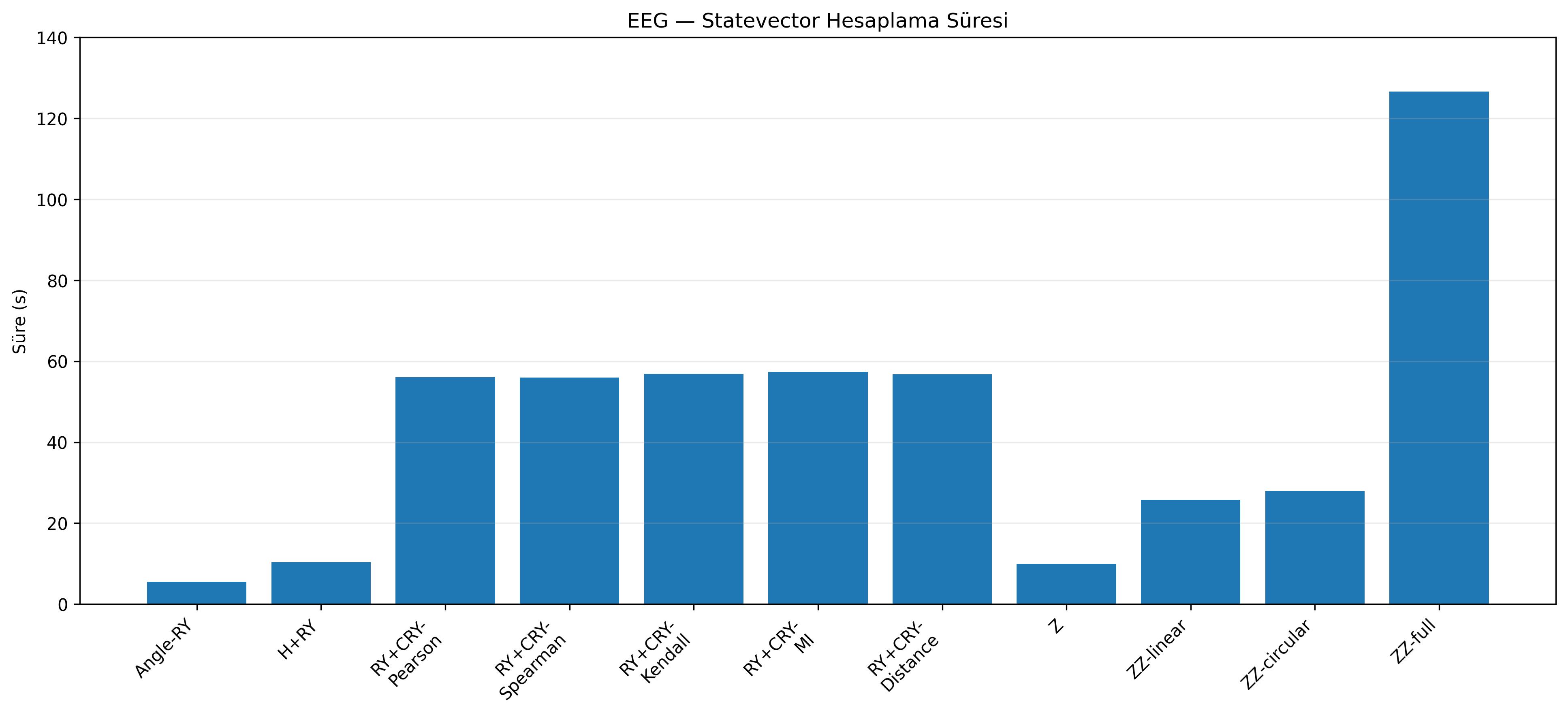}
\caption{state vector computation times for the EEG Eye State
Classification Dataset.}
\label{fig:eeg_state vector_time}
\end{figure}

Table~\ref{tab:eeg_time} and Fig.~\ref{fig:eeg_state vector_time} show that
differences in circuit complexity are also reflected in computational time.
Angle-RY had the lowest state vector computation time at approximately
5.58 seconds. The corresponding times for H+RY and Z feature Map were
approximately 10.35 and 9.96 seconds, respectively.

The RY+CRY methods required approximately 56--57 seconds for state vector
computation. The similar execution times of the five methods are consistent
with their identical circuit topology and equal numbers of RY and CRY gates.

The state vector computation times for ZZ linear and ZZ circular were
approximately 25.8 and 28.0 seconds, respectively, whereas the time for
ZZ full increased to 126.6 seconds. Thus, ZZ full required approximately

\begin{equation}
\frac{126.6}{5.58}\approx22.7
\end{equation}

times the state vector computation cost of Angle-RY.

An even more informative comparison can be made between ZZ circular and
ZZ full. ZZ full required approximately 126.6 seconds, whereas ZZ circular
required approximately 28.0 seconds. Thus, full connectivity required
approximately 4.5 times more state vector computation time while providing
an additional test accuracy of only 0.004.


\begin{figure}[!htbp]
\centering
\includegraphics[width=0.90\columnwidth]{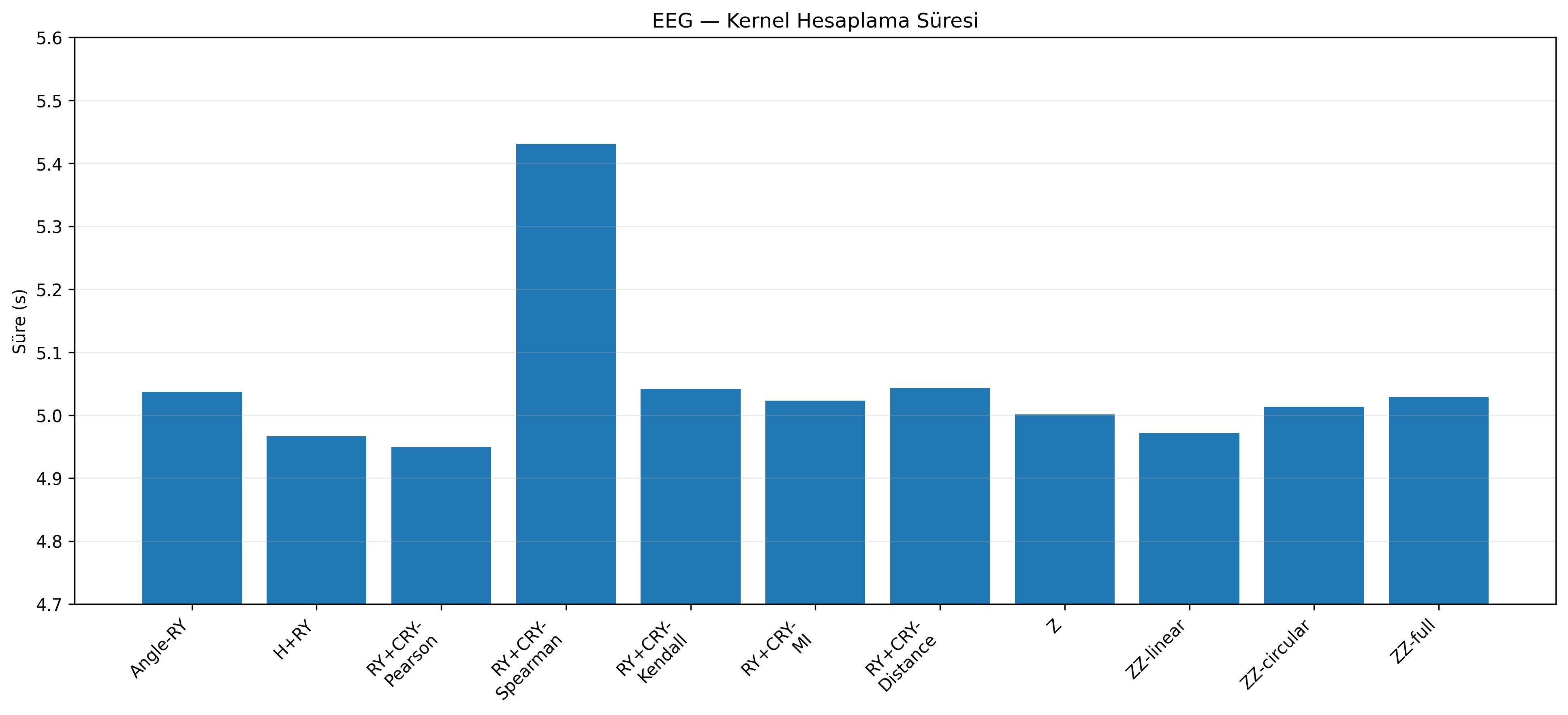}
\caption{Kernel computation times for the EEG Eye State Classification
Dataset.}
\label{fig:eeg_kernel_time}
\end{figure}

In contrast, Fig.~\ref{fig:eeg_kernel_time} shows that Kernel Time remained
largely constant across the encoding methods, ranging from approximately
4.95 to 5.43 seconds. Despite the substantial differences in circuit
complexity, the similar kernel computation times indicate that the main
differences in total computational cost did not originate from the kernel
calculation stage.


\begin{figure}[!htbp]
\centering
\includegraphics[width=0.90\columnwidth]{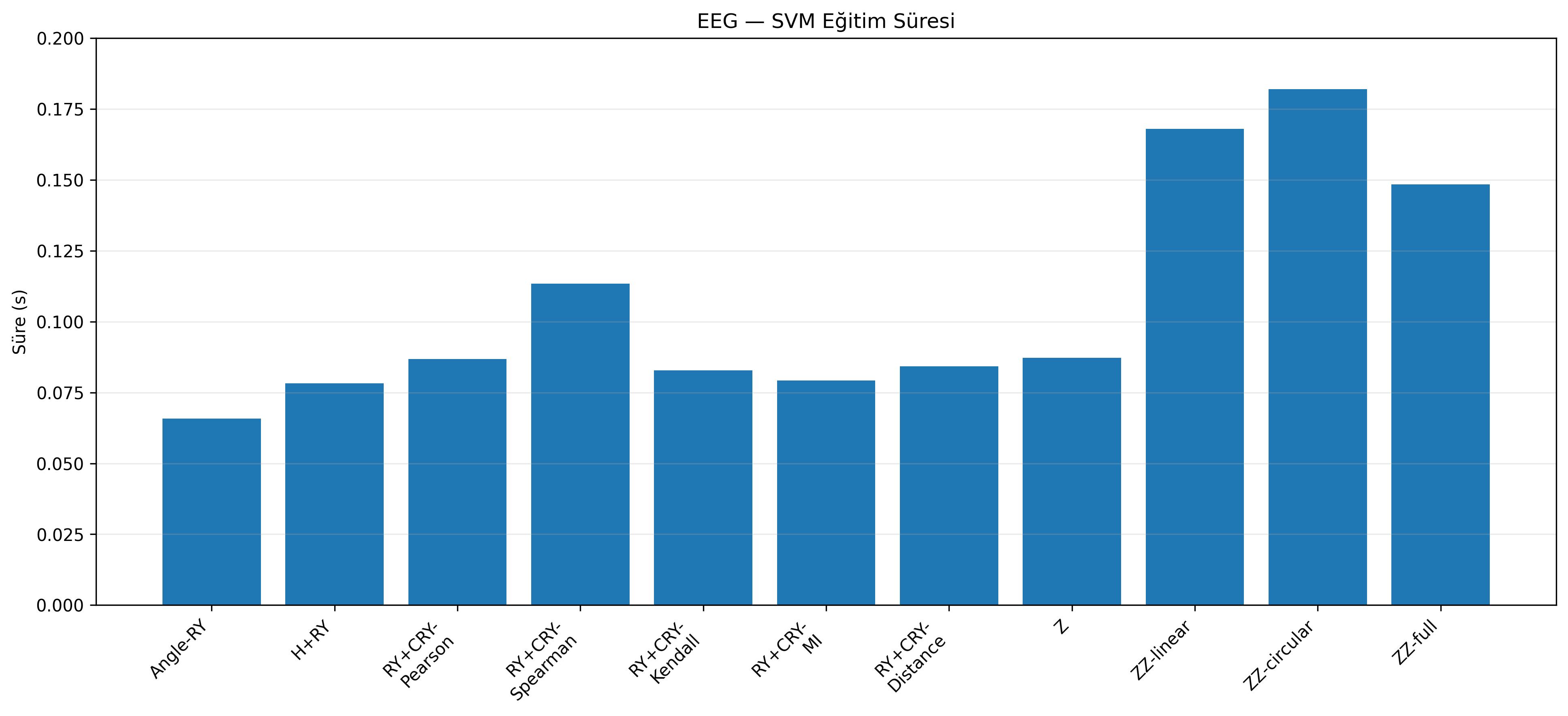}
\caption{QSVM training times for the EEG Eye State Classification Dataset.}
\label{fig:eeg_training_time}
\end{figure}

Similarly, QSVM training times remained relatively low for all methods
(Fig.~\ref{fig:eeg_training_time}), ranging from approximately 0.066 to
0.182 seconds. Although the highest value, 0.182 seconds, was observed for
ZZ circular, this time is negligible compared with the computational cost
of state vector generation.

Taken together, these results indicate that the differences in computational
cost among the encoding methods arise primarily from quantum-state
preparation and state vector computation rather than from classical SVM
optimization.

\subsubsection{Overall Evaluation of the EEG Dataset}

The experiments on the EEG dataset demonstrate that the quantum feature
encoding strategy can substantially affect QSVM classification performance.
ZZ full achieved the highest test performance, with a Test Accuracy of
0.850, an F1 score of 0.817, and a ROC-AUC of 0.928. However, the same
method also exhibited a training accuracy of 1.000, an Accuracy Gap of
approximately 0.150, 301 logical gates, a circuit depth of 77, and a
state vector computation time of 126.6 seconds. Thus, the maximum test
performance was achieved at the cost of high circuit and computational
complexity, together with a pronounced training--test gap.

ZZ circular achieved nearly equivalent test performance, with a Test
Accuracy of 0.846 and a ROC-AUC of 0.927, while requiring only 70 logical
gates and approximately 28 seconds of state vector computation. Therefore,
for the EEG dataset, ZZ circular can be regarded as a more efficient
alternative to ZZ full in terms of the trade-off between classification
performance and circuit complexity.

Z feature Map is particularly notable for its substantially lower circuit
complexity. With only 28 logical gates and a circuit depth of 2, it achieved
a Test Accuracy of 0.834, an F1 score of 0.802, and a ROC-AUC of 0.910.
Furthermore, its Accuracy Gap of 0.060 indicates that Z feature Map provides
a strong balance among classification performance, generalization, and
circuit cost.

The RY+CRY results demonstrate that incorporating statistical relationships
between features into quantum encoding can provide a meaningful performance
improvement over Angle-RY. With Pearson-based RY+CRY, test accuracy
increased from 0.748 to 0.806. However, representing all 91 feature pairs
with CRY gates resulted in 182 CX gates and a decomposed circuit depth of
89.

This observation suggests that, although statistical dependency information
can be beneficial, transferring all pairwise relationships into the quantum
circuit may not be optimal from an efficiency perspective. Sparser
structures in which only the strongest or most meaningful feature
relationships are encoded using CRY gates may represent a promising
direction for reducing circuit depth and the number of entangling gates
while preserving the performance advantage of the RY+CRY approach.

Overall, the EEG experiments show that more complex quantum feature maps
can generally provide higher classification performance; however, as
connectivity density increases, the marginal performance gain may become
increasingly limited while circuit complexity and computational cost rise
rapidly. Therefore, the evaluation of quantum feature encoding strategies
should not focus exclusively on maximum classification accuracy, but rather
on the overall balance among classification performance, generalization
behavior, and quantum circuit cost.

\subsection{Heart Failure Prediction Dataset}
\label{subsec:heart_results}

The experiments conducted on the Heart Failure Prediction Dataset compared
Angle-RY, H+RY, RY+CRY encodings based on different statistical dependency
measures, Z feature Map, and ZZ feature Map with different entanglement
structures. The evaluation was performed in terms of classification
performance, generalization behavior, quantum circuit complexity, and
computational cost.

\subsubsection{Classification Performance}

\begin{table*}[!htbp]
\centering
\caption{Classification performance of quantum feature encoding strategies
on the Heart Failure Prediction Dataset.}
\label{tab:heart_performance}

\footnotesize
\setlength{\tabcolsep}{4pt}
\renewcommand{\arraystretch}{1.10}

\begin{tabular}{lccccccc}
\toprule
\textbf{Encoding} &
\makecell{\textbf{Train}\\\textbf{Acc.}} &
\makecell{\textbf{Test}\\\textbf{Acc.}} &
\makecell{\textbf{Acc.}\\\textbf{Gap}} &
\makecell{\textbf{Test}\\\textbf{Prec.}} &
\makecell{\textbf{Test}\\\textbf{Rec.}} &
\makecell{\textbf{Test}\\\textbf{F1}} &
\makecell{\textbf{Test}\\\textbf{AUC}} \\
\midrule
Angle-RY        & 0.903 & 0.874 & 0.029 & 0.850 & 0.937 & 0.891 & 0.909 \\
H+RY            & 0.903 & 0.874 & 0.029 & 0.850 & 0.937 & 0.891 & 0.909 \\
RY+CRY-Pearson  & 0.914 & 0.883 & 0.032 & 0.873 & 0.921 & 0.897 & 0.927 \\
RY+CRY-Spearman & 0.914 & 0.878 & 0.036 & 0.867 & 0.921 & 0.893 & 0.928 \\
RY+CRY-Kendall  & 0.910 & 0.878 & 0.032 & 0.867 & 0.921 & 0.893 & 0.925 \\
RY+CRY-MI       & 0.919 & 0.874 & 0.045 & 0.866 & 0.913 & 0.889 & 0.930 \\
RY+CRY-Distance & 0.916 & 0.883 & 0.033 & 0.873 & 0.921 & 0.897 & 0.929 \\
Z               & 0.894 & 0.878 & 0.016 & 0.851 & 0.945 & 0.896 & 0.919 \\
ZZ linear       & 0.991 & 0.804 & 0.187 & 0.781 & 0.898 & 0.835 & 0.887 \\
ZZ circular     & 0.994 & 0.757 & 0.238 & 0.726 & 0.898 & 0.803 & 0.856 \\
ZZ full         & 1.000 & 0.843 & 0.157 & 0.810 & 0.937 & 0.869 & 0.911 \\
\bottomrule
\end{tabular}
\end{table*}


\begin{figure}[!htbp]
\centering
\includegraphics[width=0.95\columnwidth]{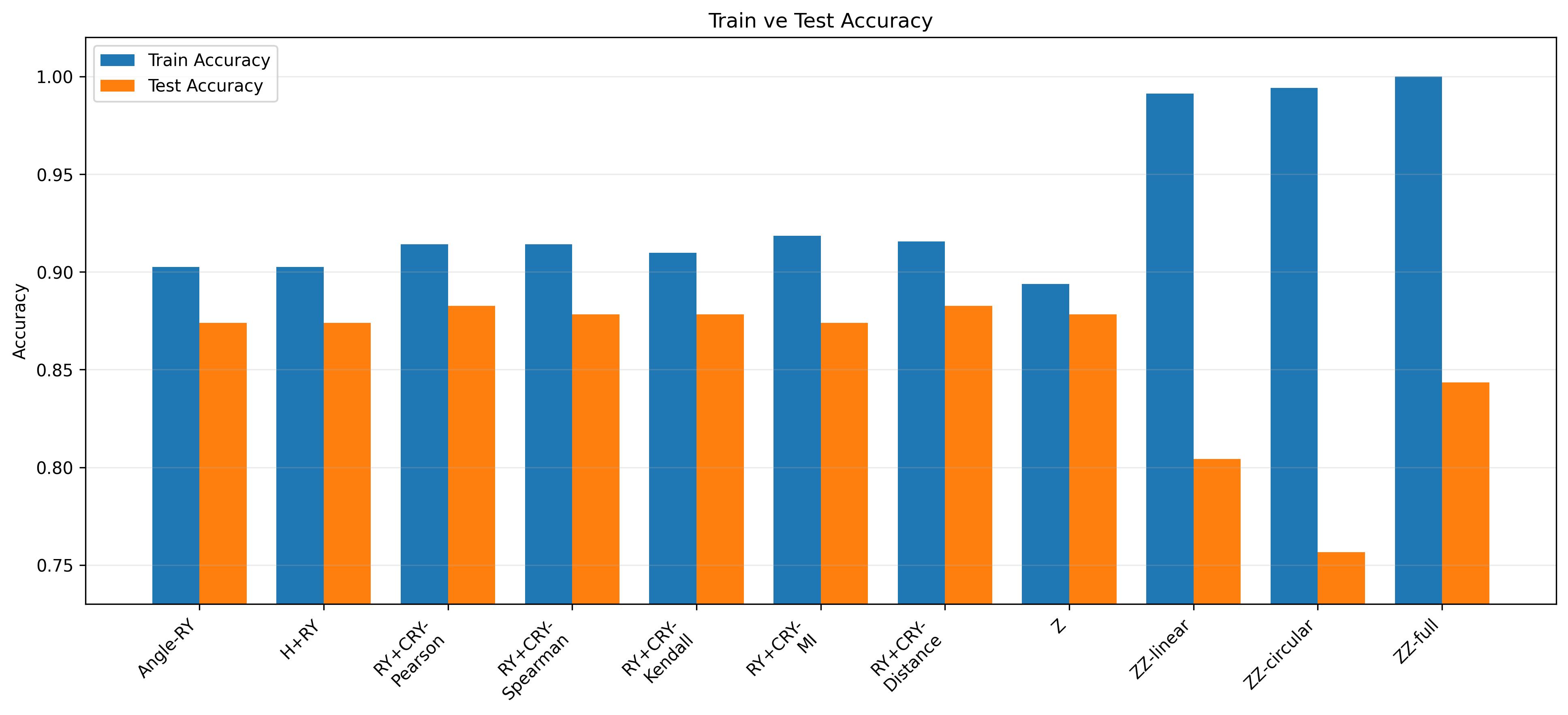}
\caption{Comparison of training and test accuracy across encoding strategies
for the Heart Failure Prediction Dataset.}
\label{fig:heart_accuracy}
\end{figure}

Table~\ref{tab:heart_performance} and Fig.~\ref{fig:heart_accuracy} show
that the highest test accuracy for the Heart Failure Prediction Dataset was
obtained with the Pearson- and Distance Correlation-based RY+CRY encodings,
both achieving 0.883. These methods were followed by Spearman, Kendall, and
Z feature Map, each achieving a test accuracy of 0.878. Angle-RY and H+RY
both achieved a test accuracy of 0.874.

Compared with Angle-RY, the absolute increase in test accuracy obtained by
the Pearson- and Distance Correlation-based RY+CRY encodings was

\begin{equation}
0.883-0.874=0.009,
\end{equation}

corresponding to approximately 0.9 percentage points. Therefore, encoding
inter-feature relationships through CRY gates improved performance over the
basic RY encoding for this dataset, although the improvement was more limited
than that observed for the EEG dataset.

It is also noteworthy that H+RY produced exactly the same training and test
performance as Angle-RY. Although the addition of the Hadamard layer
increased circuit complexity, it did not provide a measurable improvement
in classification performance in the present experiment.

A different behavior was observed for the ZZ-based feature maps. Although
the training accuracies reached 0.991 for ZZ linear, 0.994 for ZZ circular,
and 1.000 for ZZ full, their test accuracies decreased to 0.804, 0.757, and
0.843, respectively. These results indicate that denser ZZ connectivity did
not directly translate into improved test performance for the Heart Failure
Prediction Dataset.


\begin{figure}[!htbp]
\centering
\includegraphics[width=0.95\columnwidth]
{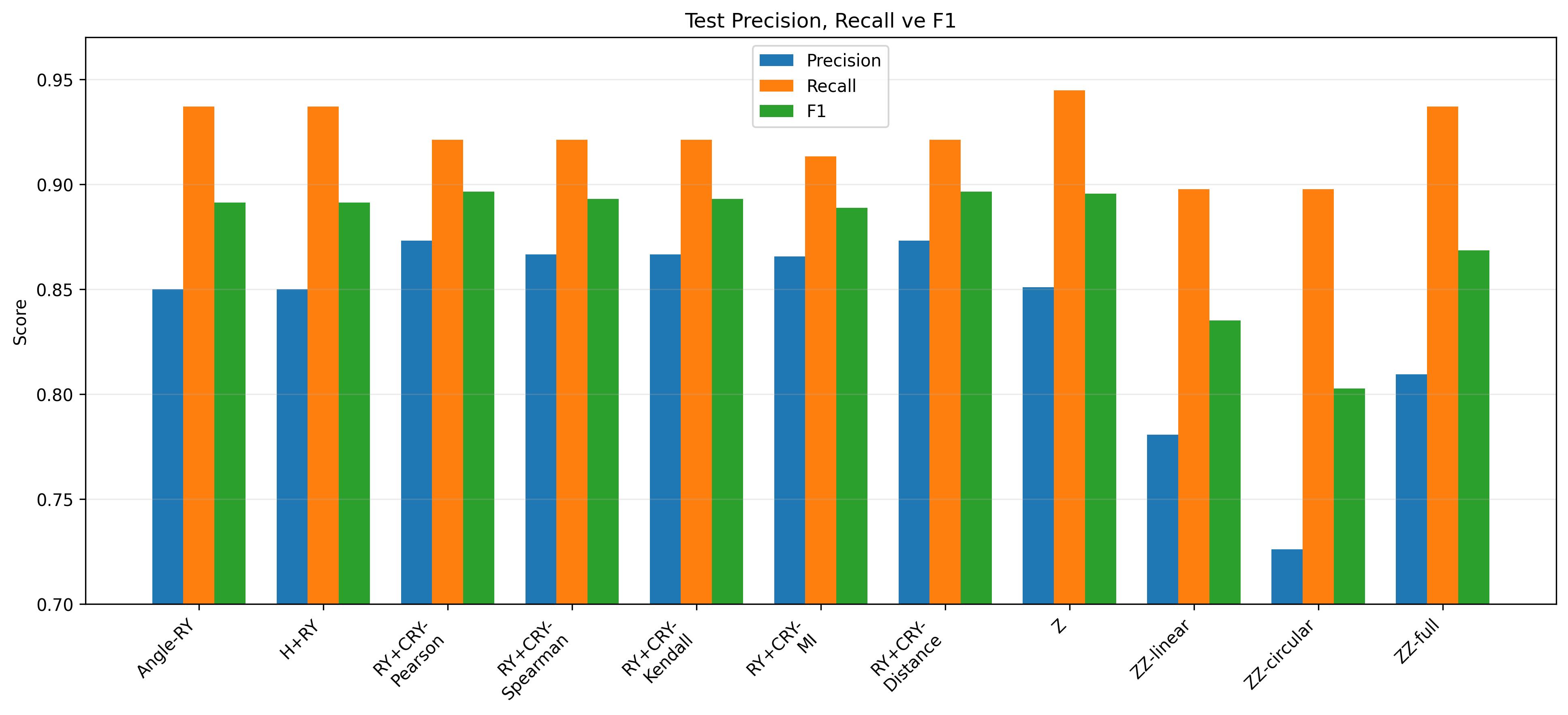}
\caption{Comparison of test Precision, Recall, and F1 scores for the Heart
Failure Prediction Dataset.}
\label{fig:heart_prf}
\end{figure}

Figure~\ref{fig:heart_prf} presents the Precision, Recall, and F1 results
together. The highest F1 score, 0.897, was obtained with the Pearson- and
Distance Correlation-based RY+CRY encodings. Z feature Map produced a very
similar result, with an F1 score of 0.896.

Z feature Map achieved a Recall of 0.945, the highest test Recall among all
methods. In contrast, the Pearson- and Distance Correlation-based RY+CRY
methods exhibited a more balanced distribution, with a Precision of 0.873
and a Recall of 0.921, resulting in an F1 score of 0.897.

The decrease in the Precision of ZZ circular to 0.726 and its F1 score to
0.803 indicates that the very high performance obtained on the training set
was not transferred to the test set to the same extent.


\begin{figure}[!htbp]
\centering
\includegraphics[width=0.90\columnwidth]{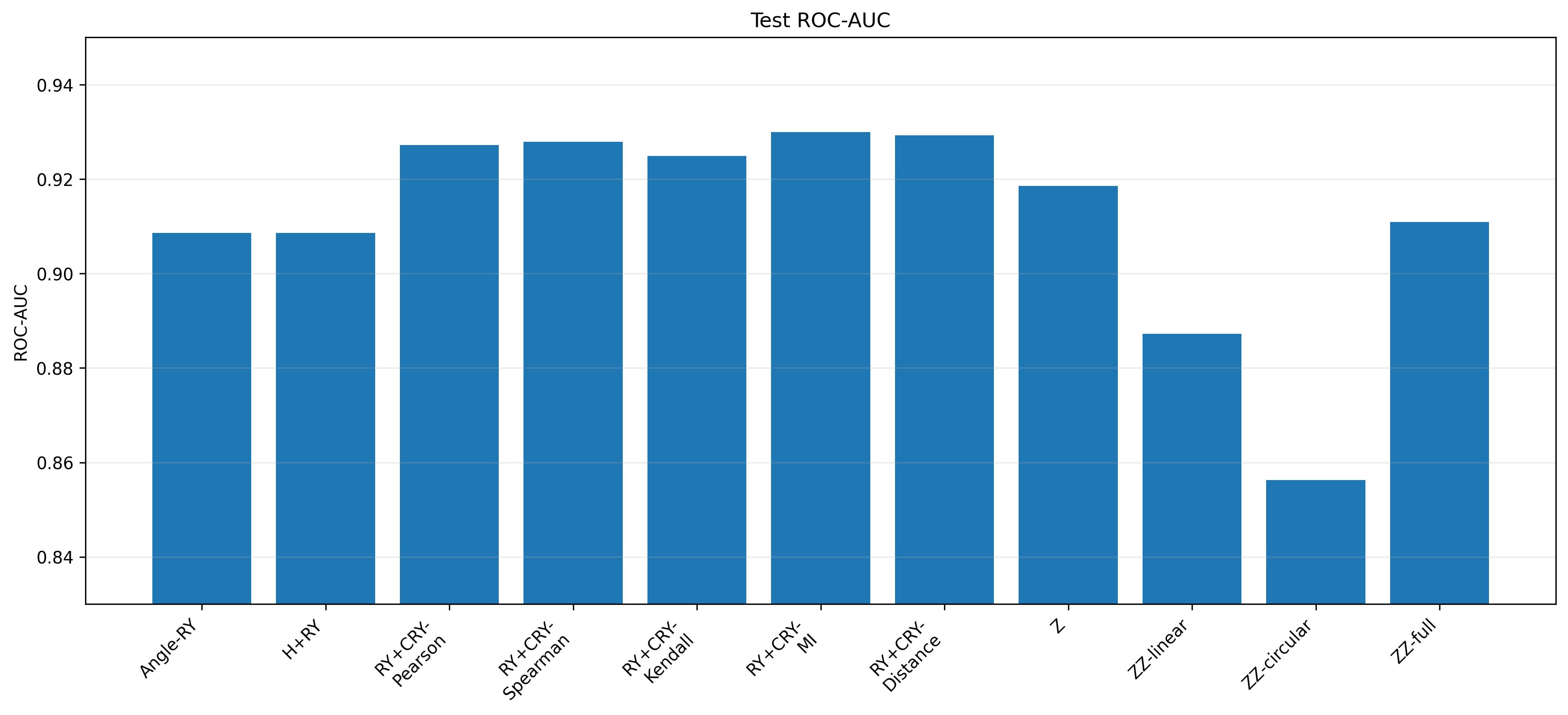}
\caption{Comparison of test ROC-AUC values for the Heart Failure Prediction
Dataset.}
\label{fig:heart_auc}
\end{figure}

In terms of ROC-AUC, the highest value of 0.930 was obtained with the
MI-based RY+CRY encoding. Distance Correlation, Spearman, and Pearson
closely followed with values of 0.929, 0.928, and 0.927, respectively.

This result is noteworthy because MI achieved the highest ROC-AUC while its
test accuracy of 0.874 remained below those of Pearson and Distance
Correlation. Thus, different performance metrics may emphasize different
aspects of the quantum feature maps.

ZZ linear and particularly ZZ circular remained behind the other methods,
with ROC-AUC values of 0.887 and 0.856, respectively. Therefore, denser ZZ
connectivity did not result in greater class-separation capability for this
dataset.

\subsubsection{Generalization Behavior and Log-Loss}

\begin{table*}[!htbp]
\centering
\caption{Training--test performance gaps and Log-Loss values for the Heart
Failure Prediction Dataset.}
\label{tab:heart_generalization}

\footnotesize
\setlength{\tabcolsep}{3.8pt}
\renewcommand{\arraystretch}{1.10}

\begin{tabular}{lccccccc}
\toprule
\textbf{Encoding} &
\makecell{\textbf{Acc.}\\\textbf{Gap}} &
\makecell{\textbf{Prec.}\\\textbf{Gap}} &
\makecell{\textbf{Recall}\\\textbf{Gap}} &
\makecell{\textbf{F1}\\\textbf{Gap}} &
\makecell{\textbf{AUC}\\\textbf{Gap}} &
\makecell{\textbf{Train}\\\textbf{Loss}} &
\makecell{\textbf{Test}\\\textbf{Loss}} \\
\midrule
Angle-RY        & 0.029 & 0.031 & 0.016  & 0.024 & 0.061 & 0.251 & 0.345 \\
H+RY            & 0.029 & 0.031 & 0.016  & 0.024 & 0.061 & 0.251 & 0.345 \\
RY+CRY-Pearson  & 0.032 & 0.021 & 0.037  & 0.029 & 0.049 & 0.231 & 0.319 \\
RY+CRY-Spearman & 0.036 & 0.028 & 0.037  & 0.032 & 0.048 & 0.232 & 0.319 \\
RY+CRY-Kendall  & 0.032 & 0.021 & 0.037  & 0.029 & 0.050 & 0.237 & 0.321 \\
RY+CRY-MI       & 0.045 & 0.034 & 0.047  & 0.040 & 0.047 & 0.227 & 0.317 \\
RY+CRY-Distance & 0.033 & 0.022 & 0.039  & 0.030 & 0.046 & 0.228 & 0.316 \\
Z               & 0.016 & 0.026 & -0.005 & 0.012 & 0.047 & 0.273 & 0.345 \\
ZZ linear       & 0.187 & 0.204 & 0.102  & 0.157 & 0.112 & 0.047 & 0.425 \\
ZZ circular     & 0.238 & 0.263 & 0.102  & 0.192 & 0.144 & 0.031 & 0.472 \\
ZZ full         & 0.157 & 0.190 & 0.063  & 0.131 & 0.089 & 0.001 & 0.372 \\
\bottomrule
\end{tabular}
\end{table*}


\begin{figure}[!htbp]
\centering
\includegraphics[width=0.90\columnwidth]
{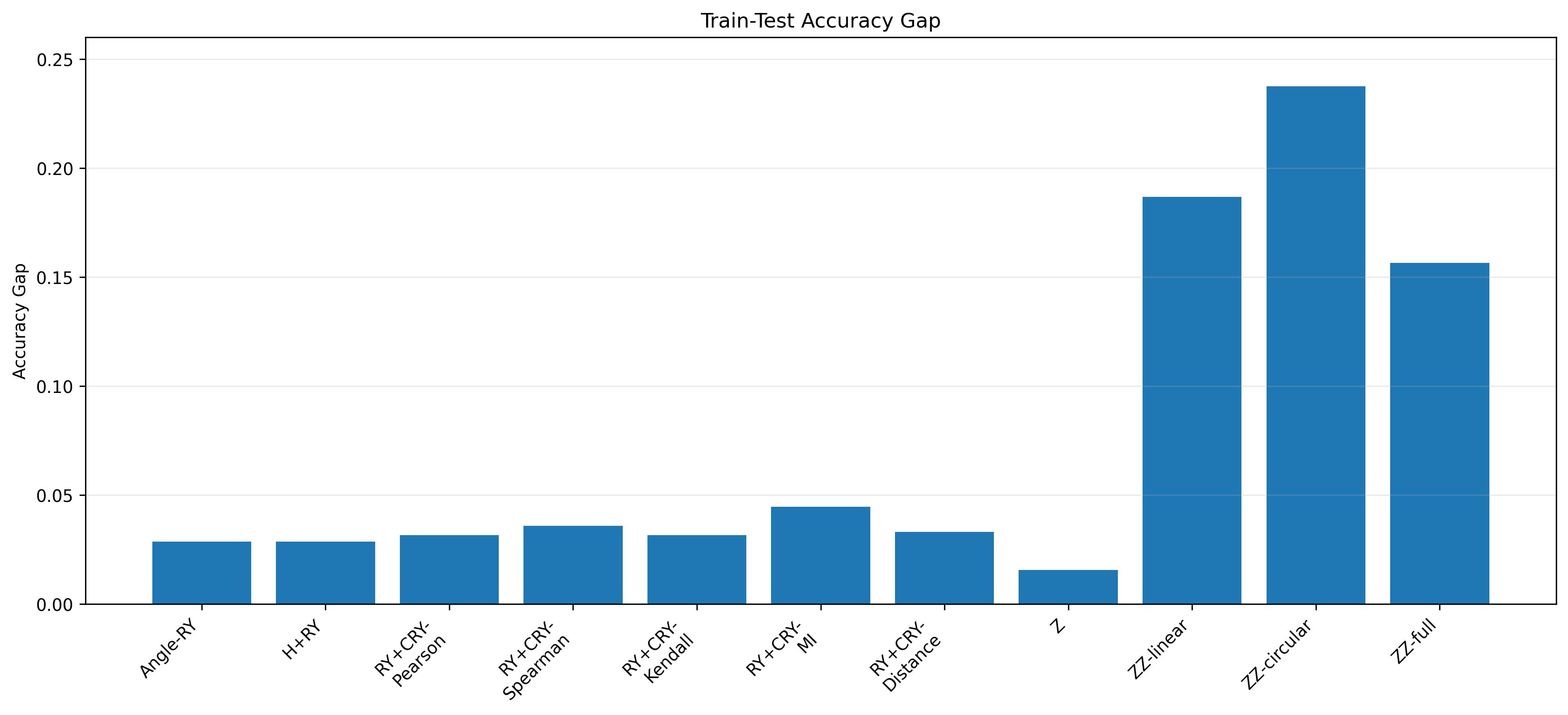}
\caption{Training--test Accuracy Gap across encoding strategies for the
Heart Failure Prediction Dataset.}
\label{fig:heart_accuracy_gap}
\end{figure}

Figure~\ref{fig:heart_accuracy_gap} clearly illustrates the generalization
behavior of the different methods. The lowest Accuracy Gap, \textbf{0.016},
was obtained with Z feature Map. The corresponding value was 0.029 for
Angle-RY and H+RY, whereas the RY+CRY methods remained within the range of
approximately 0.032--0.045.

These results indicate relatively small differences between the training
and test performance of the RY+CRY methods. In particular, Pearson and
Distance Correlation produced the highest test accuracies while maintaining
Accuracy Gap values of approximately 0.03. Therefore, the performance
improvement achieved by the RY+CRY methods was not accompanied by a
pronounced loss of generalization.

In contrast, the Accuracy Gap values for ZZ linear, ZZ circular, and
ZZ full were 0.187, 0.238, and 0.157, respectively. The largest gap was
observed for ZZ circular. Its training accuracy of 0.994 decreased to a
test accuracy of 0.757, indicating a pronounced separation between training
and test performance.


\begin{figure}[!htbp]
\centering
\includegraphics[width=0.90\columnwidth]{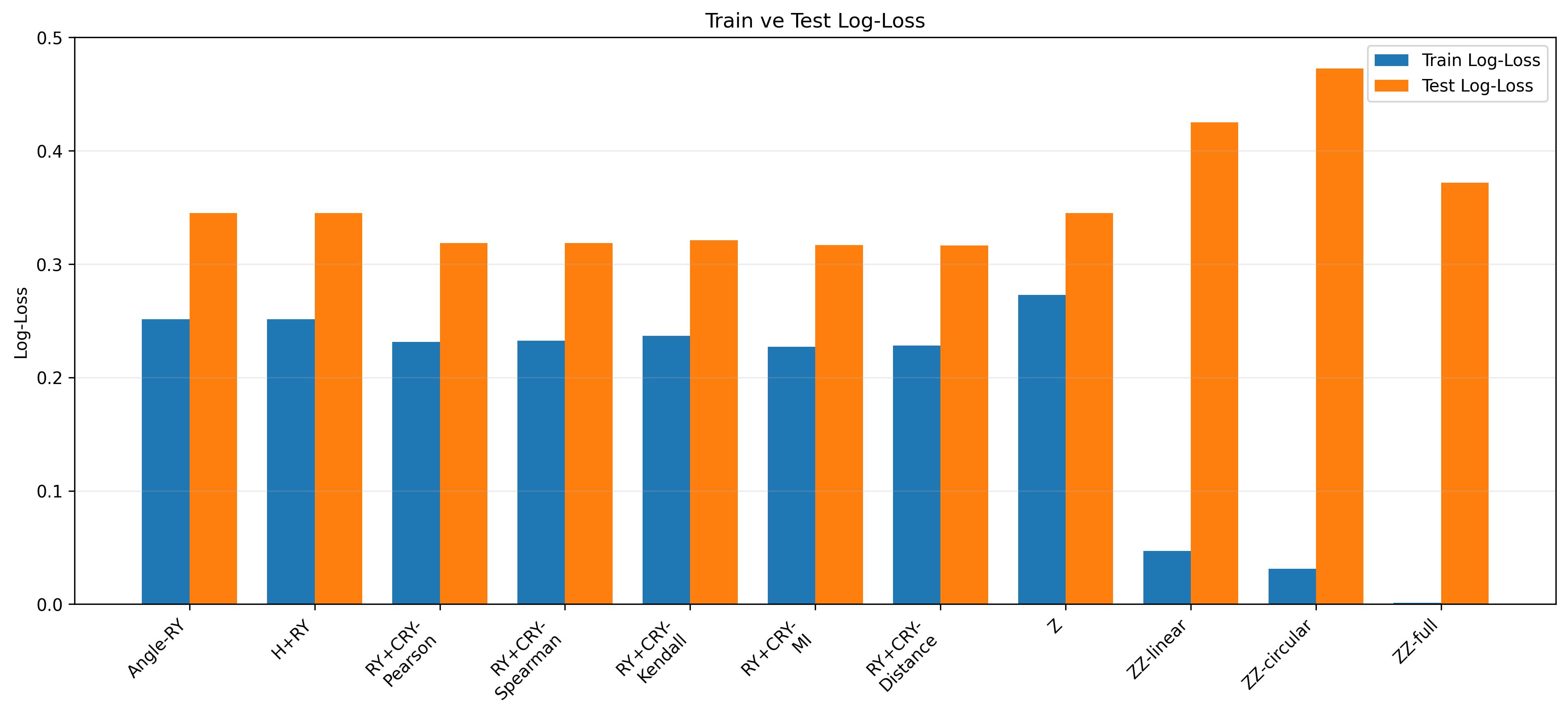}
\caption{Comparison of training and test Log-Loss values for the Heart
Failure Prediction Dataset.}
\label{fig:heart_logloss}
\end{figure}

The Log-Loss results support the findings obtained from the Accuracy Gap.
RY+CRY-Distance Correlation achieved the lowest Test Log-Loss of 0.316,
while MI, Pearson, and Spearman produced very similar values of 0.317,
0.319, and 0.319, respectively.

In contrast, although the Train Log-Loss values for ZZ linear and
ZZ circular were only 0.047 and 0.031, respectively, their Test Log-Loss
values increased to 0.425 and 0.472. In particular, ZZ circular produced
the highest test loss among all methods despite its very low training loss,
clearly demonstrating the divergence between its training and test behavior.

For ZZ full, the Train Log-Loss was only 0.001, whereas the Test Log-Loss
was 0.372. Therefore, for all three ZZ-based methods, the very strong fit to
the training data was not transferred to test performance to the same extent.

\subsubsection{Quantum Circuit Complexity}

\begin{table*}[!htbp]
\centering
\caption{Circuit and gate complexities of quantum feature encoding strategies
for the Heart Failure Prediction Dataset.}
\label{tab:heart_circuit}

\footnotesize
\setlength{\tabcolsep}{3.5pt}
\renewcommand{\arraystretch}{1.10}

\begin{tabular}{lrrrrrrrrr}
\toprule
\textbf{Encoding} &
\textbf{H} &
\textbf{RY} &
\textbf{CRY} &
\textbf{CX} &
\makecell{\textbf{Total}\\\textbf{Gates}} &
\textbf{Depth} &
\makecell{\textbf{Dec.}\\\textbf{CX}} &
\makecell{\textbf{Dec.}\\\textbf{Gates}} &
\makecell{\textbf{Dec.}\\\textbf{Depth}} \\
\midrule
Angle-RY        & 0  & 11 & 0  & 0   & 11  & 1  & 0   & 11  & 1 \\
H+RY            & 11 & 11 & 0  & 0   & 22  & 2  & 0   & 22  & 2 \\
RY+CRY-Pearson  & 0  & 11 & 55 & 0   & 66  & 20 & 110 & 231 & 68 \\
RY+CRY-Spearman & 0  & 11 & 55 & 0   & 66  & 20 & 110 & 231 & 68 \\
RY+CRY-Kendall  & 0  & 11 & 55 & 0   & 66  & 20 & 110 & 231 & 68 \\
RY+CRY-MI       & 0  & 11 & 55 & 0   & 66  & 20 & 110 & 231 & 68 \\
RY+CRY-Distance & 0  & 11 & 55 & 0   & 66  & 20 & 110 & 231 & 68 \\
Z               & 11 & 0  & 0  & 0   & 22  & 2  & 0   & 22  & 2 \\
ZZ linear       & 11 & 0  & 0  & 20  & 52  & 32 & 20  & 52  & 32 \\
ZZ circular     & 11 & 0  & 0  & 22  & 55  & 35 & 22  & 55  & 35 \\
ZZ full         & 11 & 0  & 0  & 110 & 187 & 59 & 110 & 187 & 59 \\
\bottomrule
\end{tabular}
\end{table*}


\begin{figure*}[!htbp]
\centering
\begin{subfigure}{0.48\textwidth}
\centering
\includegraphics[width=\linewidth]{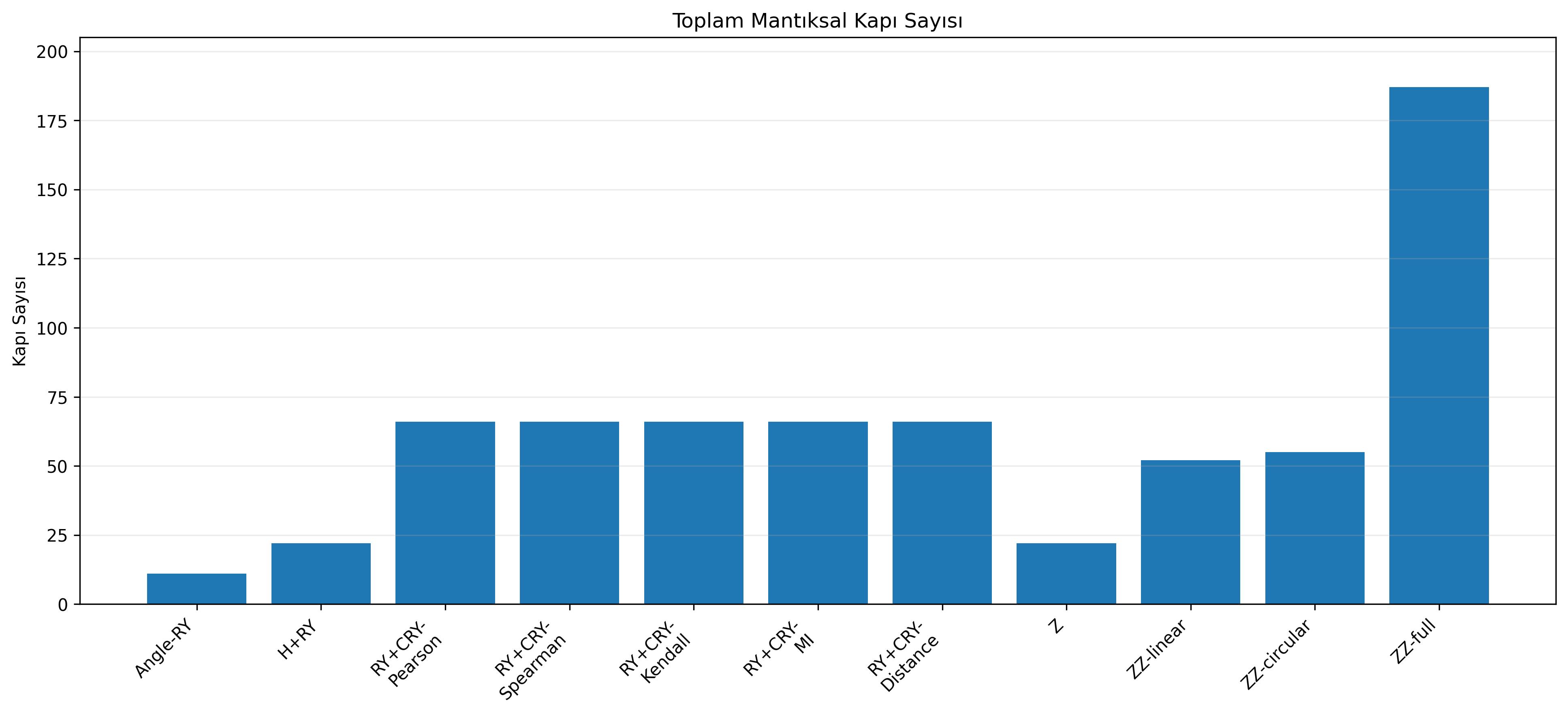}
\caption{Total number of logical gates.}
\end{subfigure}
\hfill
\begin{subfigure}{0.48\textwidth}
\centering
\includegraphics[width=\linewidth]{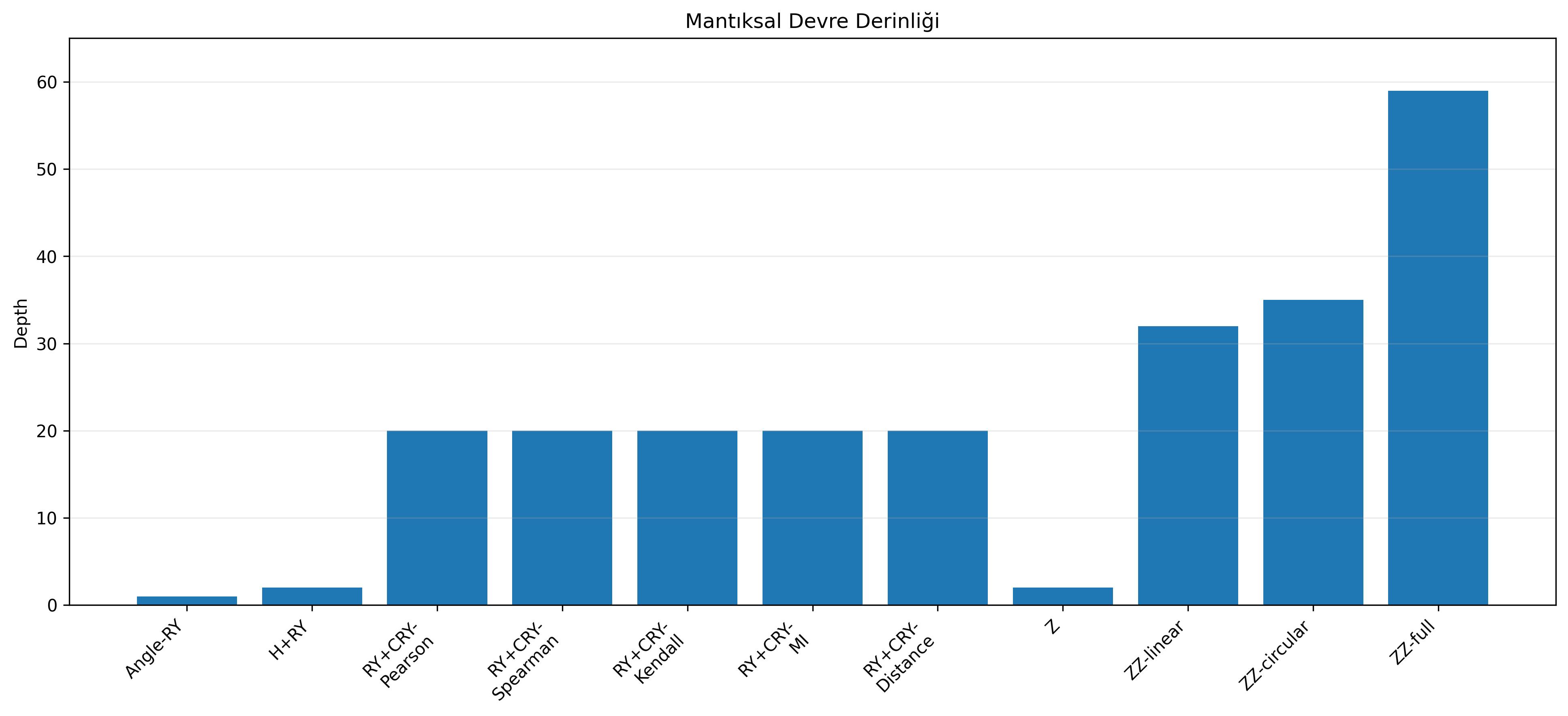}
\caption{Logical circuit depth.}
\end{subfigure}
\caption{Comparison of total gate count and circuit depth for the Heart
Failure Prediction Dataset.}
\label{fig:heart_gate_depth}
\end{figure*}

Since 11 features were used for the Heart Failure Prediction Dataset, the
basic RY-based encoding consists of 11 RY gates. The number of all unique
feature pairs is

\begin{equation}
\binom{11}{2}=55.
\end{equation}

Therefore, each RY+CRY method contains 55 CRY gates. These circuits have a
total logical gate count of 66 and a circuit depth of 20.

It is important to note that the Pearson, Spearman, Kendall, MI, and
Distance Correlation methods have identical circuit complexities.
Consequently, differences in classification performance among these methods
do not arise from circuit size but from the statistical dependency values
used to determine the CRY rotation angles.

Z feature Map achieved a test accuracy of 0.878 with only 22 gates and a
circuit depth of 2. This makes Z feature Map a particularly efficient option
in terms of circuit cost for the Heart Failure Prediction Dataset.


\begin{figure}[!htbp]
\centering
\includegraphics[width=0.95\columnwidth]
{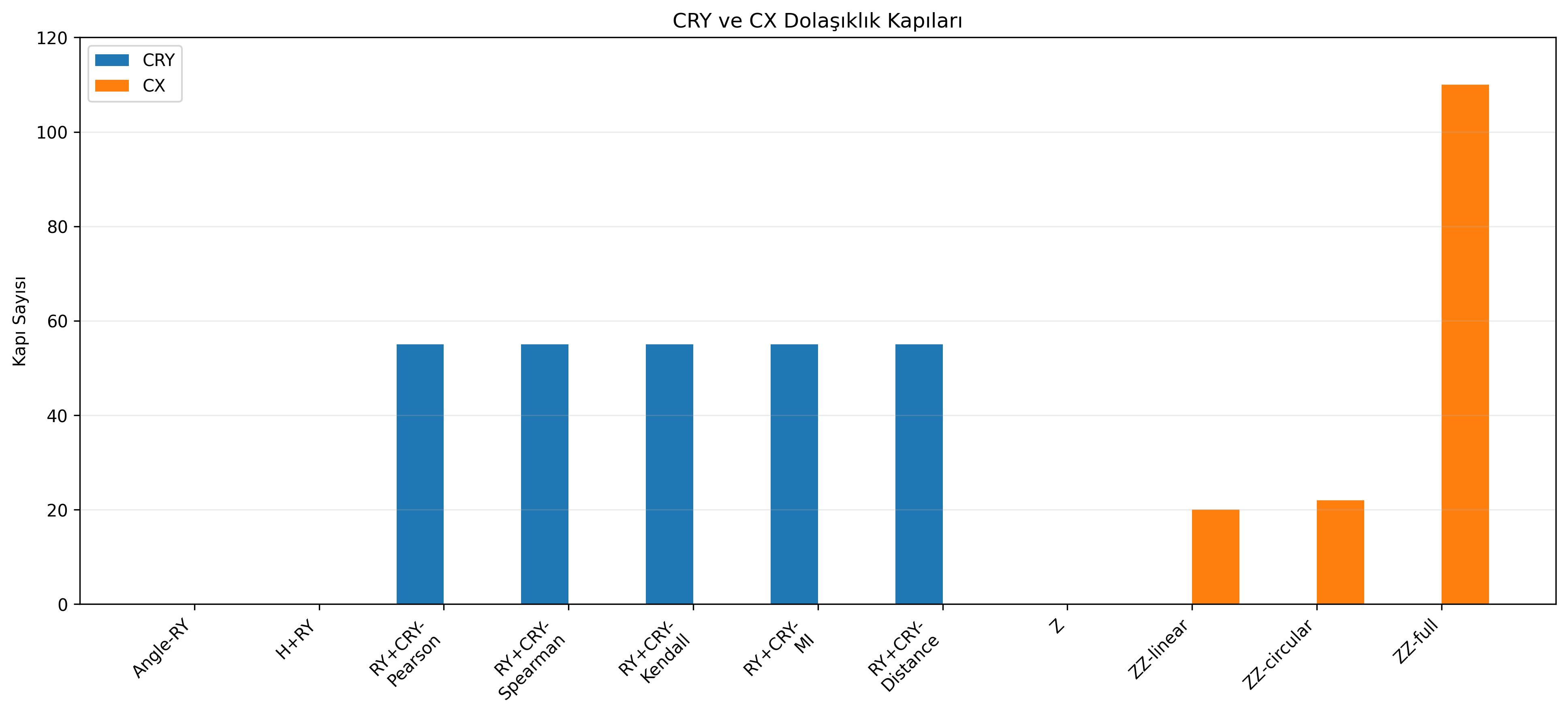}
\caption{Comparison of CRY and CX entangling gates for the Heart Failure
Prediction Dataset.}
\label{fig:heart_entangling}
\end{figure}

As the density of ZZ connectivity increases, the number of CX gates also
increases substantially. ZZ linear, ZZ circular, and ZZ full contain 20,
22, and 110 CX gates, respectively. Nevertheless, the test accuracy of
ZZ full was only 0.843, remaining below the results obtained with RY+CRY
and Z feature Map.

This finding is important for the Heart Failure Prediction Dataset: a denser
entanglement structure alone does not guarantee higher classification
performance. The compatibility between the structure of the feature space
and the selected entanglement topology appears to be more important than
the number of entangling gates alone.


\begin{figure}[!htbp]
\centering
\includegraphics[width=0.95\columnwidth]
{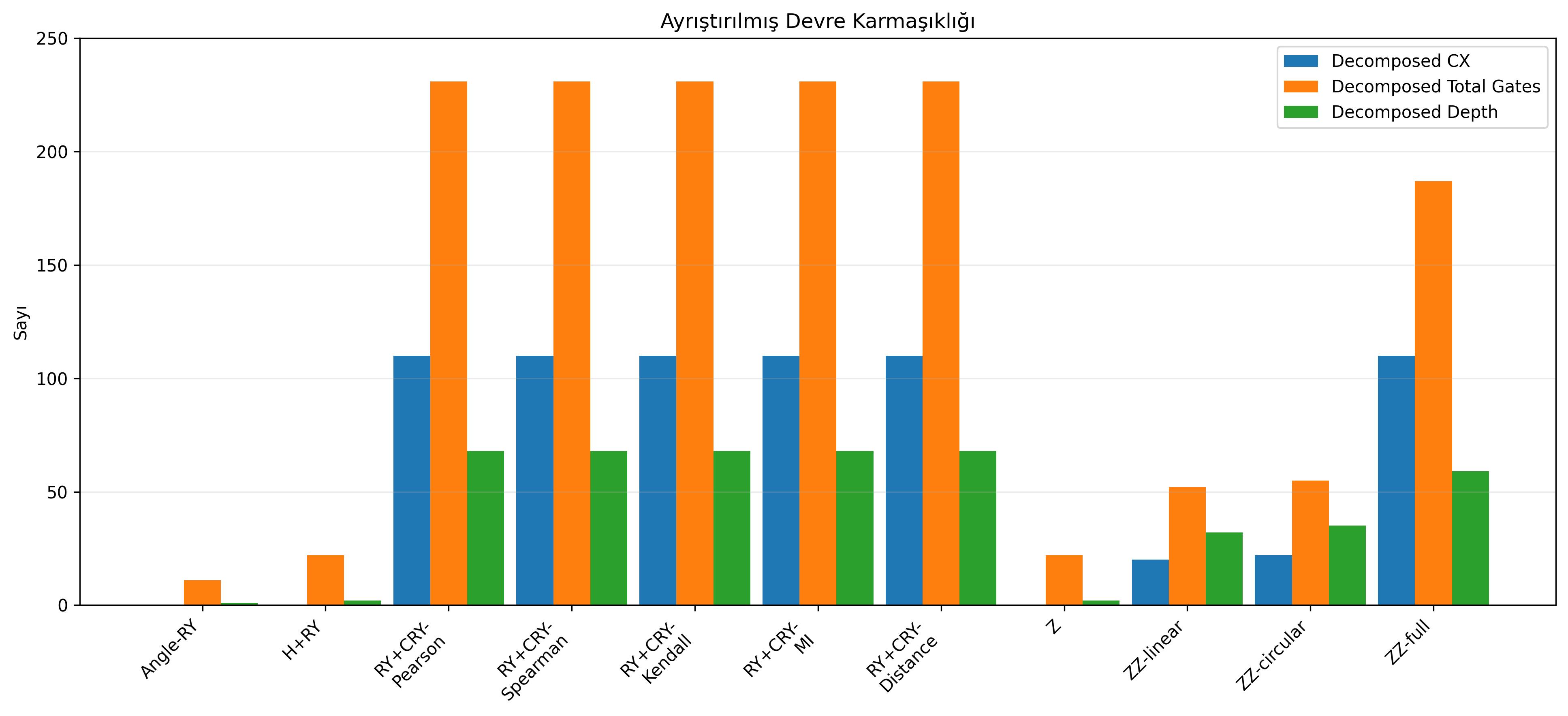}
\caption{Comparison of decomposed circuit complexity for the Heart Failure
Prediction Dataset.}
\label{fig:heart_decomposed}
\end{figure}

After decomposition, the RY+CRY circuits contain 231 total gates, 110 CX
gates, and a circuit depth of 68. These values indicate a substantial
circuit cost compared with Angle-RY and Z feature Map.

Nevertheless, the fact that the Pearson- and Distance Correlation-based
RY+CRY methods achieved the highest test accuracy of 0.883 indicates that
incorporating dependency information into quantum encoding can be beneficial
for classification performance. However, an accuracy improvement of
approximately 0.9 percentage points requires 231 decomposed gates and
110 CX gates, emphasizing the need to evaluate the trade-off between
performance and circuit cost.

\subsubsection{Computational Cost}

\begin{table}[!htbp]
\centering
\caption{Computation times for the Heart Failure Prediction Dataset
(seconds).}
\label{tab:heart_time}

\footnotesize
\setlength{\tabcolsep}{4pt}
\renewcommand{\arraystretch}{1.10}

\begin{tabular}{lrrr}
\toprule
\textbf{Encoding} &
\makecell{\textbf{State vector}\\\textbf{Time}} &
\makecell{\textbf{Kernel}\\\textbf{Time}} &
\makecell{\textbf{Training}\\\textbf{Time}} \\
\midrule
Angle-RY        & 1.044 & 0.1510 & 0.0208 \\
H+RY            & 1.757 & 0.1500 & 0.0194 \\
RY+CRY-Pearson  & 6.594 & 0.1524 & 0.0200 \\
RY+CRY-Spearman & 6.549 & 0.1515 & 0.0208 \\
RY+CRY-Kendall  & 6.702 & 0.1525 & 0.0206 \\
RY+CRY-MI       & 6.668 & 0.1509 & 0.0207 \\
RY+CRY-Distance & 6.939 & 0.1521 & 0.0213 \\
Z               & 1.910 & 0.1518 & 0.0228 \\
ZZ linear       & 3.837 & 0.1495 & 0.0314 \\
ZZ circular     & 4.173 & 0.1525 & 0.0324 \\
ZZ full         & 13.633 & 0.1510 & 0.0311 \\
\bottomrule
\end{tabular}
\end{table}


\begin{figure}[!htbp]
\centering
\includegraphics[width=0.90\columnwidth]
{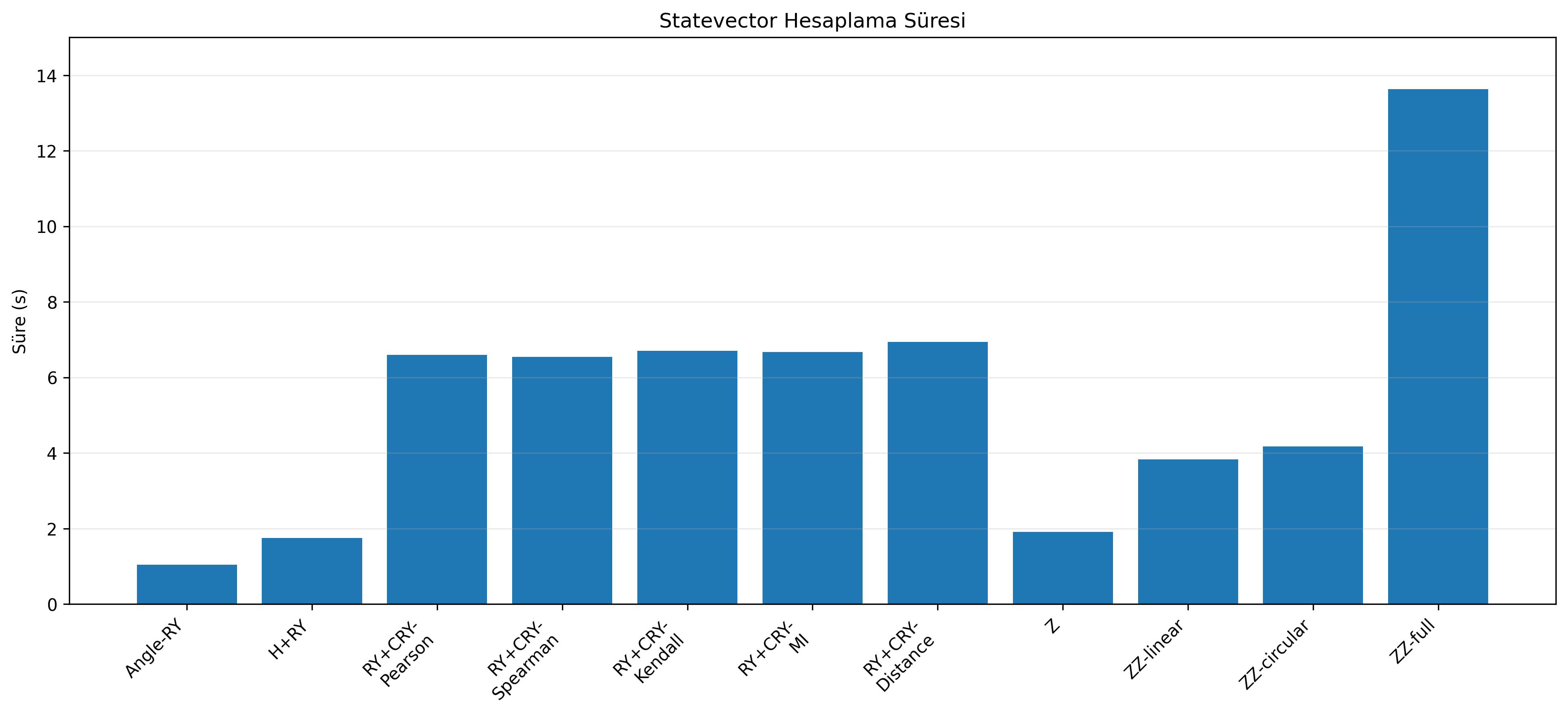}
\caption{State vector computation times for the Heart Failure Prediction
Dataset.}
\label{fig:heart_state vector}
\end{figure}

The computation times generally follow the differences in circuit
complexity. Angle-RY had the lowest state vector computation time at
approximately 1.04 seconds, whereas the corresponding value for Z feature Map
was approximately 1.91 seconds.

The state vector computation times of the RY+CRY methods ranged from
approximately 6.55 to 6.94 seconds. The similarity of these values is
consistent with the identical circuit topology used by these methods.

The highest state vector computation time, 13.63 seconds, was obtained with
ZZ full. Compared with Angle-RY, this corresponds to approximately

\begin{equation}
\frac{13.633}{1.044}\approx13.1
\end{equation}

times higher computational cost. Nevertheless, because the test accuracy of
ZZ full was lower than that of Angle-RY, the increased circuit cost did not
translate into improved performance for this dataset.


\begin{figure}[!htbp]
\centering
\includegraphics[width=0.90\columnwidth]
{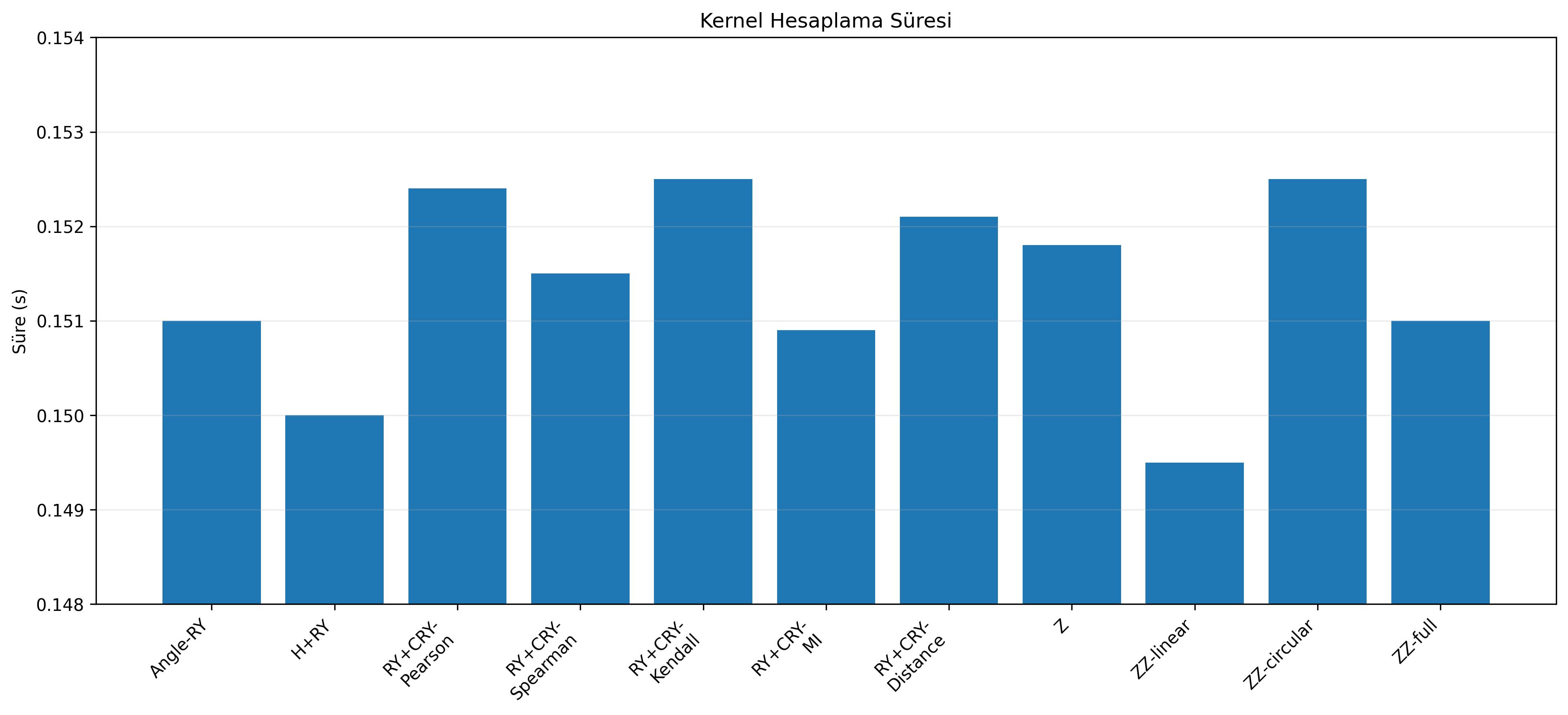}
\caption{Kernel computation times for the Heart Failure Prediction Dataset.}
\label{fig:heart_kernel}
\end{figure}

Kernel computation times remained highly stable across the methods, ranging
from approximately 0.150 to 0.153 seconds. Despite substantial differences
in circuit complexity, the similar kernel times indicate that the main
differences in total computational cost did not originate from the kernel
computation stage.


\begin{figure}[!htbp]
\centering
\includegraphics[width=0.90\columnwidth]
{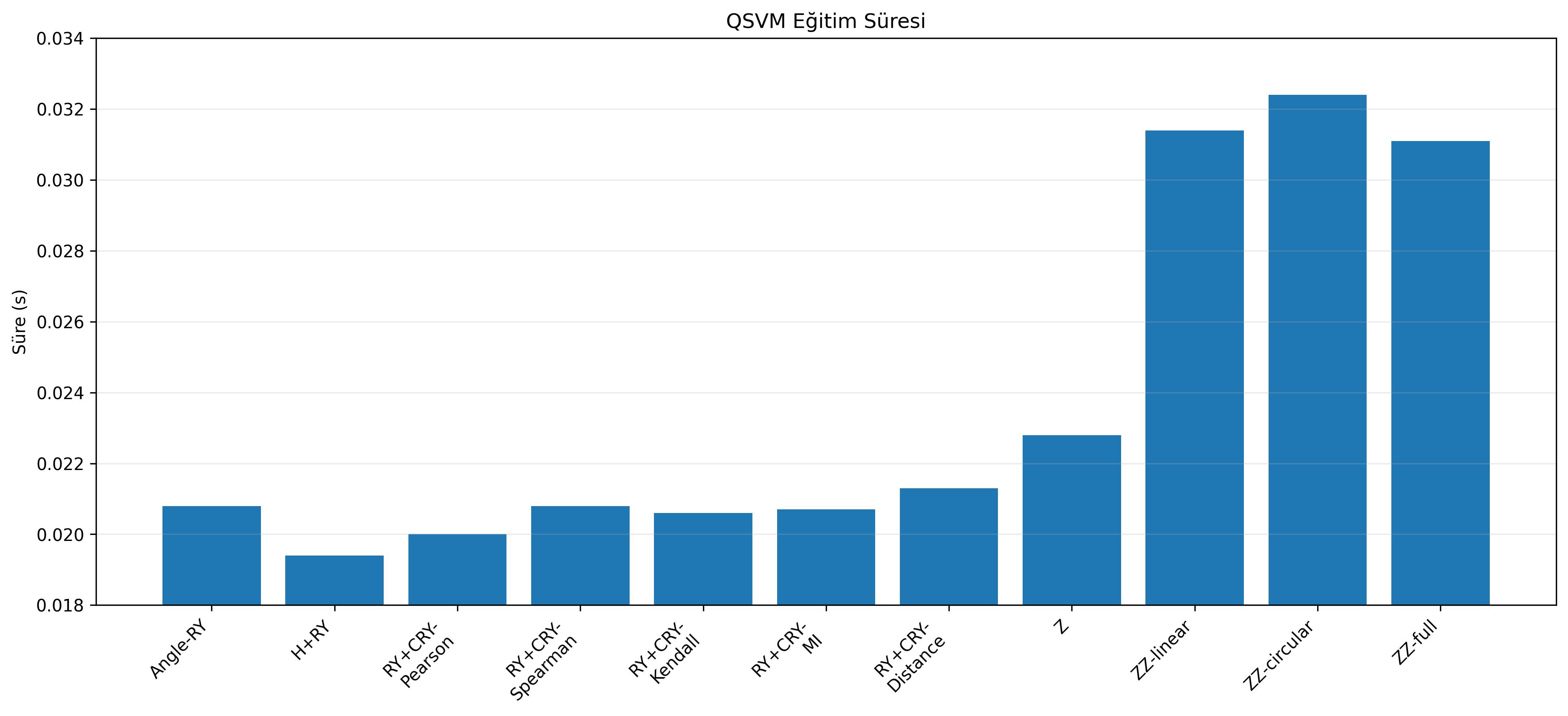}
\caption{QSVM training times for the Heart Failure Prediction Dataset.}
\label{fig:heart_training}
\end{figure}

QSVM training times were also very low for all methods. The training time
was approximately 0.021 seconds for Angle-RY and approximately
0.031--0.032 seconds for ZZ linear, ZZ circular, and ZZ full.

Therefore, the primary source of the computational cost differences among
the methods in the Heart Failure Prediction experiments was the preparation
of quantum states and state vector computation rather than the classical SVM
training stage.

\subsubsection{Overall Evaluation of the Heart Failure Prediction Dataset}

The results obtained for the Heart Failure Prediction Dataset clearly show
that more complex quantum feature maps do not necessarily provide higher
classification performance. The highest test accuracy, 0.883, was achieved
with the Pearson- and Distance Correlation-based RY+CRY encodings. These
methods also exhibited relatively balanced training and test behavior, with
Accuracy Gap values of approximately 0.03.

Z feature Map is particularly noteworthy in terms of the
performance--complexity trade-off. With only 22 logical gates and a circuit
depth of 2, it achieved a test accuracy of 0.878, an F1 score of 0.896, and
a ROC-AUC of 0.919. Furthermore, it produced the lowest Accuracy Gap among
all methods, at 0.016. Therefore, when maximum test accuracy is not the sole
objective, Z feature Map can be considered a highly efficient option for the
Heart Failure Prediction Dataset.

The RY+CRY methods provided a limited but consistent improvement over
Angle-RY. In particular, Pearson and Distance Correlation achieved the
highest test accuracy. However, encoding all 55 feature pairs using CRY
gates resulted in 231 gates, 110 CX gates, and a decomposed circuit depth of
68. Therefore, the observed performance improvement should be evaluated
together with the associated circuit cost.

The most pronounced behavior was observed for the ZZ-based feature maps.
Although these methods achieved nearly perfect performance on the training
set, their test performance decreased substantially and the training--test
gaps increased. In particular, ZZ circular achieved a training accuracy of
0.994 but only a test accuracy of 0.757, resulting in an Accuracy Gap of
0.238.

Overall, the Heart Failure Prediction experiments demonstrate that the
effectiveness of a quantum feature encoding strategy is not determined
solely by the density of circuit connectivity. Greater entanglement and more
complex feature maps may fail to produce higher classification performance
when they are not well matched to the underlying data structure. Therefore,
the evaluation of feature encoding strategies for QSVM should jointly
consider test performance, generalization behavior, and quantum circuit
complexity.

\subsection{Credit Risk Dataset}
\label{subsec:credit_results}

The experiments conducted on the Credit Risk Dataset compared Angle-RY,
H+RY, RY+CRY encodings based on different statistical dependency measures,
Z feature Map, and ZZ feature Map with linear, circular, and full entanglement
structures. The evaluation was performed in terms of classification
performance, generalization behavior, quantum circuit complexity, and
computational cost.

\subsubsection{Classification Performance}

\begin{table*}[!htbp]
\centering
\caption{Classification performance of quantum feature encoding strategies
on the Credit Risk Dataset.}
\label{tab:credit_performance}

\footnotesize
\setlength{\tabcolsep}{4pt}
\renewcommand{\arraystretch}{1.10}

\begin{tabular}{lccccccc}
\toprule
\textbf{Encoding} &
\makecell{\textbf{Train}\\\textbf{Acc.}} &
\makecell{\textbf{Test}\\\textbf{Acc.}} &
\makecell{\textbf{Acc.}\\\textbf{Gap}} &
\makecell{\textbf{Test}\\\textbf{Prec.}} &
\makecell{\textbf{Test}\\\textbf{Rec.}} &
\makecell{\textbf{Test}\\\textbf{F1}} &
\makecell{\textbf{Test}\\\textbf{AUC}} \\
\midrule
Angle-RY        & 0.907 & 0.908 & -0.001 & 0.824 & 0.908 & 0.864 & 0.654 \\
H+RY            & 0.907 & 0.908 & -0.001 & 0.824 & 0.908 & 0.864 & 0.654 \\
RY+CRY-Pearson  & 0.907 & 0.908 & -0.001 & 0.824 & 0.908 & 0.864 & 0.658 \\
RY+CRY-Spearman & 0.907 & 0.908 & -0.001 & 0.824 & 0.908 & 0.864 & 0.659 \\
RY+CRY-Kendall  & 0.907 & 0.908 & -0.001 & 0.824 & 0.908 & 0.864 & 0.654 \\
RY+CRY-MI       & 0.907 & 0.908 & -0.001 & 0.824 & 0.908 & 0.864 & 0.654 \\
RY+CRY-Distance & 0.908 & 0.908 &  0.000 & 0.824 & 0.908 & 0.864 & 0.663 \\
Z               & 0.915 & 0.908 &  0.007 & 0.824 & 0.908 & 0.864 & 0.464 \\
ZZ linear       & 0.933 & 0.908 &  0.025 & 0.824 & 0.908 & 0.864 & 0.619 \\
ZZ circular     & 0.933 & 0.908 &  0.025 & 0.824 & 0.908 & 0.864 & 0.606 \\
ZZ full         & 0.945 & 0.908 &  0.037 & 0.824 & 0.908 & 0.864 & 0.613 \\
\bottomrule
\end{tabular}
\end{table*}


\begin{figure}[!htbp]
\centering
\includegraphics[width=0.95\columnwidth]{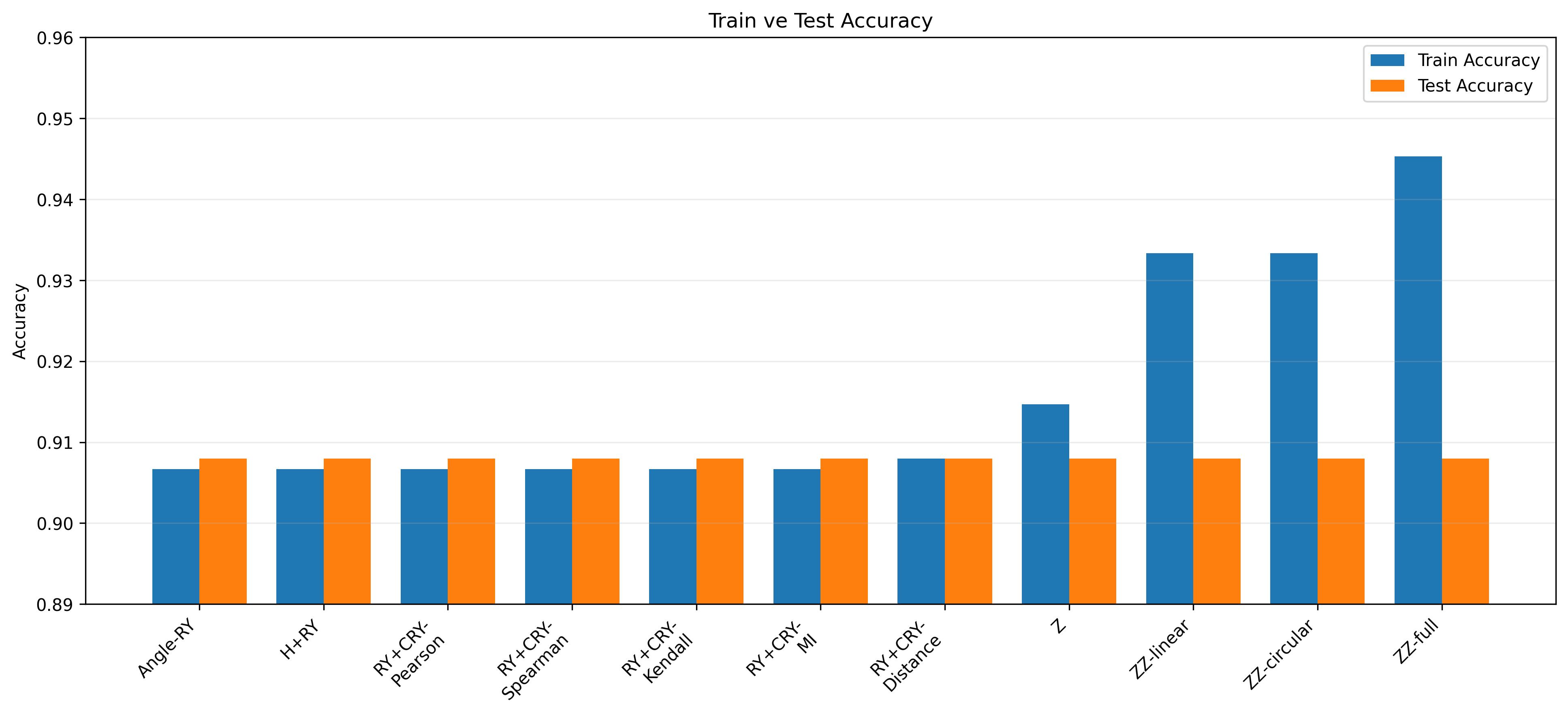}
\caption{Comparison of training and test accuracy across encoding strategies
for the Credit Risk Dataset.}
\label{fig:credit_accuracy}
\end{figure}

Table~\ref{tab:credit_performance} and Fig.~\ref{fig:credit_accuracy}
show that all encoding methods produced the same test accuracy
(\textbf{0.908}) on the Credit Risk Dataset. Therefore, changing the quantum
feature encoding strategy did not alter the final classification accuracy
under the decision threshold used in this study.

Differences were nevertheless observed in training accuracy. Angle-RY,
H+RY, and the Pearson-, Spearman-, Kendall-, and MI-based RY+CRY methods
produced training accuracies of approximately 0.907. The training accuracy
increased to 0.915 for Z feature Map, 0.933 for both ZZ linear and
ZZ circular, and 0.945 for ZZ full.

These results indicate that more complex feature maps provided a stronger
fit to the training data; however, this additional fit did not translate
into an improvement in test accuracy.


\begin{figure}[!htbp]
\centering
\includegraphics[width=0.95\columnwidth]
{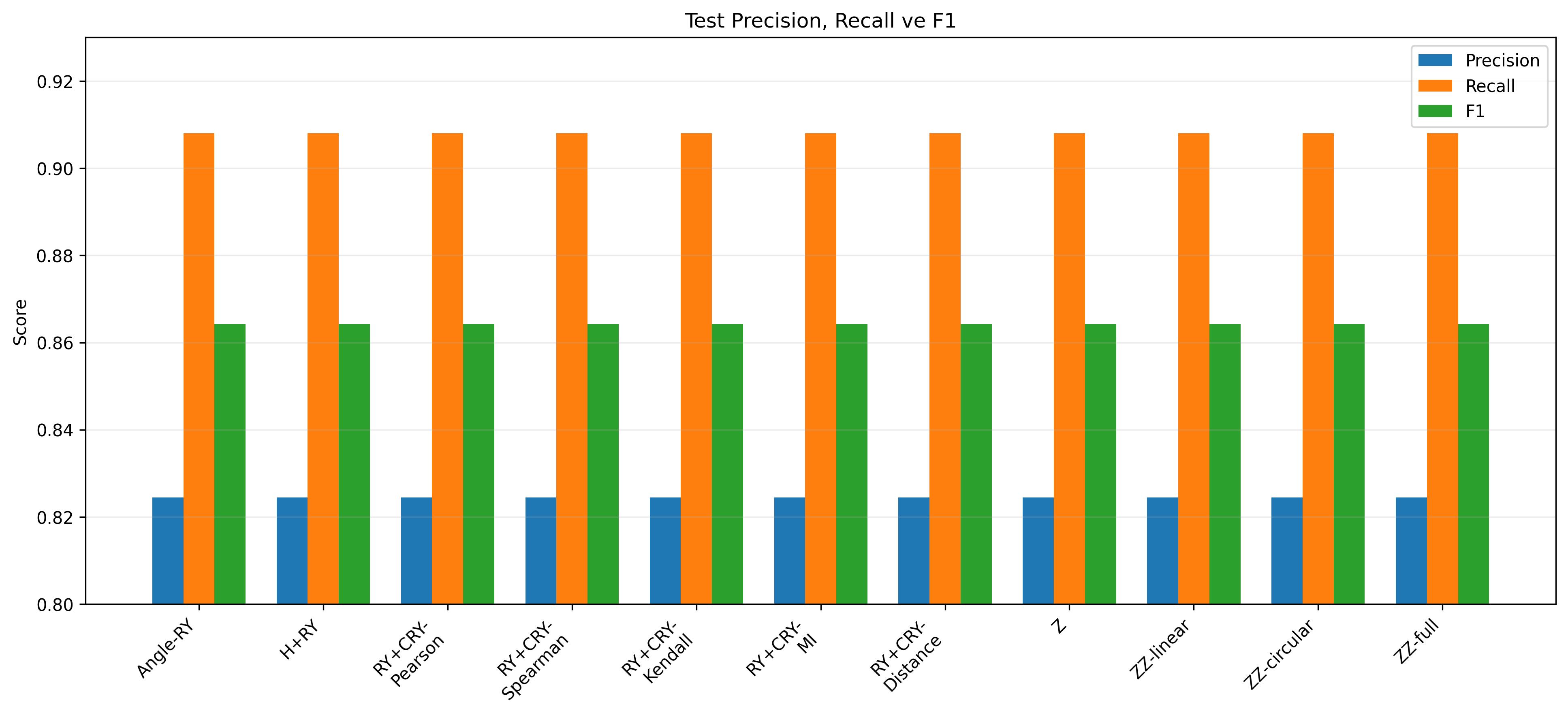}
\caption{Comparison of test Precision, Recall, and F1 scores for the Credit
Risk Dataset.}
\label{fig:credit_prf}
\end{figure}

As shown in Fig.~\ref{fig:credit_prf}, not only Accuracy but also Precision,
Recall, and F1 were identical across all methods. Test Precision was 0.824,
Recall was 0.908, and F1 was 0.864.

This result highlights an important observation. Although different quantum
feature maps may generate different similarity structures in the kernel
space, they produced the same class labels under the classification
threshold used in this experiment. Consequently, threshold-dependent
metrics such as Accuracy, Precision, Recall, and F1 were unable to
differentiate the methods in this case.

Therefore, ROC-AUC and probability-based measures such as Log-Loss provide
more informative criteria for distinguishing the encoding strategies on
this dataset.


\begin{figure}[!htbp]
\centering
\includegraphics[width=0.90\columnwidth]{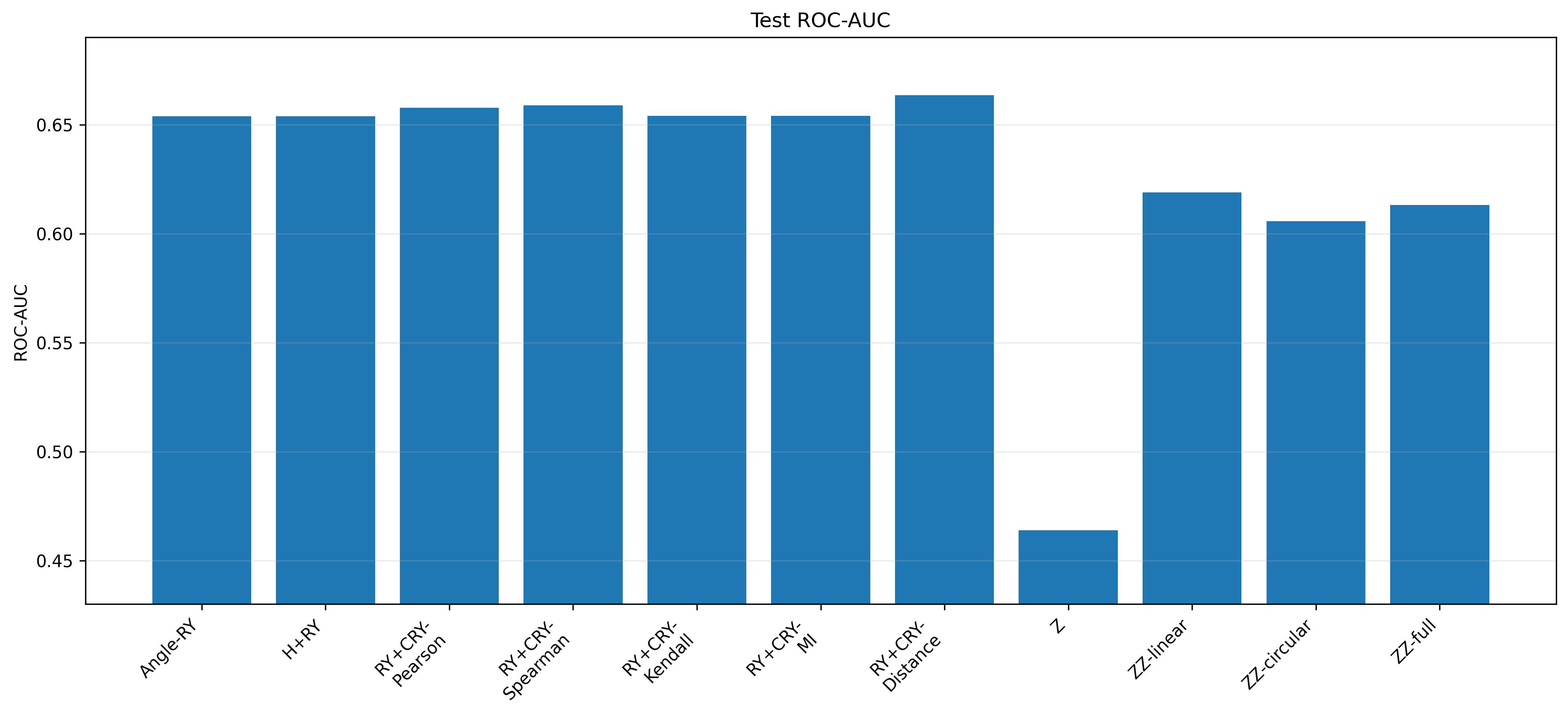}
\caption{Comparison of test ROC-AUC values for the Credit Risk Dataset.}
\label{fig:credit_auc}
\end{figure}

Although the test Accuracy was identical across all methods, the ROC-AUC
results differed considerably, as illustrated in
Fig.~\ref{fig:credit_auc}.

The highest test ROC-AUC, 0.663, was obtained with the Distance
Correlation-based RY+CRY encoding. It was followed by Spearman at 0.659
and Pearson at 0.658. Angle-RY and H+RY both produced ROC-AUC values of
approximately 0.654.

These results indicate that although the dependency-based controlled
rotations did not change the test Accuracy, they could provide a limited
improvement in the relative ranking or separability of the classes.

Notably, Z feature Map produced a ROC-AUC of only 0.464. Despite its test
accuracy of 0.908, this low value indicates weak class discrimination in
terms of the underlying decision scores.

The ROC-AUC values for ZZ linear, ZZ circular, and ZZ full were 0.619,
0.606, and 0.613, respectively. Thus, the higher circuit complexity of
these feature maps did not provide an advantage in terms of ROC-AUC.

\subsubsection{Generalization Behavior and Log-Loss}

\begin{table*}[!htbp]
\centering
\caption{Training--test performance gaps and Log-Loss values for the Credit
Risk Dataset.}
\label{tab:credit_generalization}

\footnotesize
\setlength{\tabcolsep}{3.8pt}
\renewcommand{\arraystretch}{1.10}

\begin{tabular}{lccccccc}
\toprule
\textbf{Encoding} &
\makecell{\textbf{Acc.}\\\textbf{Gap}} &
\makecell{\textbf{Prec.}\\\textbf{Gap}} &
\makecell{\textbf{Recall}\\\textbf{Gap}} &
\makecell{\textbf{F1}\\\textbf{Gap}} &
\makecell{\textbf{AUC}\\\textbf{Gap}} &
\makecell{\textbf{Train}\\\textbf{Loss}} &
\makecell{\textbf{Test}\\\textbf{Loss}} \\
\midrule
Angle-RY        & -0.001 & -0.002 & -0.001 & -0.002 & -0.109 & 0.365 & 0.361 \\
H+RY            & -0.001 & -0.002 & -0.001 & -0.002 & -0.109 & 0.365 & 0.361 \\
RY+CRY-Pearson  & -0.001 & -0.002 & -0.001 & -0.002 & -0.118 & 0.368 & 0.363 \\
RY+CRY-Spearman & -0.001 & -0.002 & -0.001 & -0.002 & -0.119 & 0.367 & 0.363 \\
RY+CRY-Kendall  & -0.001 & -0.002 & -0.001 & -0.002 & -0.112 & 0.366 & 0.362 \\
RY+CRY-MI       & -0.001 & -0.002 & -0.001 & -0.002 & -0.115 & 0.368 & 0.362 \\
RY+CRY-Distance &  0.000 &  0.040 &  0.000 &  0.001 & -0.130 & 0.373 & 0.363 \\
Z               &  0.007 &  0.098 &  0.007 &  0.016 & -0.154 & 0.414 & 0.373 \\
ZZ linear       &  0.025 &  0.113 &  0.025 &  0.052 &  0.369 & 0.282 & 0.369 \\
ZZ circular     &  0.025 &  0.113 &  0.025 &  0.052 &  0.353 & 0.283 & 0.370 \\
ZZ full         &  0.037 &  0.124 &  0.037 &  0.071 & -0.056 & 0.341 & 0.369 \\
\bottomrule
\end{tabular}
\end{table*}


\begin{figure}[!htbp]
\centering
\includegraphics[width=0.90\columnwidth]
{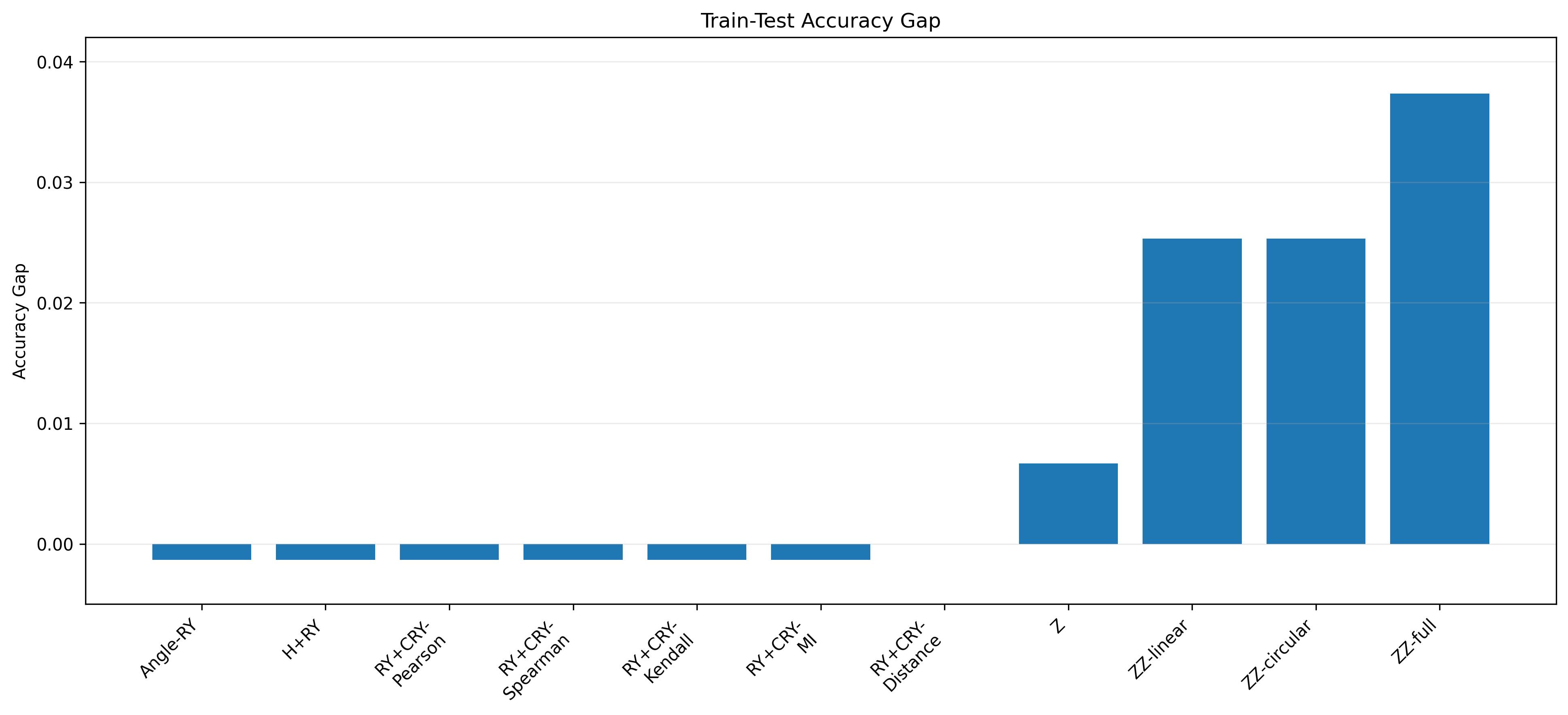}
\caption{Training--test Accuracy Gap across encoding strategies for the
Credit Risk Dataset.}
\label{fig:credit_accuracy_gap}
\end{figure}

Figure~\ref{fig:credit_accuracy_gap} demonstrates differences in
generalization behavior among the methods. For Angle-RY, H+RY, and the
first four RY+CRY methods, the Accuracy Gap was approximately -0.001.
This small negative value indicates that test accuracy was approximately
0.001 higher than training accuracy and, in practical terms, shows that
training and test performance were nearly identical.

For the Distance Correlation-based RY+CRY encoding, the training and test
accuracies were identical, resulting in an Accuracy Gap of zero.

The gap increased to 0.007 for Z feature Map, 0.025 for ZZ linear and
ZZ circular, and 0.037 for ZZ full. Nevertheless, these values remain
considerably smaller than the large generalization gaps observed for the
ZZ-based methods on some of the other datasets.

The important observation is that training accuracy increased with circuit
complexity, whereas test accuracy remained fixed at 0.908. Thus, the
additional fit of the ZZ encodings to the training data did not improve
test performance on this dataset.


\begin{figure}[!htbp]
\centering
\includegraphics[width=0.90\columnwidth]{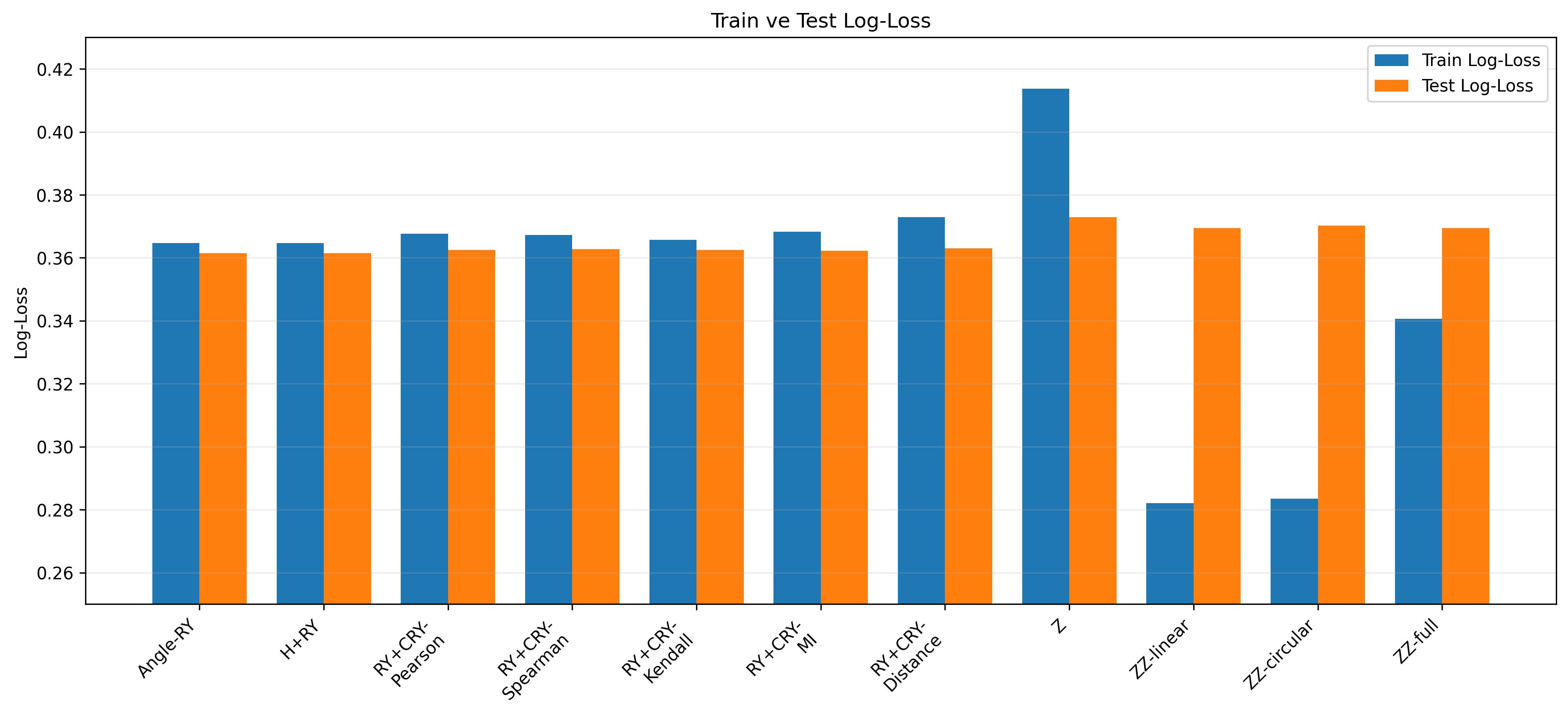}
\caption{Comparison of training and test Log-Loss values for the Credit
Risk Dataset.}
\label{fig:credit_logloss}
\end{figure}

The Log-Loss results reveal differences among the methods that are not
apparent from Accuracy alone.

The Test Log-Loss was approximately 0.361 for Angle-RY and H+RY. For the
RY+CRY methods, the values were very similar and ranged from approximately
0.362 to 0.363.

Z feature Map produced a higher Test Log-Loss of 0.373. The corresponding
values for ZZ linear, ZZ circular, and ZZ full were 0.369, 0.370, and
0.369, respectively.

In particular, although the Train Log-Loss values of ZZ linear and
ZZ circular decreased to approximately 0.282, their Test Log-Loss remained
close to 0.370. This indicates that the stronger fit obtained on the
training data was not transferred to the test data to the same extent.

Therefore, although Accuracy was identical for all methods on the Credit
Risk Dataset, the Log-Loss and ROC-AUC results reveal meaningful
differences among the quantum feature maps.

\subsubsection{Quantum Circuit Complexity}

\begin{table*}[!htbp]
\centering
\caption{Circuit and gate complexities of quantum feature encoding strategies
for the Credit Risk Dataset.}
\label{tab:credit_circuit}

\footnotesize
\setlength{\tabcolsep}{3.5pt}
\renewcommand{\arraystretch}{1.10}

\begin{tabular}{lrrrrrrrrr}
\toprule
\textbf{Encoding} &
\textbf{H} &
\textbf{RY} &
\textbf{CRY} &
\textbf{CX} &
\makecell{\textbf{Total}\\\textbf{Gates}} &
\textbf{Depth} &
\makecell{\textbf{Dec.}\\\textbf{CX}} &
\makecell{\textbf{Dec.}\\\textbf{Gates}} &
\makecell{\textbf{Dec.}\\\textbf{Depth}} \\
\midrule
Angle-RY        & 0 & 9 & 0  & 0  & 9   & 1  & 0  & 9   & 1 \\
H+RY            & 9 & 9 & 0  & 0  & 18  & 2  & 0  & 18  & 2 \\
RY+CRY-Pearson  & 0 & 9 & 36 & 0  & 45  & 16 & 72 & 153 & 54 \\
RY+CRY-Spearman & 0 & 9 & 36 & 0  & 45  & 16 & 72 & 153 & 54 \\
RY+CRY-Kendall  & 0 & 9 & 36 & 0  & 45  & 16 & 72 & 153 & 54 \\
RY+CRY-MI       & 0 & 9 & 36 & 0  & 45  & 16 & 72 & 153 & 54 \\
RY+CRY-Distance & 0 & 9 & 36 & 0  & 45  & 16 & 72 & 153 & 54 \\
Z               & 9 & 0 & 0  & 0  & 18  & 2  & 0  & 18  & 2 \\
ZZ linear       & 9 & 0 & 0  & 16 & 42  & 26 & 16 & 42  & 26 \\
ZZ circular     & 9 & 0 & 0  & 18 & 45  & 29 & 18 & 45  & 29 \\
ZZ full         & 9 & 0 & 0  & 72 & 126 & 47 & 72 & 126 & 47 \\
\bottomrule
\end{tabular}
\end{table*}


\begin{figure*}[!htbp]
\centering
\begin{subfigure}{0.48\textwidth}
\centering
\includegraphics[width=\linewidth]{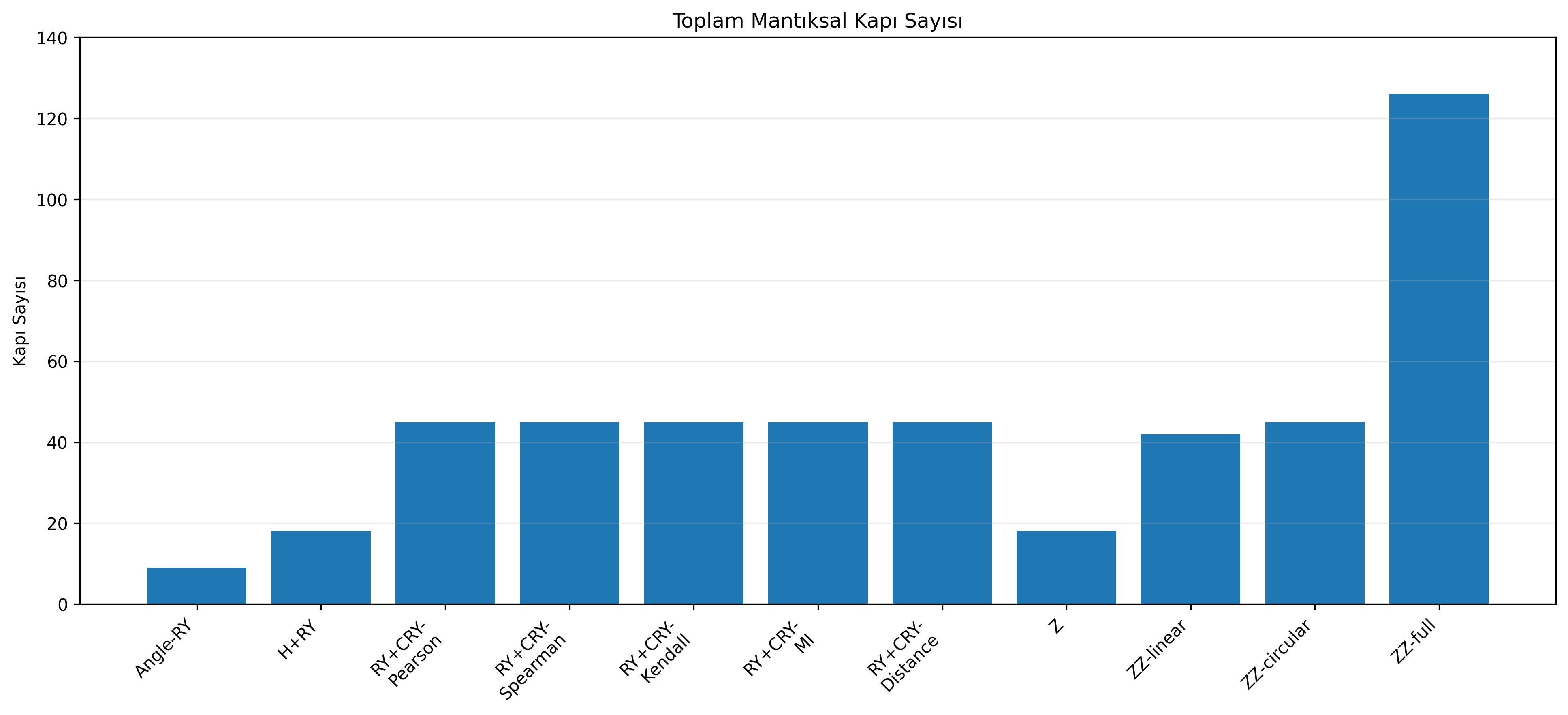}
\caption{Total number of logical gates.}
\end{subfigure}
\hfill
\begin{subfigure}{0.48\textwidth}
\centering
\includegraphics[width=\linewidth]{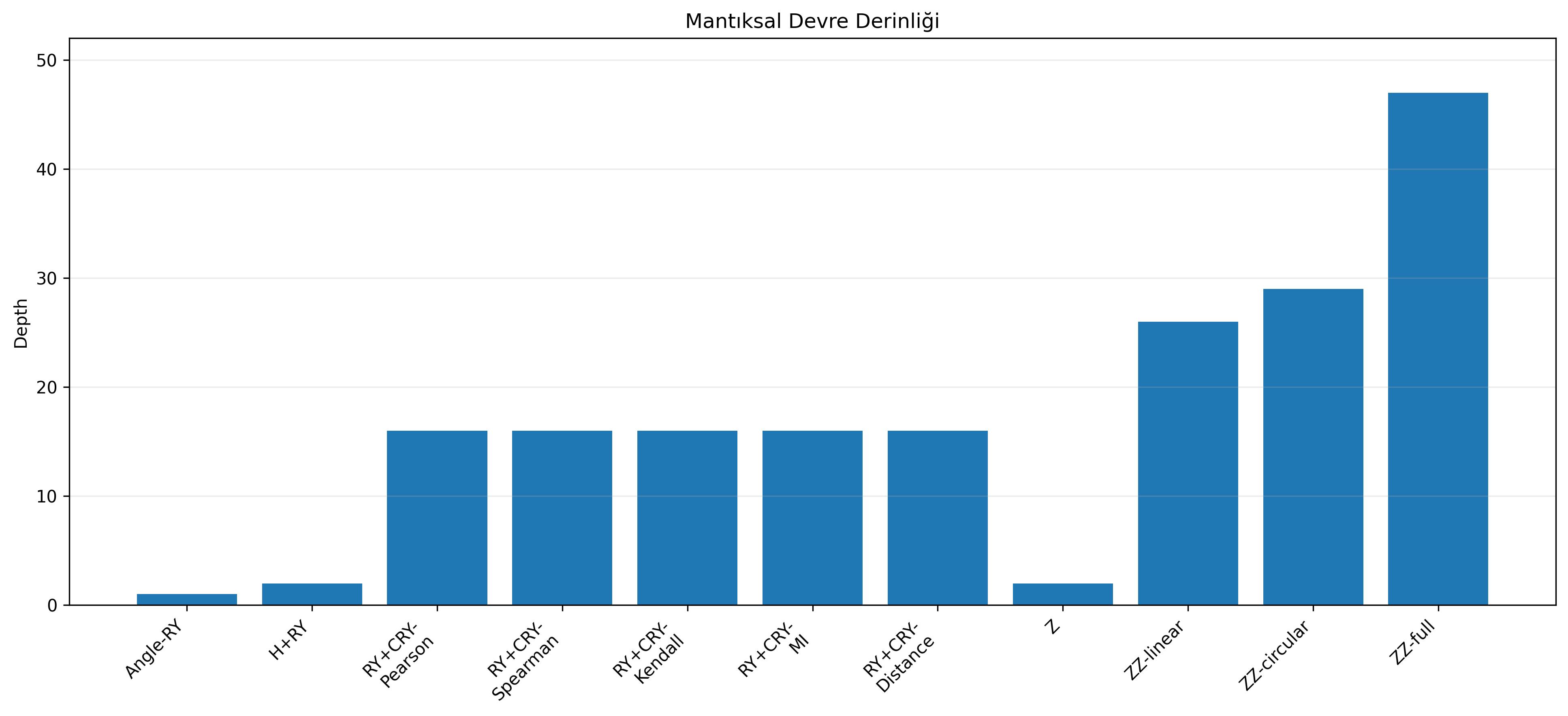}
\caption{Logical circuit depth.}
\end{subfigure}
\caption{Comparison of total gate count and circuit depth for the Credit
Risk Dataset.}
\label{fig:credit_gate_depth}
\end{figure*}

Nine features were used in the Credit Risk experiment. Therefore, the
Angle-RY encoding consists of nine RY gates, while the number of all
possible pairwise feature relationships is

\begin{equation}
\binom{9}{2}=36.
\end{equation}

Consequently, each RY+CRY method contains 36 CRY gates.

The RY+CRY circuits have a total logical gate count of 45 and a circuit
depth of 16. Since the circuit structure is identical for all five
dependency measures, the differences in ROC-AUC cannot be attributed to
circuit size but instead arise from the dependency measures used to
determine the CRY rotation angles.

Z feature Map achieved the same test accuracy of 0.908 with only 18 logical
gates and a circuit depth of 2. However, its low ROC-AUC demonstrates that
an evaluation based solely on Accuracy could overlook important differences
between the methods.


\begin{figure}[!htbp]
\centering
\includegraphics[width=0.95\columnwidth]
{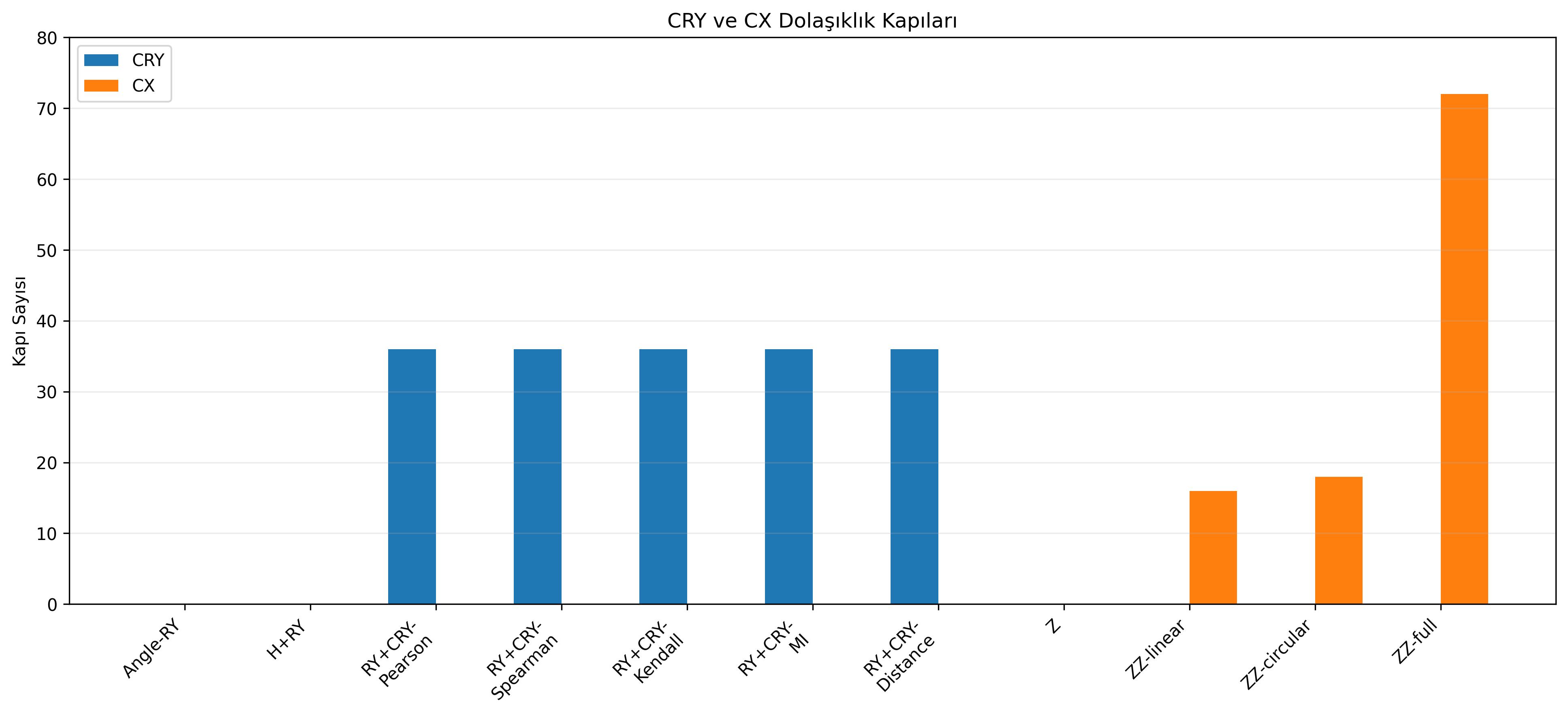}
\caption{Comparison of CRY and CX entangling gates for the Credit Risk
Dataset.}
\label{fig:credit_entangling}
\end{figure}

In terms of entangling gates, the RY+CRY methods contain 36 logical CRY
operations, which result in 72 CX gates after decomposition.

The numbers of CX gates for ZZ linear, ZZ circular, and ZZ full are 16,
18, and 72, respectively. Nevertheless, all three methods achieved the same
test accuracy of 0.908.

This result clearly demonstrates that increasing entanglement density did
not provide an additional improvement in classification accuracy on the
Credit Risk Dataset.


\begin{figure}[!htbp]
\centering
\includegraphics[width=0.95\columnwidth]
{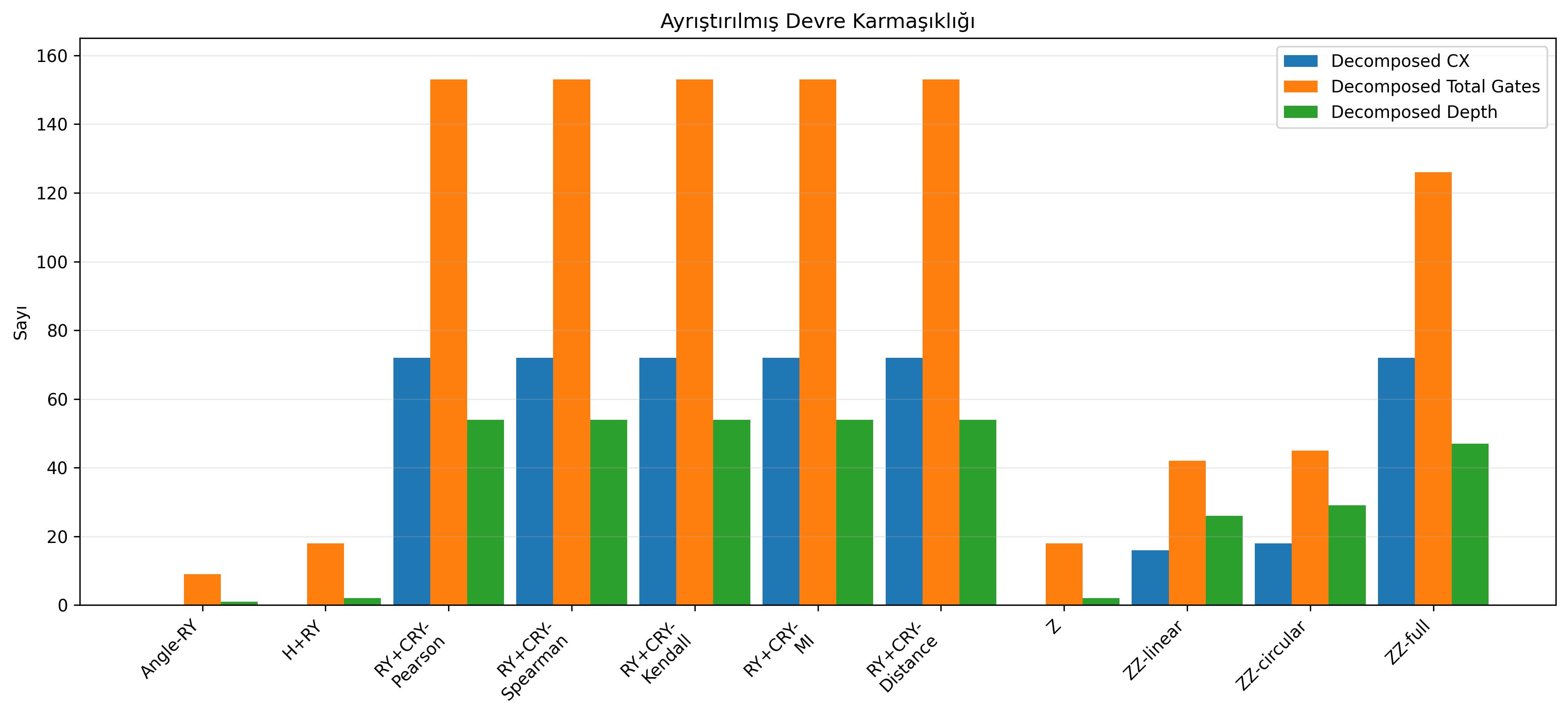}
\caption{Comparison of decomposed circuit complexity for the Credit Risk
Dataset.}
\label{fig:credit_decomposed}
\end{figure}

After decomposition, the RY+CRY circuits contain 153 total gates, 72 CX
gates, and a circuit depth of 54. In contrast, Angle-RY contains only
9 gates and has a circuit depth of 1.

Despite this substantial increase in circuit complexity, Angle-RY and the
RY+CRY methods achieved identical test accuracies. Therefore, the increase
in circuit complexity did not translate into an improvement in
threshold-dependent classification performance for this dataset.

The potential advantage of RY+CRY is more apparent in the ROC-AUC results.
For example, the ROC-AUC increased from 0.654 for Angle-RY to 0.663 for
the Distance Correlation-based RY+CRY method. However, this limited
improvement should be considered together with the cost of 153 decomposed
gates and 72 CX gates.

ZZ full is the most complex ZZ-based structure, with 126 logical gates,
72 CX gates, and a circuit depth of 47. Nevertheless, its test Accuracy,
Precision, Recall, and F1 values were identical to those obtained with the
simplest Angle-RY encoding.

\subsubsection{Computational Cost}

\begin{table}[!htbp]
\centering
\caption{Computation times for the Credit Risk Dataset (seconds).}
\label{tab:credit_time}

\footnotesize
\setlength{\tabcolsep}{4pt}
\renewcommand{\arraystretch}{1.10}

\begin{tabular}{lrrr}
\toprule
\textbf{Encoding} &
\makecell{\textbf{state vector}\\\textbf{Time}} &
\makecell{\textbf{Kernel}\\\textbf{Time}} &
\makecell{\textbf{Training}\\\textbf{Time}} \\
\midrule
Angle-RY        & 0.774 & 0.0547 & 0.0234 \\
H+RY            & 1.254 & 0.0546 & 0.0231 \\
RY+CRY-Pearson  & 3.761 & 0.0538 & 0.0238 \\
RY+CRY-Spearman & 3.935 & 0.0543 & 0.0235 \\
RY+CRY-Kendall  & 4.072 & 0.0554 & 0.0235 \\
RY+CRY-MI       & 4.033 & 0.0540 & 0.0236 \\
RY+CRY-Distance & 3.652 & 0.0607 & 0.0247 \\
Z               & 1.403 & 0.0540 & 0.0268 \\
ZZ linear       & 2.768 & 0.0547 & 0.0425 \\
ZZ circular     & 2.899 & 0.0547 & 0.0442 \\
ZZ full         & 7.396 & 0.0548 & 0.0493 \\
\bottomrule
\end{tabular}
\end{table}


\begin{figure}[!htbp]
\centering
\includegraphics[width=0.90\columnwidth]
{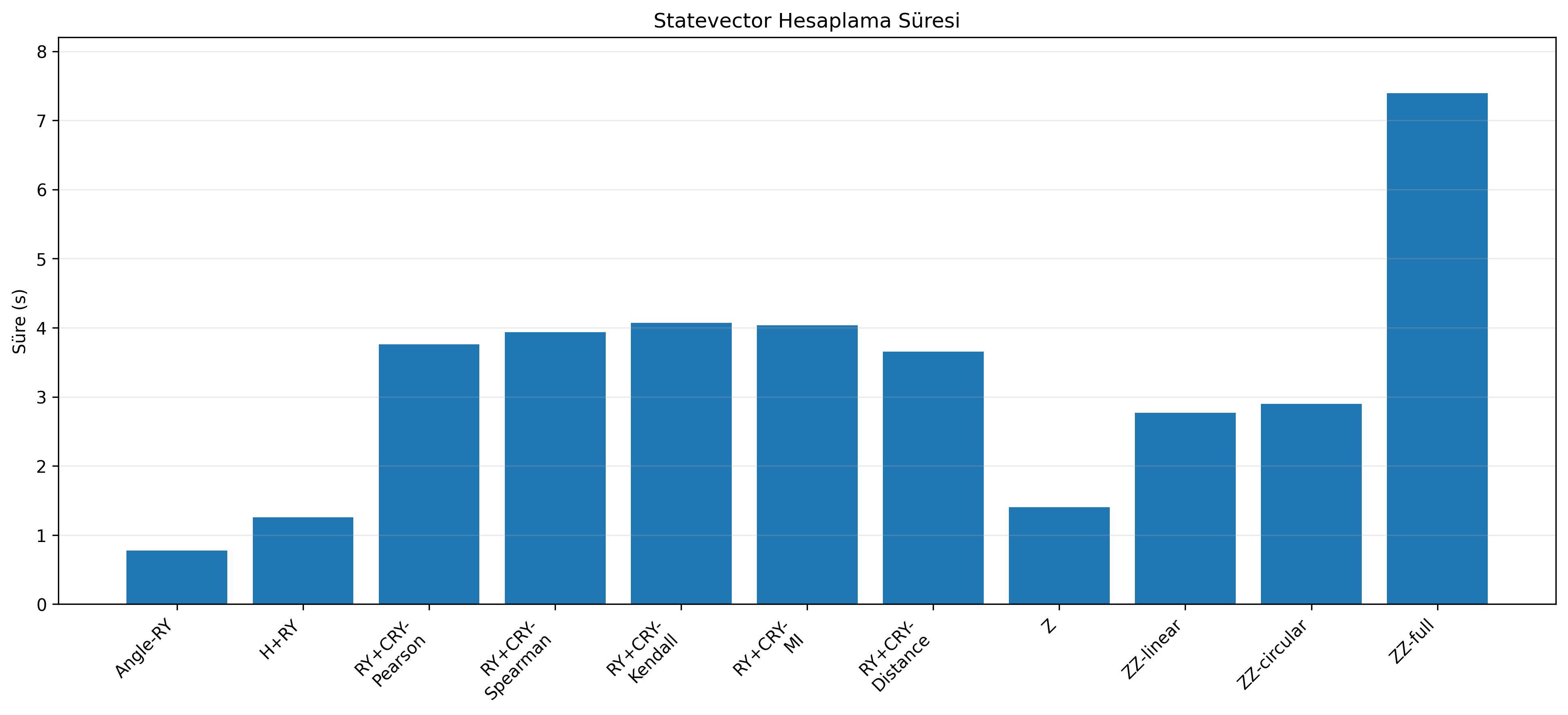}
\caption{state vector computation times for the Credit Risk Dataset.}
\label{fig:credit_state vector}
\end{figure}

The state vector computation time generally followed the circuit complexity.
The lowest computation time, 0.774 seconds, was obtained with Angle-RY.
H+RY and Z feature Map required approximately 1.25 and 1.40 seconds,
respectively.

The state vector computation times of the RY+CRY methods ranged from
approximately 3.65 to 4.07 seconds, whereas ZZ full had the highest
state vector cost at 7.40 seconds.

The state vector computational cost of ZZ full relative to Angle-RY was
approximately

\begin{equation}
\frac{7.396}{0.774}\approx9.6.
\end{equation}

Thus, ZZ full required approximately 9.6 times more state vector computation
time than Angle-RY. Nevertheless, the two methods produced identical test
Accuracy, Precision, Recall, and F1 values. This result strongly indicates
that the increase in circuit complexity did not directly translate into
improved classification output for the Credit Risk Dataset.


\begin{figure}[!htbp]
\centering
\includegraphics[width=0.90\columnwidth]
{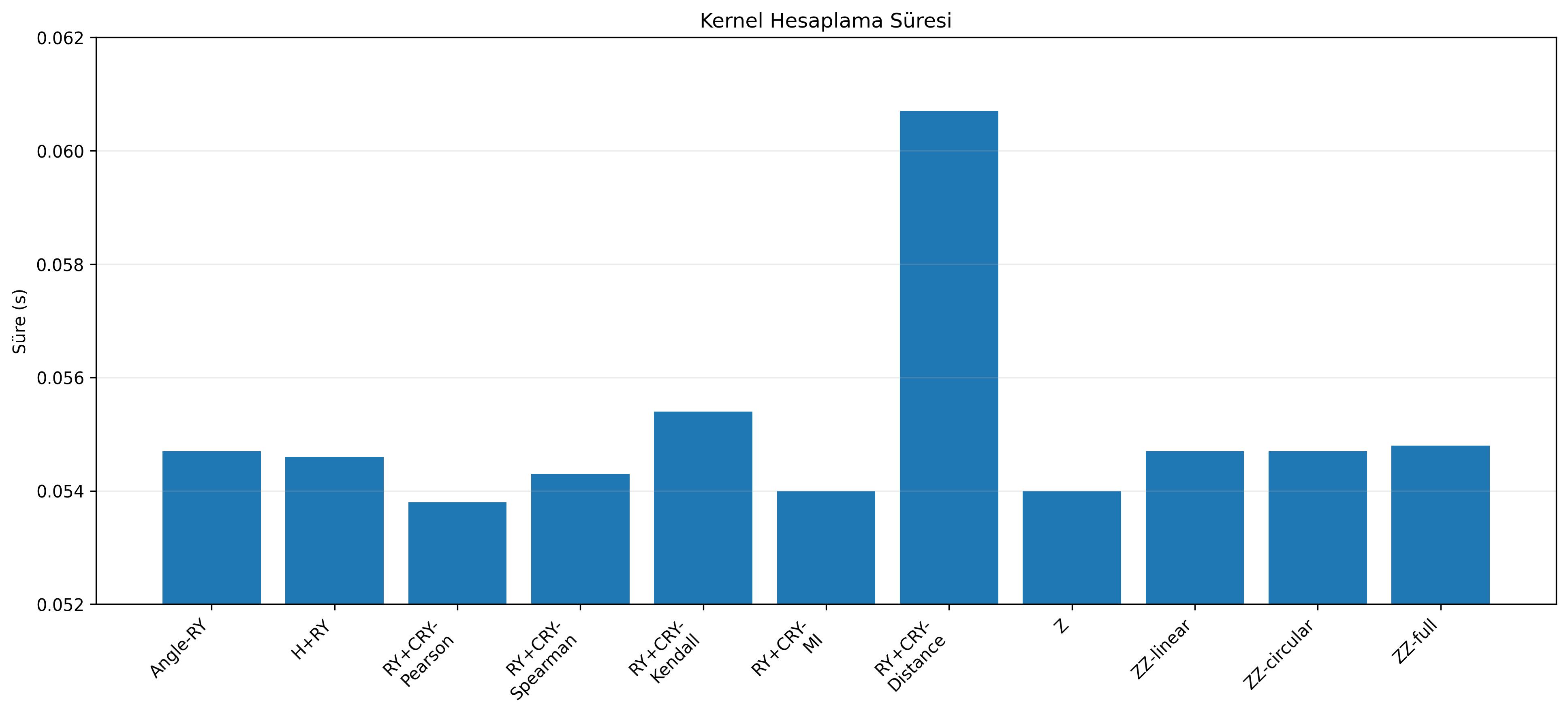}
\caption{Kernel computation times for the Credit Risk Dataset.}
\label{fig:credit_kernel}
\end{figure}

Kernel computation times generally remained within approximately
0.054--0.055 seconds. Although a value of approximately 0.061 seconds was
observed for the Distance Correlation-based RY+CRY method, the differences
among the methods were generally small.

This result indicates that the differences in total computational cost
primarily arose during quantum-state preparation rather than from the
subsequent processing of the kernel matrix.


\begin{figure}[!htbp]
\centering
\includegraphics[width=0.90\columnwidth]
{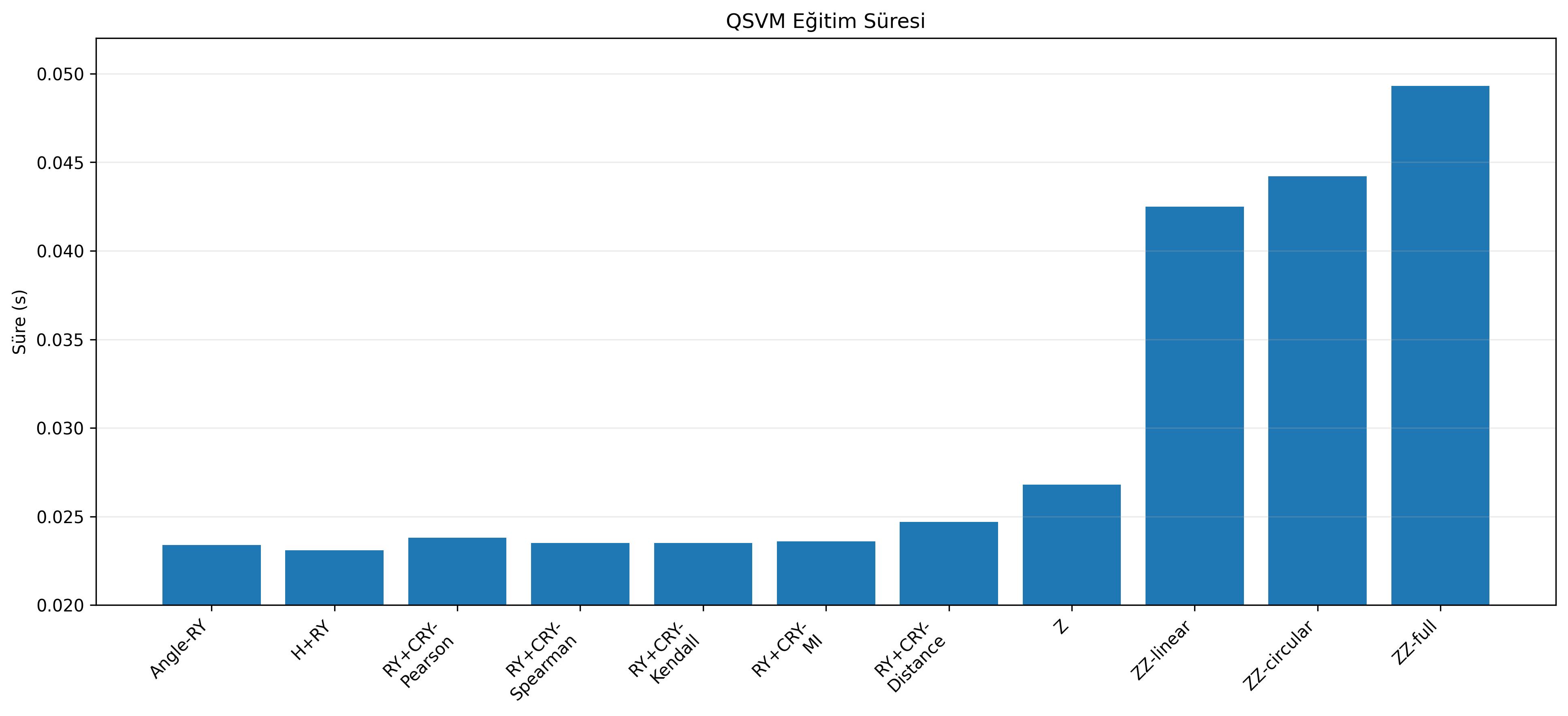}
\caption{QSVM training times for the Credit Risk Dataset.}
\label{fig:credit_training}
\end{figure}

The training times for Angle-RY and the RY+CRY methods were approximately
0.023--0.025 seconds. They increased to approximately 0.043--0.044 seconds
for ZZ linear and ZZ circular and to approximately 0.049 seconds for
ZZ full.

Nevertheless, all training times were substantially smaller than the
state vector computation times. Therefore, the primary source of the
computational cost differences among the methods was again the preparation
of the quantum feature states.

\subsubsection{Overall Evaluation of the Credit Risk Dataset}

The Credit Risk Dataset exhibited a different behavior from the other
datasets. All quantum feature encoding strategies produced the same test
Accuracy of 0.908, Precision of 0.824, Recall of 0.908, and F1 score of
0.864. Therefore, when only these metrics are considered, no performance
difference among the methods is observed.

However, the ROC-AUC results demonstrate that the quantum feature maps are
not entirely equivalent. The highest ROC-AUC, 0.663, was obtained with the
Distance Correlation-based RY+CRY method, whereas Z feature Map produced only
0.464. Thus, despite identical final classification accuracy, the ranking
of samples according to their decision scores may differ substantially.

In terms of generalization, the training--test accuracy gaps for Angle-RY
and the RY+CRY methods were nearly zero. In contrast, as the ZZ connectivity
density increased, training accuracy improved while test accuracy remained
unchanged. This indicates that greater feature-map complexity improved the
fit to the training data without providing an additional benefit in
generalization performance.

A similar conclusion can be drawn from circuit cost. Angle-RY achieved a
test accuracy of 0.908 using only 9 logical gates and a circuit depth of 1,
whereas ZZ full produced the same test accuracy using 126 gates and a
circuit depth of 47. The state vector computation time also increased from
approximately 0.77 seconds to 7.40 seconds.

Therefore, the use of more complex encodings does not provide a clear
advantage in terms of threshold-dependent classification performance for
the Credit Risk Dataset. The limited advantage of the RY+CRY methods is
primarily observed in threshold-independent discrimination metrics such as
ROC-AUC.

Overall, the Credit Risk experiments demonstrate that higher circuit
complexity does not necessarily translate into higher classification
performance for every dataset. For this dataset in particular, quantum
feature encoding strategies should be evaluated not only in terms of
Accuracy but also by jointly considering ROC-AUC, Log-Loss, generalization
gap, and circuit cost.

\subsection{Student Performance Dataset}
\label{subsec:student_results}

The experiments conducted on the Student Performance Dataset compared
Angle-RY, H+RY, RY+CRY encodings based on different statistical dependency
measures, Z feature Map, and ZZ feature Map with different entanglement
structures. The evaluation was performed in terms of classification
performance, generalization behavior, quantum circuit complexity, and
computational cost.

\subsubsection{Classification Performance}

\begin{table*}[!htbp]
\centering
\caption{Classification performance of quantum feature encoding strategies
on the Student Performance Dataset.}
\label{tab:student_performance}

\footnotesize
\setlength{\tabcolsep}{4pt}
\renewcommand{\arraystretch}{1.10}

\begin{tabular}{lccccccc}
\toprule
\textbf{Encoding} &
\makecell{\textbf{Train}\\\textbf{Acc.}} &
\makecell{\textbf{Test}\\\textbf{Acc.}} &
\makecell{\textbf{Acc.}\\\textbf{Gap}} &
\makecell{\textbf{Test}\\\textbf{Prec.}} &
\makecell{\textbf{Test}\\\textbf{Rec.}} &
\makecell{\textbf{Test}\\\textbf{F1}} &
\makecell{\textbf{Test}\\\textbf{AUC}} \\
\midrule
Angle-RY        & 0.995 & 0.882 & 0.113 & 0.900 & 0.673 & 0.770 & 0.953 \\
H+RY            & 0.995 & 0.882 & 0.113 & 0.900 & 0.673 & 0.770 & 0.953 \\
RY+CRY-Pearson  & 0.998 & 0.890 & 0.108 & 0.918 & 0.687 & 0.786 & 0.962 \\
RY+CRY-Spearman & 0.998 & 0.890 & 0.108 & 0.918 & 0.687 & 0.786 & 0.962 \\
RY+CRY-Kendall  & 0.997 & 0.890 & 0.107 & 0.918 & 0.687 & 0.786 & 0.961 \\
RY+CRY-MI       & 0.997 & 0.886 & 0.111 & 0.902 & 0.687 & 0.780 & 0.960 \\
RY+CRY-Distance & 0.998 & 0.890 & 0.108 & 0.918 & 0.687 & 0.786 & 0.963 \\
Z               & 0.968 & 0.944 & 0.024 & 0.947 & 0.857 & 0.900 & 0.975 \\
ZZ linear       & 0.997 & 0.908 & 0.089 & 0.932 & 0.741 & 0.826 & 0.960 \\
ZZ circular     & 0.996 & 0.888 & 0.108 & 0.925 & 0.673 & 0.780 & 0.948 \\
ZZ full         & 1.000 & 0.712 & 0.288 & 1.000 & 0.020 & 0.040 & 0.909 \\
\bottomrule
\end{tabular}
\end{table*}


\begin{figure}[!htbp]
\centering
\includegraphics[width=0.95\columnwidth]{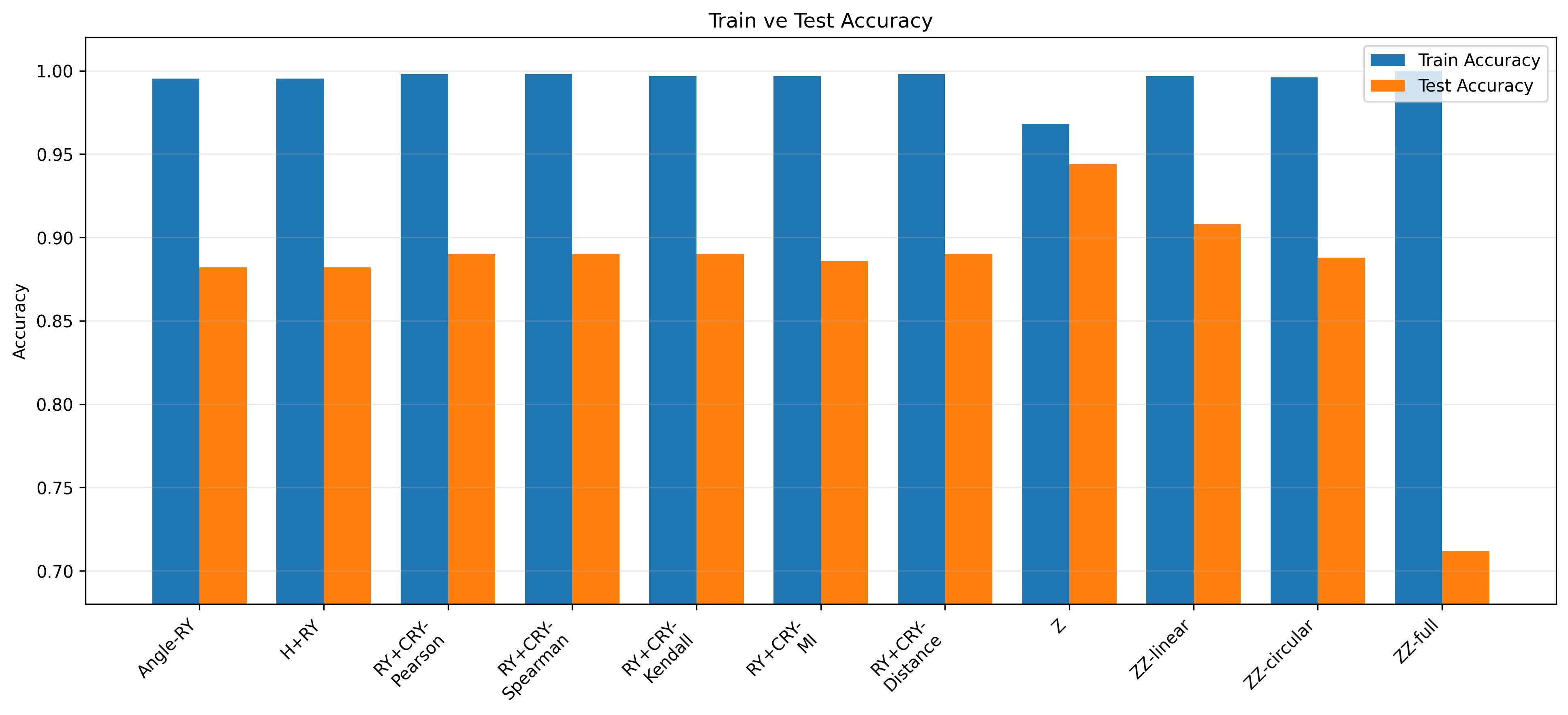}
\caption{Comparison of training and test accuracy across encoding strategies
for the Student Performance Dataset.}
\label{fig:student_accuracy}
\end{figure}

Table~\ref{tab:student_performance} and Fig.~\ref{fig:student_accuracy}
show that the highest test accuracy on the Student Performance Dataset was
obtained with Z feature Map, which achieved 0.944. ZZ linear produced the
second-highest test accuracy at 0.908, whereas the Pearson-, Spearman-,
Kendall-, and Distance Correlation-based RY+CRY methods achieved 0.890.

Angle-RY and H+RY both achieved a test accuracy of 0.882. Although the
improvement provided by the RY+CRY encodings over this baseline was limited,
it was consistent. For example, the absolute increase in accuracy for the
Pearson-based RY+CRY method was

\begin{equation}
0.890-0.882=0.008,
\end{equation}

corresponding to approximately 0.8 percentage points.

In contrast, the improvement of Z feature Map over Angle-RY was

\begin{equation}
0.944-0.882=0.062,
\end{equation}

corresponding to approximately 6.2 percentage points. This result suggests
that the Z-based feature map provides a representation better suited to the
structure of the Student Performance Dataset than the other investigated
encodings.

ZZ full, however, achieved a training accuracy of 1.000 while its test
accuracy decreased to only 0.712. This clearly demonstrates that greater
model complexity does not guarantee improved test performance.


\begin{figure}[!htbp]
\centering
\includegraphics[width=0.95\columnwidth]
{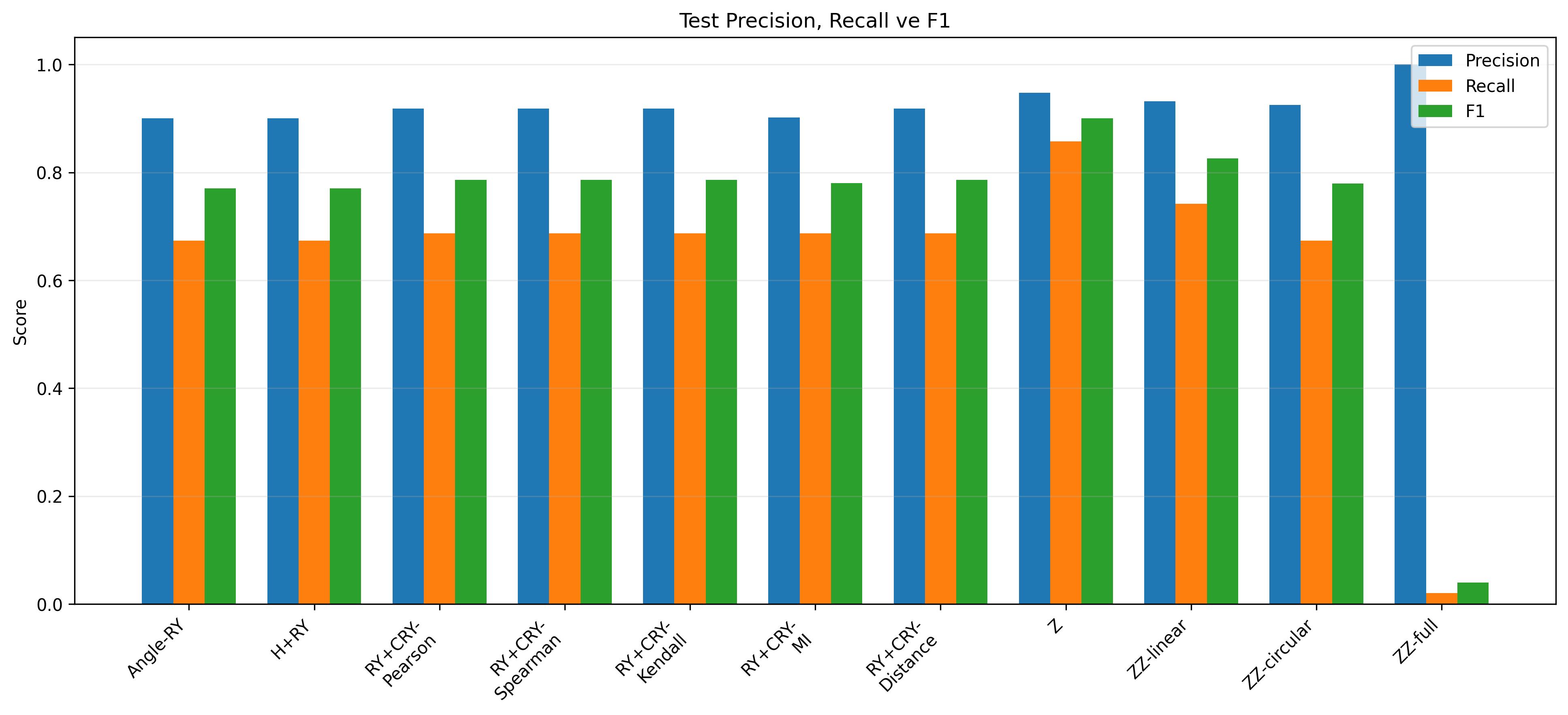}
\caption{Comparison of test Precision, Recall, and F1 scores for the Student
Performance Dataset.}
\label{fig:student_prf}
\end{figure}

Figure~\ref{fig:student_prf} further highlights the superior performance of
Z feature Map. It achieved a Precision of 0.947, Recall of 0.857, and F1 score
of 0.900, providing the most balanced test performance among all methods.

The Pearson-, Spearman-, Kendall-, and Distance Correlation-based RY+CRY
methods achieved a Precision of 0.918, Recall of 0.687, and an F1 score of
approximately 0.786. Although Precision remained high for these methods,
their lower Recall values reduced the overall F1 score.

ZZ linear achieved a Precision of 0.932 and a Recall of 0.741, resulting in
an F1 score of 0.826. This performance exceeded that of the RY+CRY methods,
but remained below the F1 score of 0.900 obtained with Z feature Map.

The most striking result was observed for ZZ full. Although its Precision
reached 1.000, its Recall was only 0.020, resulting in an F1 score of 0.040.
Thus, while ZZ full was highly accurate for the very small number of
samples it predicted as positive, it failed to identify nearly all actual
positive samples.


\begin{figure}[!htbp]
\centering
\includegraphics[width=0.90\columnwidth]{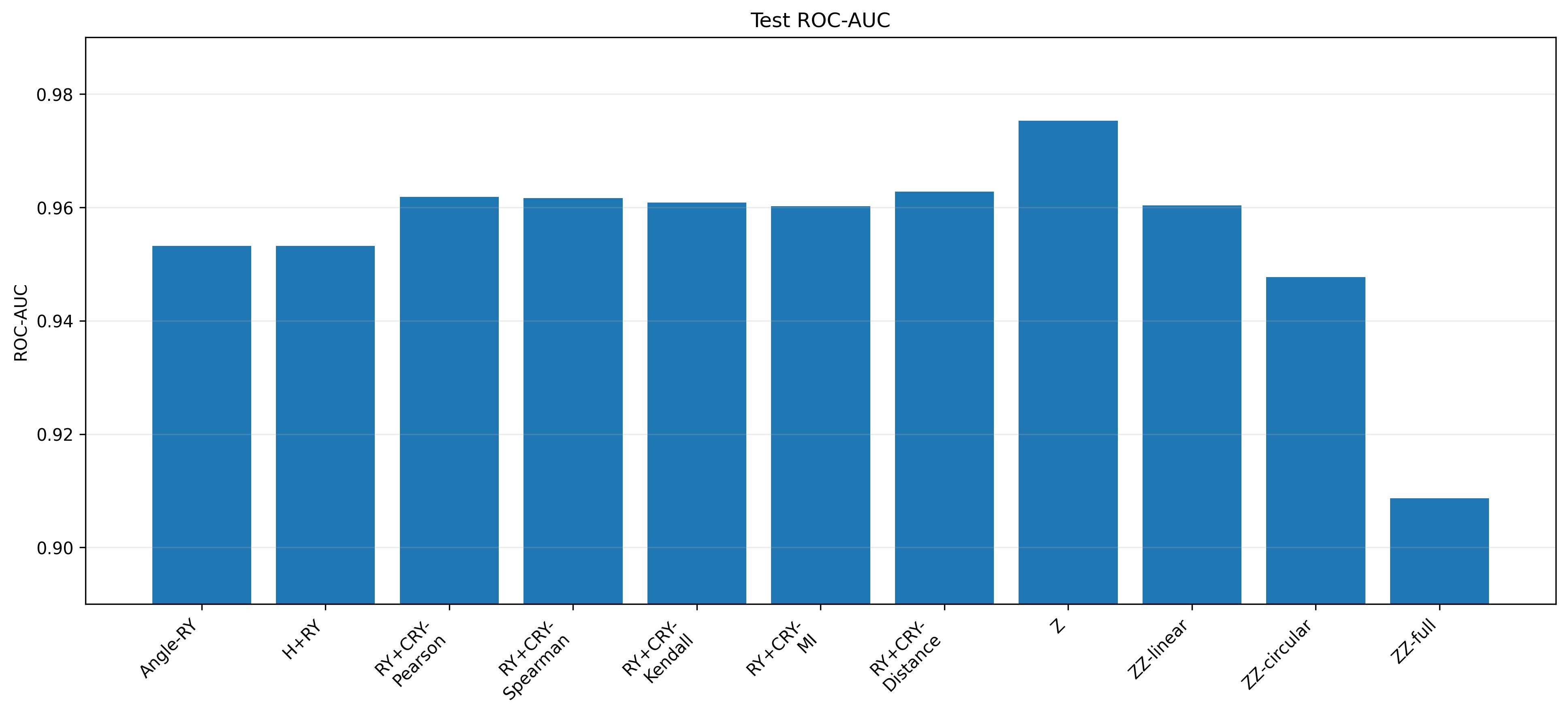}
\caption{Comparison of test ROC-AUC values for the Student Performance
Dataset.}
\label{fig:student_auc}
\end{figure}

The ROC-AUC results also support the superiority of Z feature Map. The highest
test ROC-AUC value, 0.975, was obtained with Z feature Map.

Among the RY+CRY methods, Distance Correlation achieved the highest ROC-AUC
at 0.963, followed closely by Pearson and Spearman at approximately 0.962.
This finding indicates that dependency-based CRY rotations can improve class
separability relative to Angle-RY to some extent.

ZZ linear achieved a strong ROC-AUC value of 0.960, whereas ZZ circular and
ZZ full decreased to 0.948 and 0.909, respectively. Therefore, increasing
the connectivity density did not provide a systematic advantage in terms of
ROC-AUC for this dataset.

\subsubsection{Generalization Behavior and Log-Loss}

\begin{table*}[!htbp]
\centering
\caption{Training--test performance gaps and Log-Loss values for the Student
Performance Dataset.}
\label{tab:student_generalization}

\footnotesize
\setlength{\tabcolsep}{3.8pt}
\renewcommand{\arraystretch}{1.10}

\begin{tabular}{lccccccc}
\toprule
\textbf{Encoding} &
\makecell{\textbf{Acc.}\\\textbf{Gap}} &
\makecell{\textbf{Prec.}\\\textbf{Gap}} &
\makecell{\textbf{Recall}\\\textbf{Gap}} &
\makecell{\textbf{F1}\\\textbf{Gap}} &
\makecell{\textbf{AUC}\\\textbf{Gap}} &
\makecell{\textbf{Train}\\\textbf{Loss}} &
\makecell{\textbf{Test}\\\textbf{Loss}} \\
\midrule
Angle-RY        & 0.113 & 0.093 & 0.318 & 0.222 & 0.047 & 0.022 & 0.260 \\
H+RY            & 0.113 & 0.093 & 0.318 & 0.222 & 0.047 & 0.022 & 0.260 \\
RY+CRY-Pearson  & 0.108 & 0.080 & 0.308 & 0.211 & 0.038 & 0.013 & 0.236 \\
RY+CRY-Spearman & 0.108 & 0.080 & 0.308 & 0.211 & 0.038 & 0.013 & 0.237 \\
RY+CRY-Kendall  & 0.107 & 0.077 & 0.306 & 0.208 & 0.039 & 0.015 & 0.240 \\
RY+CRY-MI       & 0.111 & 0.094 & 0.306 & 0.214 & 0.040 & 0.016 & 0.245 \\
RY+CRY-Distance & 0.108 & 0.080 & 0.308 & 0.211 & 0.037 & 0.012 & 0.233 \\
Z               & 0.024 & 0.043 & 0.044 & 0.043 & 0.021 & 0.079 & 0.160 \\
ZZ linear       & 0.089 & 0.068 & 0.247 & 0.169 & 0.040 & 0.011 & 0.217 \\
ZZ circular     & 0.108 & 0.075 & 0.313 & 0.214 & 0.052 & 0.008 & 0.252 \\
ZZ full         & 0.288 & 0.000 & 0.980 & 0.960 & 0.091 & 0.001 & 0.349 \\
\bottomrule
\end{tabular}
\end{table*}


\begin{figure}[!htbp]
\centering
\includegraphics[width=0.90\columnwidth]
{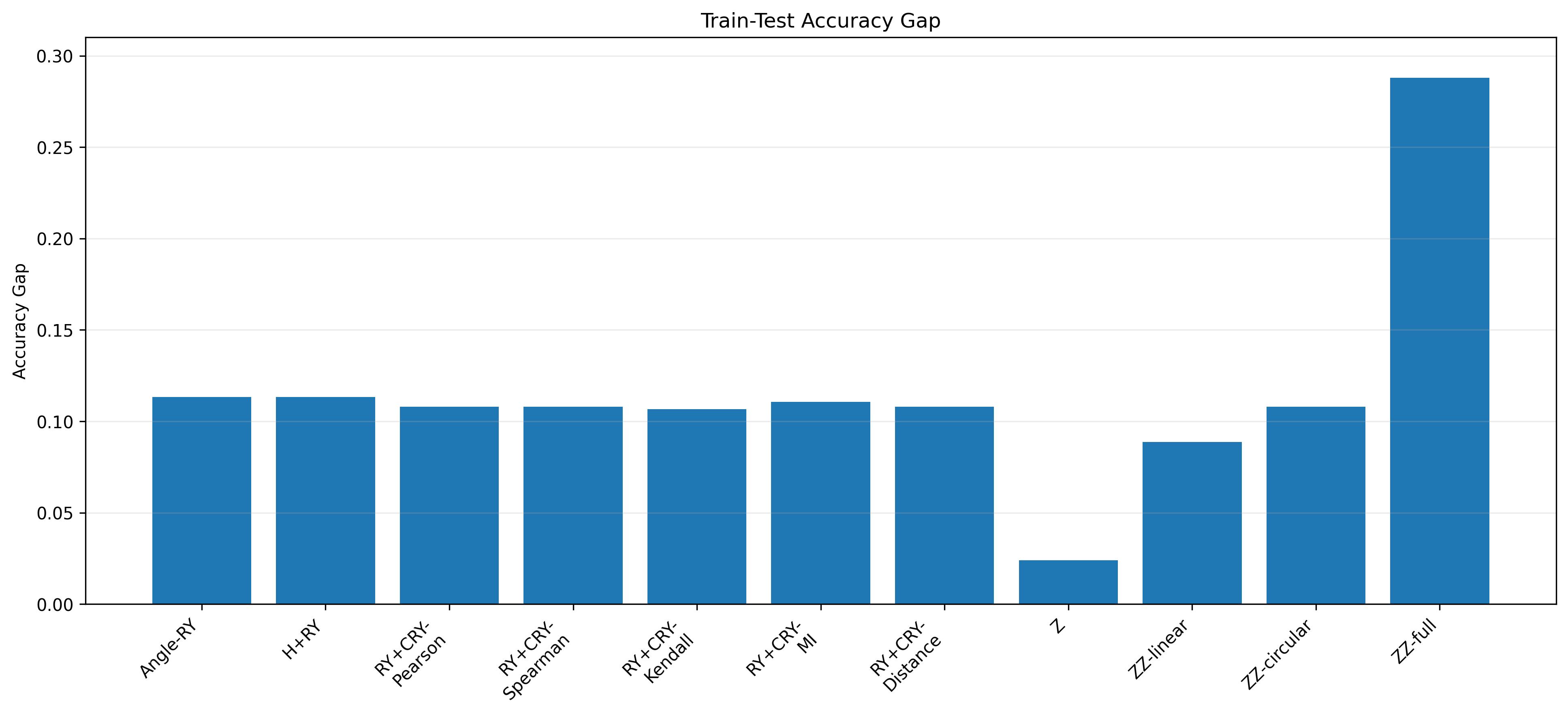}
\caption{Training--test Accuracy Gap across encoding strategies for the
Student Performance Dataset.}
\label{fig:student_accuracy_gap}
\end{figure}

In terms of generalization behavior, Z feature Map produced the most notable
result. Its training accuracy was 0.968 and its test accuracy was 0.944,
resulting in an Accuracy Gap of only 0.024.

In contrast, Angle-RY and H+RY produced an Accuracy Gap of approximately
0.113, whereas the RY+CRY methods produced values in the range of
approximately 0.107--0.111. These methods achieved training accuracies of
approximately 0.997--0.998 but remained near 0.89 in test accuracy.

The Accuracy Gap was 0.089 for ZZ linear and increased to 0.108 for
ZZ circular. The highest generalization gap, 0.288, was obtained with
ZZ full.

Although ZZ full achieved a training accuracy of 1.000, its test accuracy
decreased to 0.712, indicating a strong overfitting tendency. This behavior
is even more apparent from its Recall Gap of approximately 0.980 and F1 Gap
of 0.960.


\begin{figure}[!htbp]
\centering
\includegraphics[width=0.90\columnwidth]{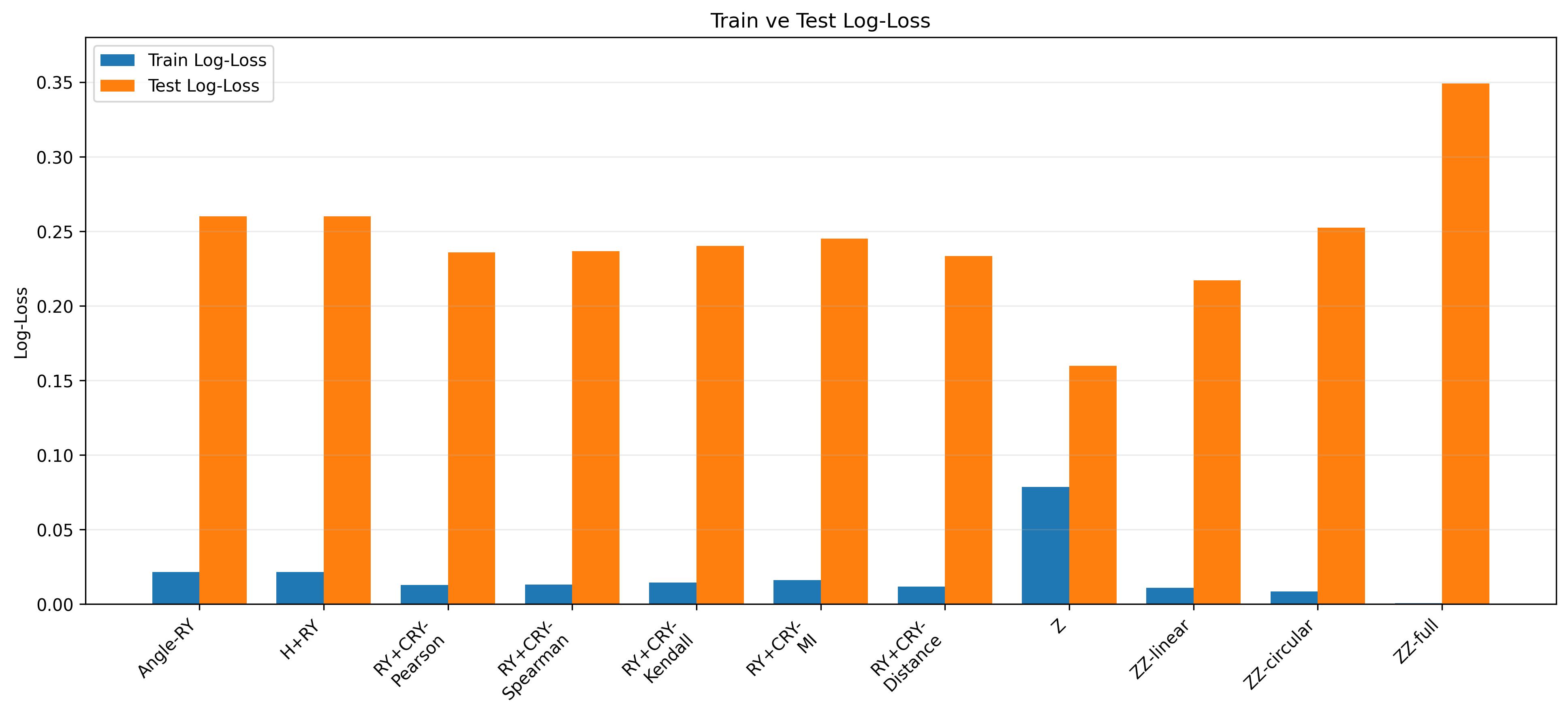}
\caption{Comparison of training and test Log-Loss values for the Student
Performance Dataset.}
\label{fig:student_logloss}
\end{figure}

The Log-Loss results support the generalization analysis. The Test Log-Loss
was approximately 0.260 for Angle-RY and H+RY, whereas it decreased to
approximately 0.236 for the Pearson- and Spearman-based RY+CRY methods.

RY+CRY-Distance Correlation achieved the lowest test loss within the
RY+CRY group, at 0.233. However, the lowest overall Test Log-Loss, 0.160,
was obtained with Z feature Map.

The fact that Z feature Map achieved both the highest test accuracy and the
lowest Test Log-Loss indicates that it not only predicted class labels more
accurately but also produced more reliable prediction confidence.

In contrast, ZZ full achieved a Train Log-Loss of only 0.001 while its
Test Log-Loss increased to 0.349. This sharp divergence clearly indicates
that the extremely strong fit to the training data did not generalize to
the test set.

\subsubsection{Quantum Circuit Complexity}

\begin{table*}[!htbp]
\centering
\caption{Circuit and gate complexities of quantum feature encoding strategies
for the Student Performance Dataset.}
\label{tab:student_circuit}

\footnotesize
\setlength{\tabcolsep}{3.5pt}
\renewcommand{\arraystretch}{1.10}

\begin{tabular}{lrrrrrrrrr}
\toprule
\textbf{Encoding} &
\textbf{H} &
\textbf{RY} &
\textbf{CRY} &
\textbf{CX} &
\makecell{\textbf{Total}\\\textbf{Gates}} &
\textbf{Depth} &
\makecell{\textbf{Dec.}\\\textbf{CX}} &
\makecell{\textbf{Dec.}\\\textbf{Gates}} &
\makecell{\textbf{Dec.}\\\textbf{Depth}} \\
\midrule
Angle-RY        & 0  & 13 & 0  & 0   & 13  & 1  & 0   & 13  & 1 \\
H+RY            & 13 & 13 & 0  & 0   & 26  & 2  & 0   & 26  & 2 \\
RY+CRY-Pearson  & 0  & 13 & 78 & 0   & 91  & 24 & 156 & 325 & 82 \\
RY+CRY-Spearman & 0  & 13 & 78 & 0   & 91  & 24 & 156 & 325 & 82 \\
RY+CRY-Kendall  & 0  & 13 & 78 & 0   & 91  & 24 & 156 & 325 & 82 \\
RY+CRY-MI       & 0  & 13 & 78 & 0   & 91  & 24 & 156 & 325 & 82 \\
RY+CRY-Distance & 0  & 13 & 78 & 0   & 91  & 24 & 156 & 325 & 82 \\
Z               & 13 & 0  & 0  & 0   & 26  & 2  & 0   & 26  & 2 \\
ZZ linear       & 13 & 0  & 0  & 24  & 62  & 38 & 24  & 62  & 38 \\
ZZ circular     & 13 & 0  & 0  & 26  & 65  & 41 & 26  & 65  & 41 \\
ZZ full         & 13 & 0  & 0  & 156 & 260 & 71 & 156 & 260 & 71 \\
\bottomrule
\end{tabular}
\end{table*}


\begin{figure*}[!htbp]
\centering
\begin{subfigure}{0.48\textwidth}
\centering
\includegraphics[width=\linewidth]{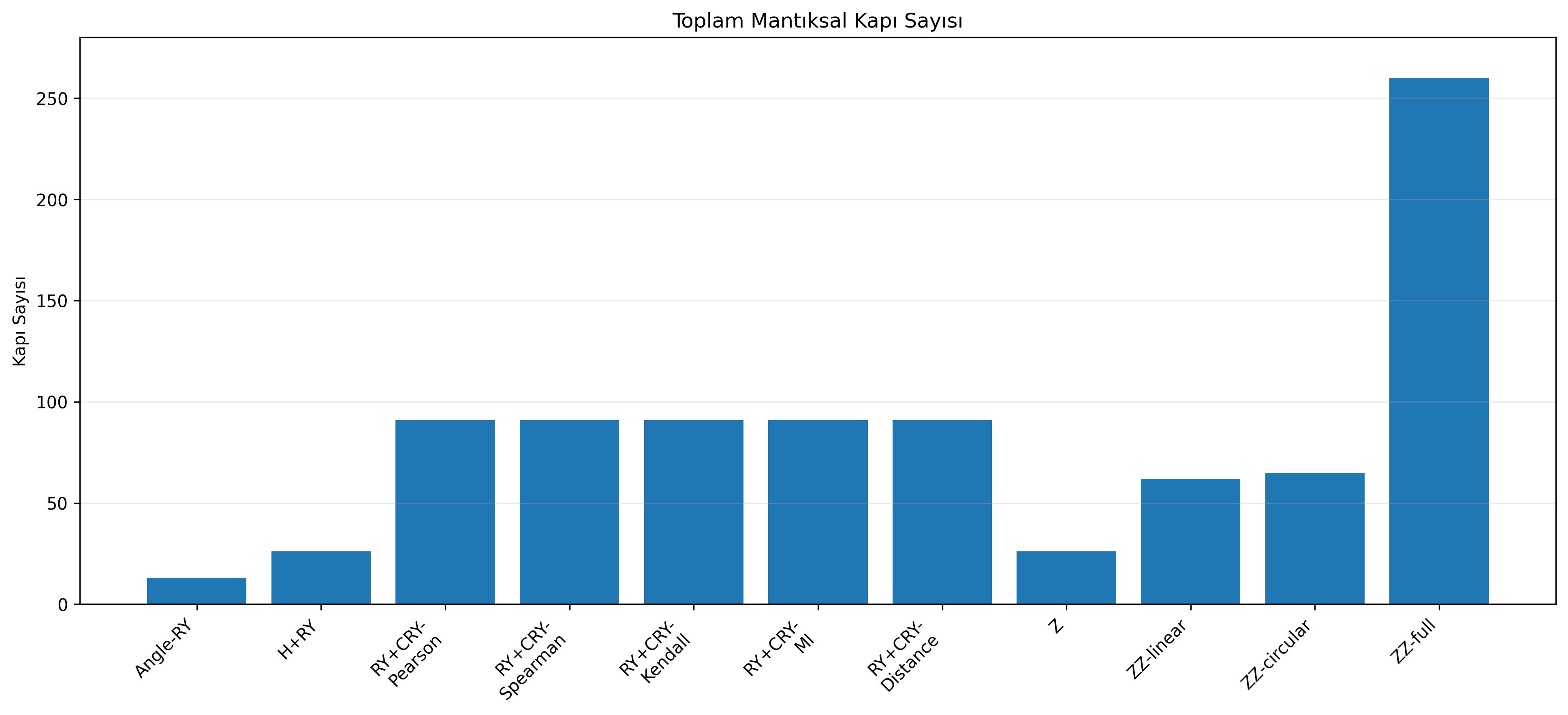}
\caption{Total number of logical gates.}
\end{subfigure}
\hfill
\begin{subfigure}{0.48\textwidth}
\centering
\includegraphics[width=\linewidth]{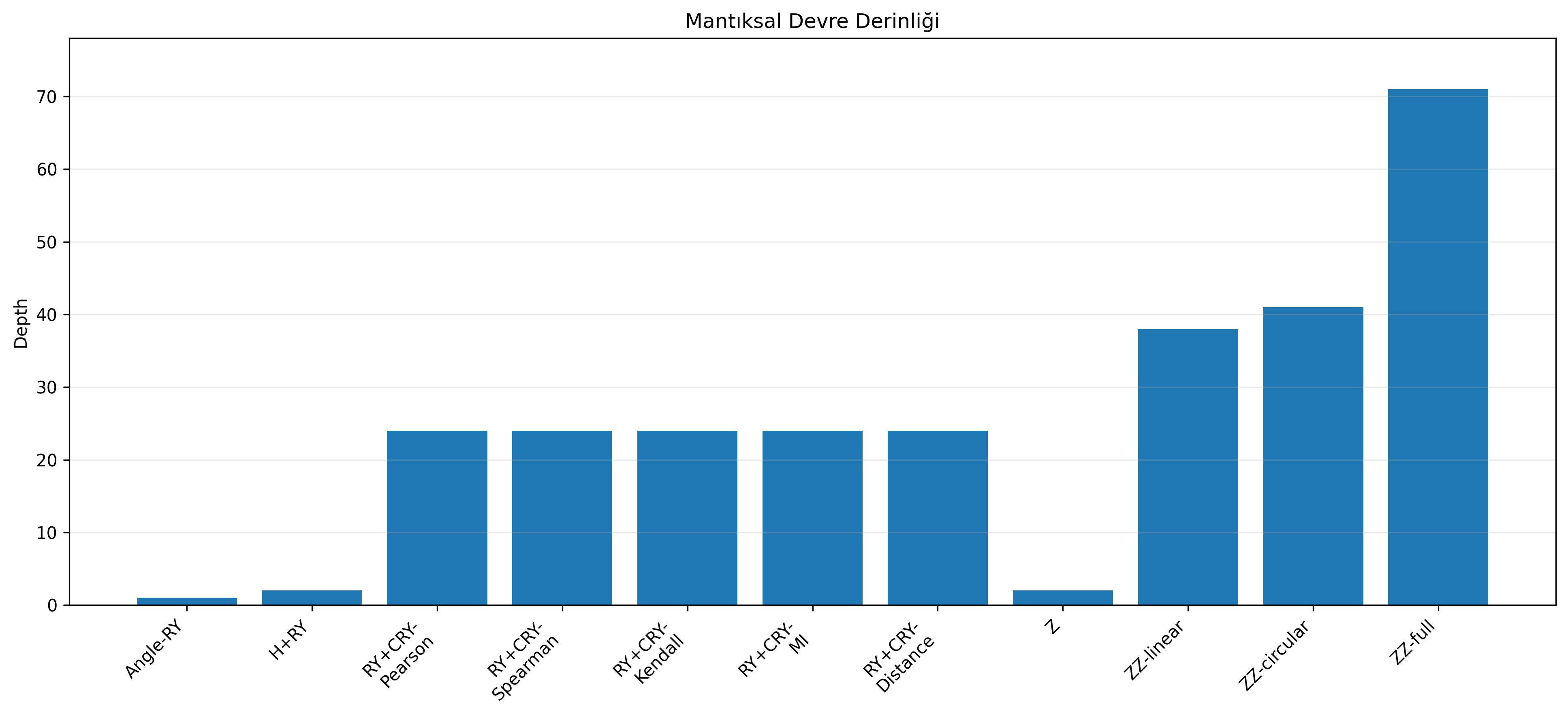}
\caption{Logical circuit depth.}
\end{subfigure}
\caption{Comparison of total gate count and circuit depth for the Student
Performance Dataset.}
\label{fig:student_gate_depth}
\end{figure*}

Since 13 features were used in the Student Performance experiment, the
Angle-RY circuit consists of 13 RY gates. The number of all unique feature
pairs is

\begin{equation}
\binom{13}{2}=78.
\end{equation}

Therefore, the RY+CRY methods contain 78 CRY gates.

At the logical level, the RY+CRY methods contain 91 gates and have a circuit
depth of 24. After decomposition, the total gate count increases to 325,
the number of CX gates to 156, and the circuit depth to 82.

In contrast, Z feature Map achieved the highest test performance among all
methods with only 26 logical gates and a circuit depth of 2. This result
demonstrates that greater circuit complexity was not required for strong
performance on the Student Performance Dataset.


\begin{figure}[!htbp]
\centering
\includegraphics[width=0.95\columnwidth]
{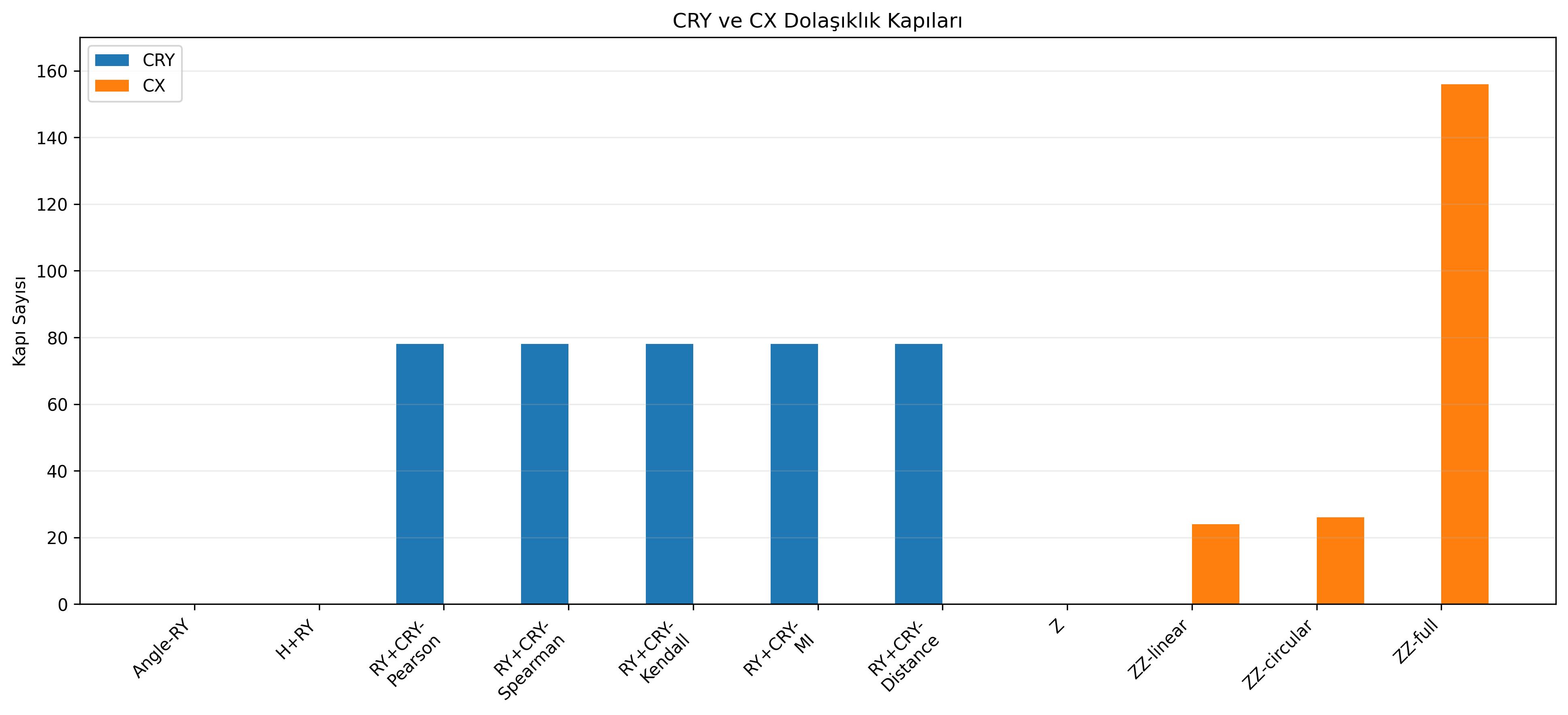}
\caption{Comparison of CRY and CX entangling gates for the Student
Performance Dataset.}
\label{fig:student_entangling}
\end{figure}

ZZ linear, ZZ circular, and ZZ full contain 24, 26, and 156 CX gates,
respectively. Nevertheless, test performance did not systematically improve
as connectivity density increased.

In particular, although ZZ full contains 156 CX gates, it achieved only
0.712 test accuracy and an F1 score of 0.040. Thus, the dense entanglement
structure did not provide a suitable representation for classification on
this dataset.


\begin{figure}[!htbp]
\centering
\includegraphics[width=0.95\columnwidth]
{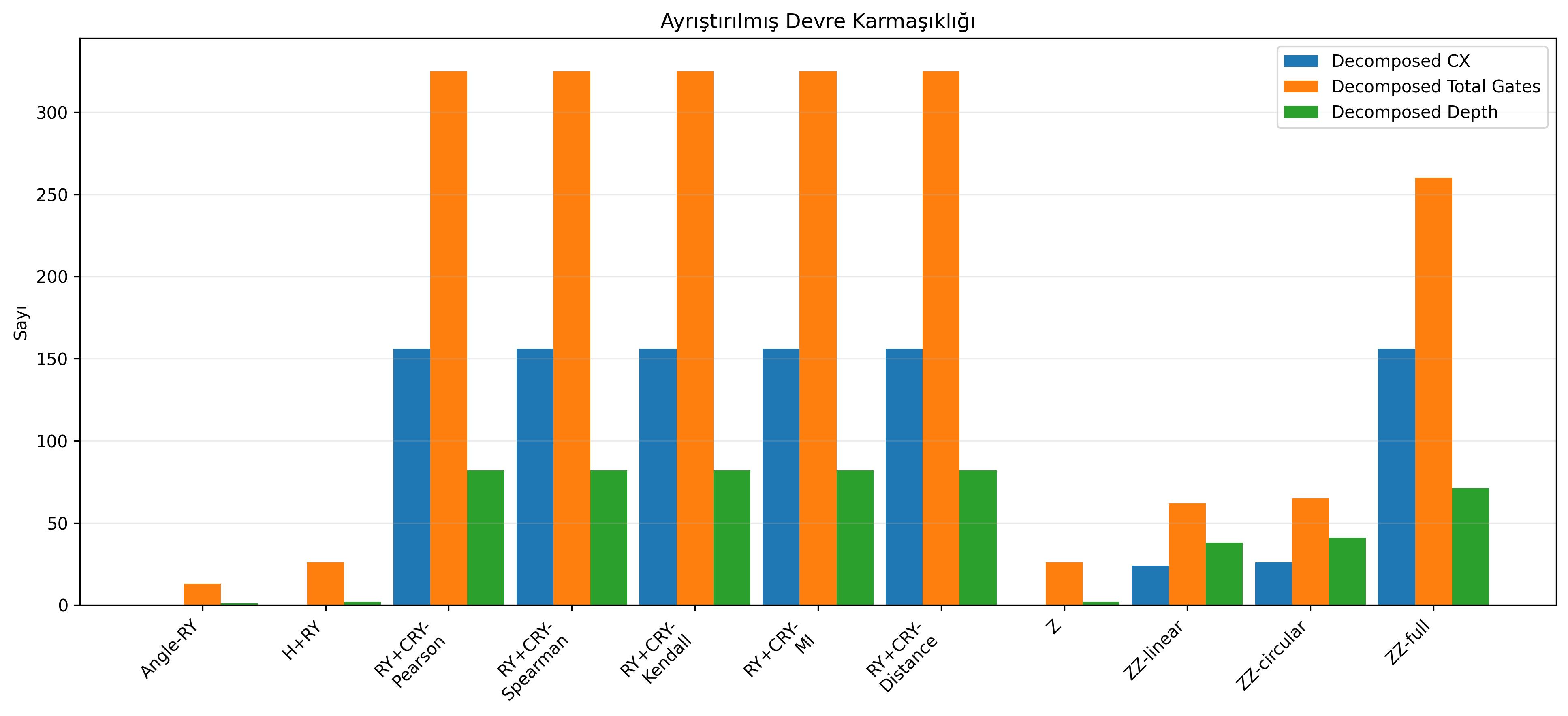}
\caption{Comparison of decomposed circuit complexity for the Student
Performance Dataset.}
\label{fig:student_decomposed}
\end{figure}

The decomposed circuit complexity shows that the RY+CRY methods require
325 total gates, 156 CX gates, and a circuit depth of 82. In contrast,
Z feature Map requires only 26 total gates and a circuit depth of 2.

The difference in classification performance between these two structures
also favors Z feature Map. Z feature Map achieved a Test Accuracy of 0.944 and
an F1 score of 0.900, whereas the RY+CRY methods remained at approximately
0.890 Test Accuracy and 0.786 F1.

Therefore, for this dataset, the more complex RY+CRY structure that connects
all feature pairs did not provide a performance advantage over the
lower-cost Z feature Map.

\subsubsection{Computational Cost}

\begin{table}[!htbp]
\centering
\caption{Computation times for the Student Performance Dataset (seconds).}
\label{tab:student_time}

\footnotesize
\setlength{\tabcolsep}{4pt}
\renewcommand{\arraystretch}{1.10}

\begin{tabular}{lrrr}
\toprule
\textbf{Encoding} &
\makecell{\textbf{state vector}\\\textbf{Time}} &
\makecell{\textbf{Kernel}\\\textbf{Time}} &
\makecell{\textbf{Training}\\\textbf{Time}} \\
\midrule
Angle-RY        & 3.584  & 2.480 & 0.135 \\
H+RY            & 6.072  & 2.461 & 0.127 \\
RY+CRY-Pearson  & 26.061 & 2.457 & 0.134 \\
RY+CRY-Spearman & 26.122 & 2.459 & 0.129 \\
RY+CRY-Kendall  & 26.533 & 2.502 & 0.127 \\
RY+CRY-MI       & 26.879 & 2.531 & 0.136 \\
RY+CRY-Distance & 26.635 & 2.500 & 0.134 \\
Z               & 6.228  & 2.500 & 0.086 \\
ZZ linear       & 13.595 & 2.482 & 0.147 \\
ZZ circular     & 14.758 & 2.465 & 0.144 \\
ZZ full         & 59.357 & 2.470 & 0.137 \\
\bottomrule
\end{tabular}
\end{table}


\begin{figure}[!htbp]
\centering
\includegraphics[width=0.90\columnwidth]
{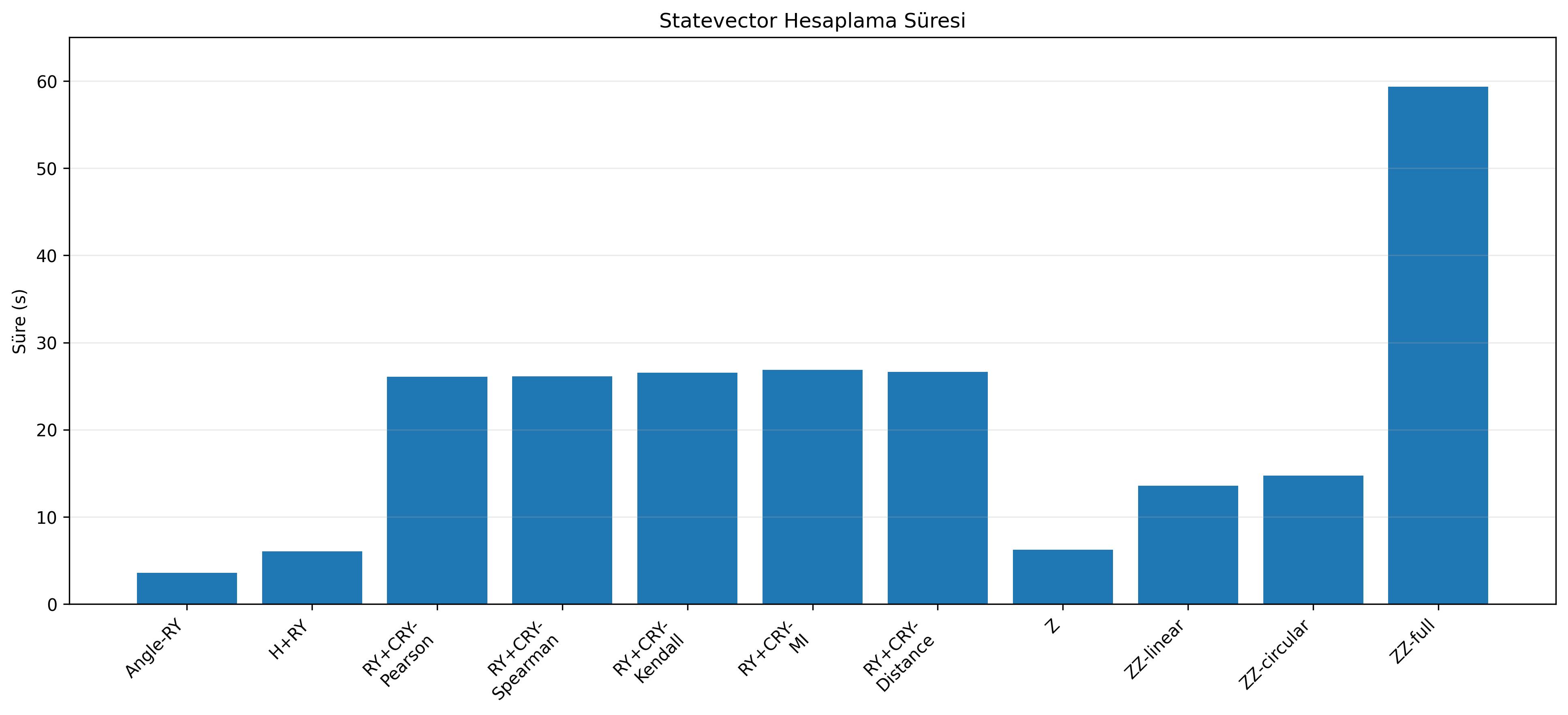}
\caption{state vector computation times for the Student Performance Dataset.}
\label{fig:student_state vector}
\end{figure}

The state vector computation times generally follow the differences in
circuit complexity. Angle-RY had the lowest computation time at
approximately 3.58 seconds, whereas Z feature Map required approximately
6.23 seconds.

The RY+CRY methods required approximately 26--27 seconds for state vector
computation. The highest value, approximately 59.36 seconds, was observed
for ZZ full.

The state vector cost of ZZ full relative to Z feature Map was approximately

\begin{equation}
\frac{59.36}{6.23}\approx9.5.
\end{equation}

Thus, although ZZ full required approximately 9.5 times more state vector
computation time than Z feature Map, its test accuracy of 0.712 remained far
below the 0.944 achieved by Z feature Map. This result demonstrates that
higher computational cost did not translate into improved performance for
this dataset.


\begin{figure}[!htbp]
\centering
\includegraphics[width=0.90\columnwidth]
{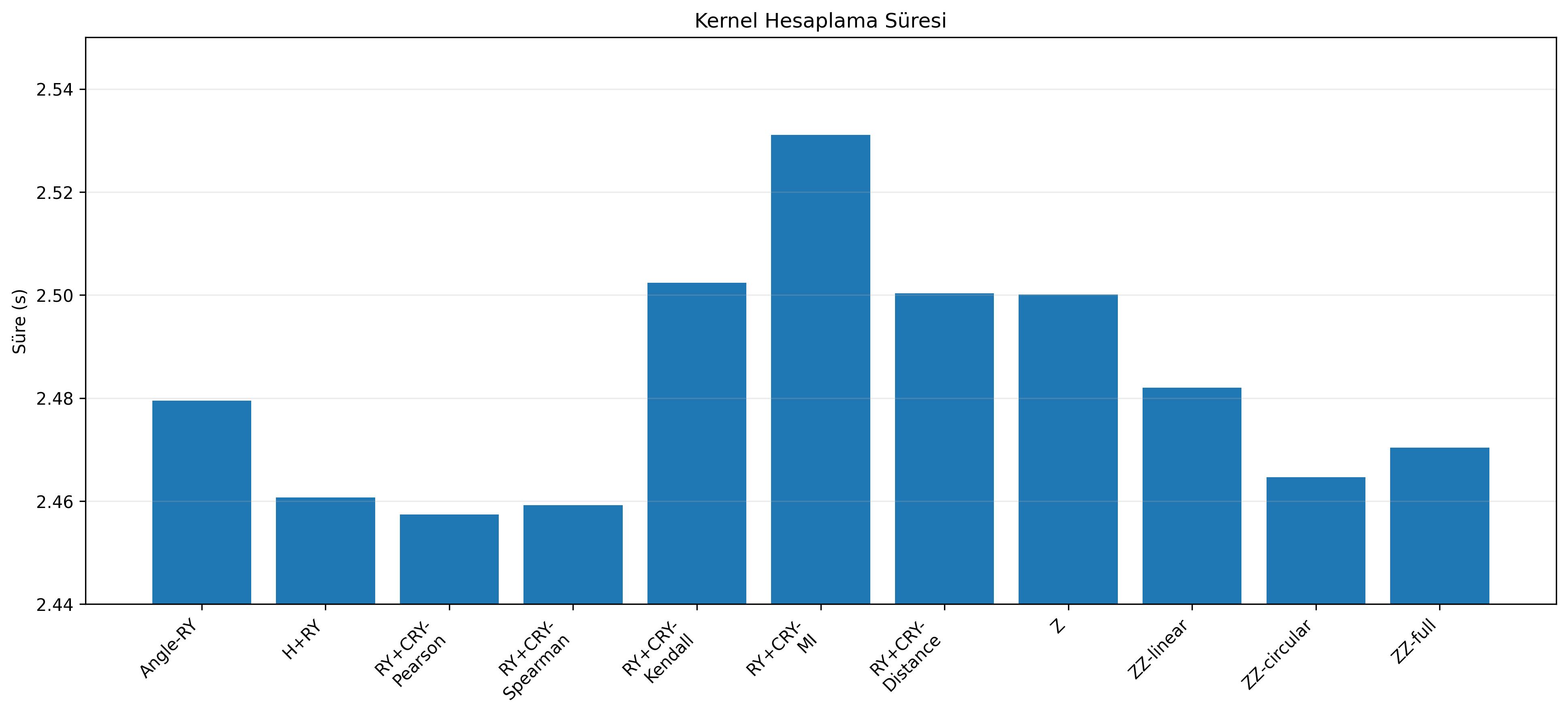}
\caption{Kernel computation times for the Student Performance Dataset.}
\label{fig:student_kernel}
\end{figure}

Kernel Time remained within approximately 2.46--2.53 seconds for all
methods. Despite substantial differences in circuit complexity, the similar
kernel computation times indicate that the primary differences in total
computational cost arose from the state vector generation stage.


\begin{figure}[!htbp]
\centering
\includegraphics[width=0.90\columnwidth]
{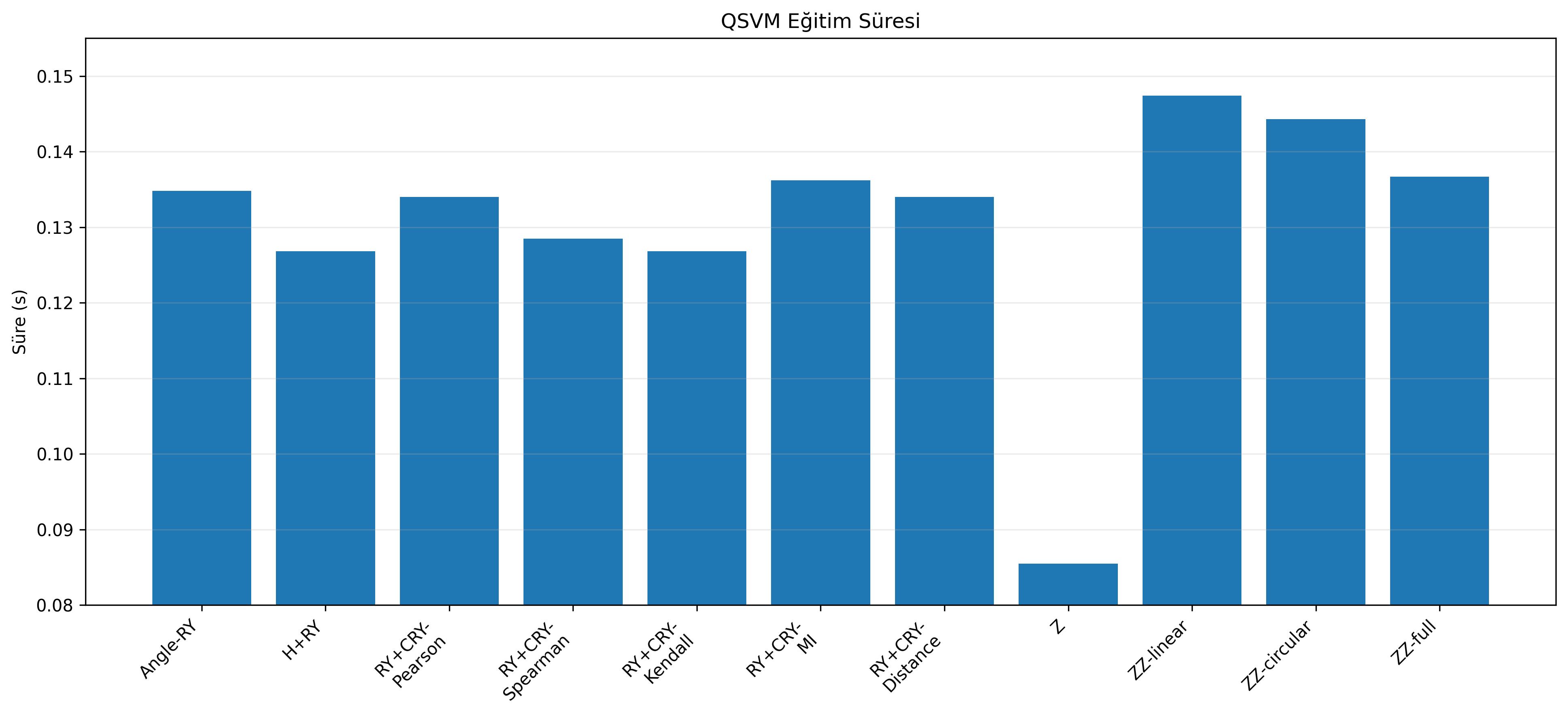}
\caption{QSVM training times for the Student Performance Dataset.}
\label{fig:student_training}
\end{figure}

Training Time ranged from approximately 0.086 to 0.147 seconds across all
methods. The lowest training time, approximately 0.086 seconds, was obtained
with Z feature Map.

These values are relatively small compared with the state vector computation
costs. Therefore, in the Student Performance experiments, the primary source
of total computational cost was again the preparation of quantum feature
states rather than classical SVM training.

\subsubsection{Overall Evaluation of the Student Performance Dataset}

The results obtained for the Student Performance Dataset clearly demonstrate
that more complex quantum feature maps do not necessarily provide better
performance.

Z feature Map achieved the highest test performance among all methods, with a
Test Accuracy of 0.944, an F1 score of 0.900, and a ROC-AUC of 0.975.
Furthermore, its Accuracy Gap was only 0.024 and its Test Log-Loss was
0.160. At the same time, the circuit required only 26 logical gates and a
depth of 2. Therefore, Z feature Map provides the most balanced encoding
strategy for this dataset in terms of performance, generalization, and
circuit cost.

The RY+CRY methods provided limited but consistent performance improvements
over Angle-RY. In particular, the Distance Correlation-based RY+CRY method
achieved the highest class-discrimination performance within this group,
with a ROC-AUC of 0.963. However, encoding all 78 feature pairs required
325 total gates, 156 CX gates, and a decomposed circuit depth of 82.

Although ZZ linear achieved the second-highest test accuracy of 0.908, it
remained behind Z feature Map in terms of Accuracy, F1, ROC-AUC, and
generalization gap.

The poorest result was obtained with ZZ full. Despite achieving a training
accuracy of 1.000, its test accuracy decreased to 0.712, Recall to 0.020,
and F1 to 0.040. Its Accuracy Gap reached 0.288 and its F1 Gap reached
0.960. These results indicate a pronounced overfitting behavior for the
fully connected ZZ feature map on the Student Performance Dataset.

Overall, the Student Performance experiments demonstrate that feature-map
complexity is not as important as compatibility with the underlying data
structure. For this dataset, the low-depth Z feature Map outperformed the more
complex RY+CRY and ZZ structures in terms of classification performance,
generalization, and computational cost.

\subsection{Alzheimer's Disease Dataset}
\label{subsec:alzheimer_results}

In the experiments conducted on the Alzheimer's Disease Dataset, Principal
Component Analysis (PCA) was applied to reduce the computational cost
associated with the high feature dimensionality in state vector-based
calculations. The first 14 principal components (PCs) were retained for
quantum feature encoding.

Following PCA, Angle-RY, H+RY, RY+CRY encodings based on different
statistical dependency measures, Z feature Map, and ZZ feature Map with
different entanglement structures were compared. The evaluation considered
classification performance, generalization behavior, quantum circuit
complexity, computational cost, and the explained variance retained by PCA.

\subsubsection{Dimensionality Reduction with PCA}

\begin{table}[!htbp]
\centering
\caption{Explained variance ratios of the first 14 principal components
for the Alzheimer's Disease Dataset.}
\label{tab:alzheimer_pca}

\footnotesize
\setlength{\tabcolsep}{5pt}
\renewcommand{\arraystretch}{1.10}

\begin{tabular}{ccc}
\toprule
\textbf{Component} &
\makecell{\textbf{Explained}\\\textbf{Variance}} &
\makecell{\textbf{Cumulative}\\\textbf{Variance}} \\
\midrule
PC1  & 0.0403 & 0.0403 \\
PC2  & 0.0387 & 0.0790 \\
PC3  & 0.0379 & 0.1169 \\
PC4  & 0.0370 & 0.1539 \\
PC5  & 0.0364 & 0.1903 \\
PC6  & 0.0363 & 0.2266 \\
PC7  & 0.0355 & 0.2621 \\
PC8  & 0.0350 & 0.2971 \\
PC9  & 0.0341 & 0.3312 \\
PC10 & 0.0337 & 0.3649 \\
PC11 & 0.0334 & 0.3983 \\
PC12 & 0.0331 & 0.4314 \\
PC13 & 0.0327 & 0.4641 \\
PC14 & 0.0320 & 0.4961 \\
\bottomrule
\end{tabular}
\end{table}

\begin{figure}[!htbp]
\centering
\includegraphics[width=0.95\columnwidth]
{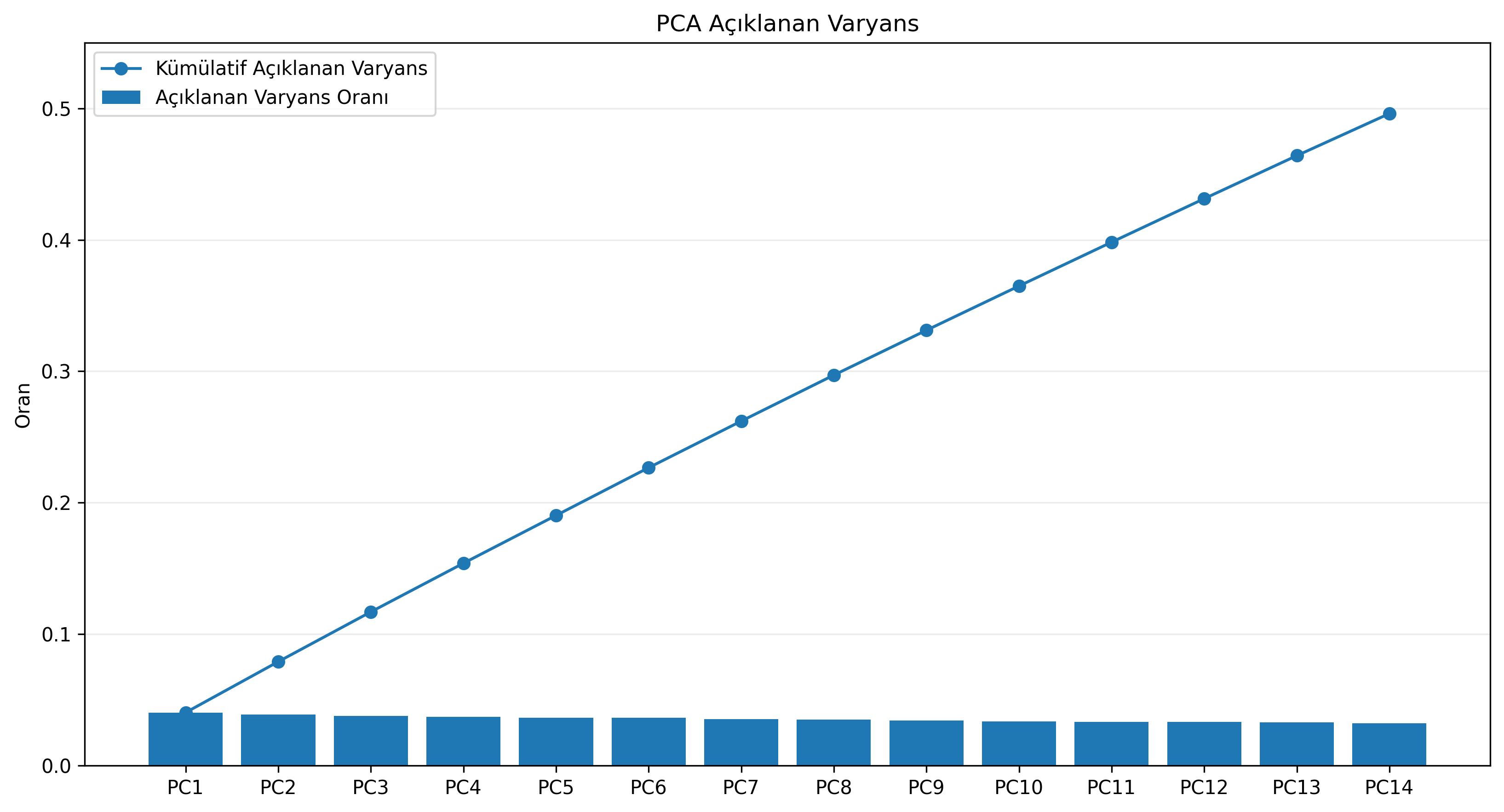}
\caption{Individual and cumulative explained variance ratios of the first
14 principal components for the Alzheimer's Disease Dataset.}
\label{fig:alzheimer_pca}
\end{figure}

Table~\ref{tab:alzheimer_pca} and Fig.~\ref{fig:alzheimer_pca} show that
the first principal component explains approximately 4.03\% of the total
variance and that no single component dominates the representation. The
cumulative explained variance of the first 14 PCs is

\begin{equation}
0.4961 \approx 49.61\%.
\end{equation}

The relatively similar explained variance ratios of the components,
approximately 3.2--4.0\%, indicate that the variance is not concentrated
in a small number of dominant components. In other words, the variance
structure of the Alzheimer's Disease Dataset is distributed over a
relatively broad feature space.

This observation is important for interpreting the results. Reducing the
data to 14 PCs made the quantum simulation computationally feasible, but
only approximately half of the variance in the original data was retained.
Therefore, the QSVM results reported below reflect not only the effects of
the quantum feature encoding strategies but also the amount of information
preserved after PCA.

Accordingly, lower test performance obtained with some complex feature maps
should not be attributed solely to their quantum circuit structures.
Information potentially useful for classification may also have been
excluded from the first 14 PCs during dimensionality reduction.

\subsubsection{Classification Performance}

\begin{table*}[!htbp]
\centering
\caption{Classification performance of quantum feature encoding strategies
on the Alzheimer's Disease Dataset.}
\label{tab:alzheimer_performance}

\footnotesize
\setlength{\tabcolsep}{4pt}
\renewcommand{\arraystretch}{1.10}

\begin{tabular}{lccccccc}
\toprule
\textbf{Encoding} &
\makecell{\textbf{Train}\\\textbf{Acc.}} &
\makecell{\textbf{Test}\\\textbf{Acc.}} &
\makecell{\textbf{Acc.}\\\textbf{Gap}} &
\makecell{\textbf{Test}\\\textbf{Prec.}} &
\makecell{\textbf{Test}\\\textbf{Rec.}} &
\makecell{\textbf{Test}\\\textbf{F1}} &
\makecell{\textbf{Test}\\\textbf{AUC}} \\
\midrule
Angle-RY        & 0.813 & 0.764 & 0.049 & 0.732 & 0.525 & 0.612 & 0.813 \\
H+RY            & 0.813 & 0.764 & 0.049 & 0.732 & 0.525 & 0.612 & 0.813 \\
RY+CRY-Pearson  & 0.813 & 0.764 & 0.049 & 0.732 & 0.525 & 0.612 & 0.813 \\
RY+CRY-Spearman & 0.814 & 0.762 & 0.052 & 0.727 & 0.525 & 0.610 & 0.812 \\
RY+CRY-Kendall  & 0.813 & 0.762 & 0.051 & 0.727 & 0.525 & 0.610 & 0.812 \\
RY+CRY-MI       & 0.871 & 0.760 & 0.111 & 0.752 & 0.480 & 0.586 & 0.803 \\
RY+CRY-Distance & 0.832 & 0.748 & 0.084 & 0.707 & 0.492 & 0.580 & 0.810 \\
Z               & 0.965 & 0.706 & 0.259 & 0.683 & 0.316 & 0.432 & 0.768 \\
ZZ linear       & 1.000 & 0.646 & 0.354 & 0.000 & 0.000 & 0.000 & 0.593 \\
ZZ circular     & 1.000 & 0.646 & 0.354 & 0.000 & 0.000 & 0.000 & 0.579 \\
ZZ full         & 1.000 & 0.650 & 0.350 & 1.000 & 0.011 & 0.022 & 0.810 \\
\bottomrule
\end{tabular}
\end{table*}

\begin{figure}[!htbp]
\centering
\includegraphics[width=0.95\columnwidth]
{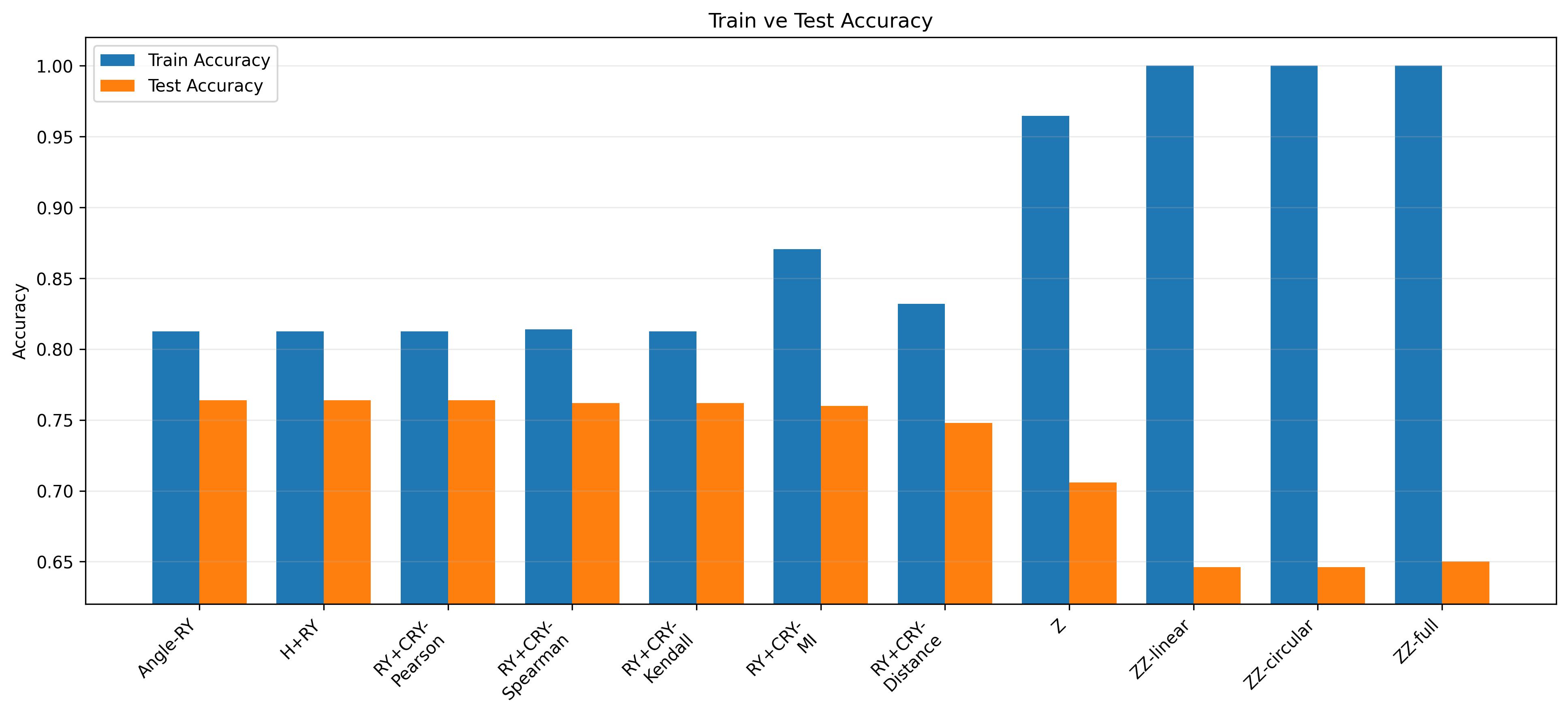}
\caption{Comparison of training and test accuracy across encoding strategies
for the Alzheimer's Disease Dataset.}
\label{fig:alzheimer_accuracy}
\end{figure}

Table~\ref{tab:alzheimer_performance} and
Fig.~\ref{fig:alzheimer_accuracy} show that the highest test accuracy,
0.764, was obtained with Angle-RY, H+RY, and the Pearson-based RY+CRY
encoding.

The Spearman- and Kendall-based RY+CRY methods achieved test accuracies of
0.762, while MI and Distance Correlation achieved 0.760 and 0.748,
respectively. Thus, incorporating statistical dependency information through
CRY gates did not produce a clear improvement in Accuracy over Angle-RY.

In particular, the identical Accuracy, Precision, Recall, F1, and ROC-AUC
values obtained with Pearson-based RY+CRY and Angle-RY indicate that the
Pearson-based controlled rotations did not alter the classification
decisions within the current PCA-transformed feature space.

In contrast, training accuracy increased substantially as circuit complexity
increased. Z feature Map achieved a training accuracy of 0.965, whereas all
three ZZ-based methods achieved 1.000. However, their test accuracies
decreased to 0.706, 0.646, 0.646, and 0.650, respectively.

These results show that the more complex quantum feature maps represented
the training data more strongly, but this increased representational
capacity did not translate into improved test performance.


\begin{figure}[!htbp]
\centering
\includegraphics[width=0.95\columnwidth]
{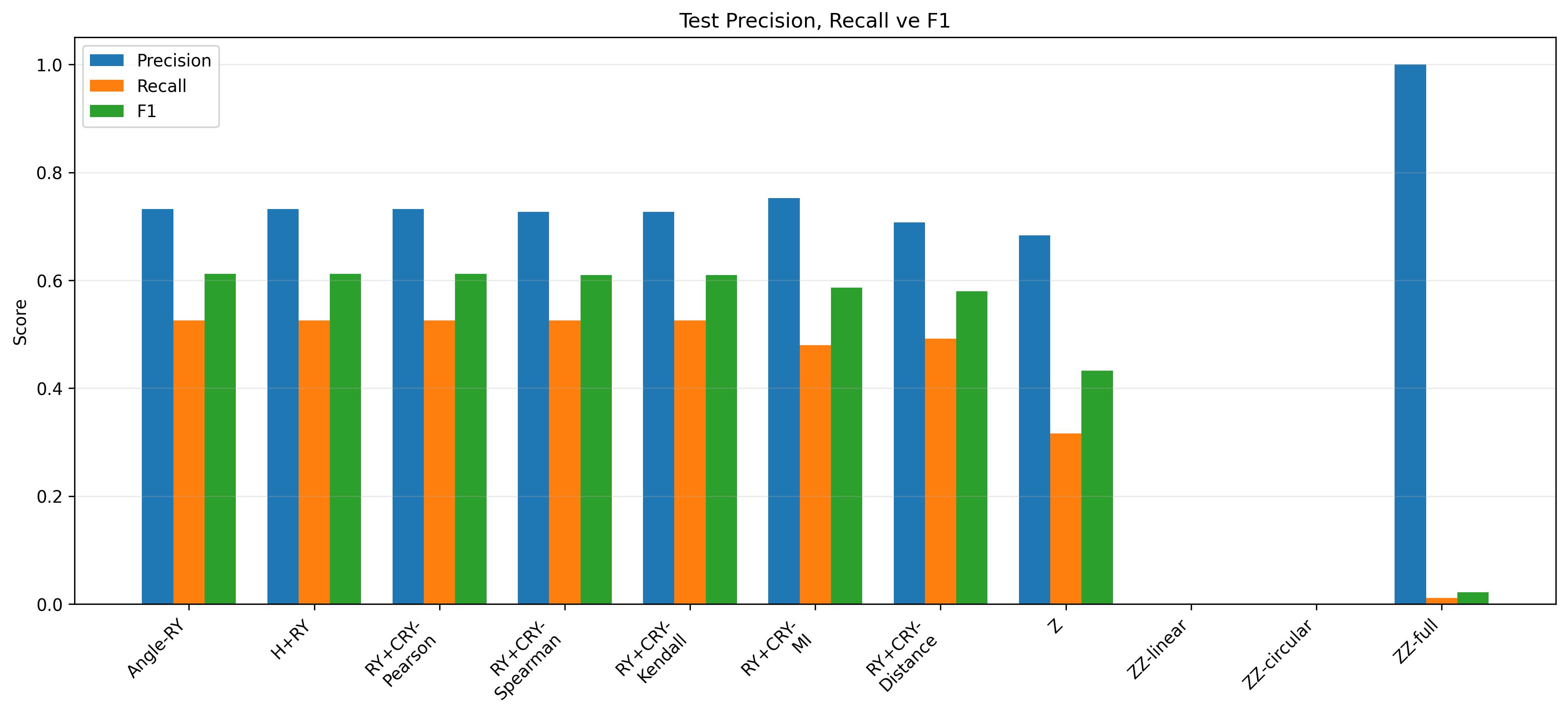}
\caption{Comparison of test Precision, Recall, and F1 scores for the
Alzheimer's Disease Dataset.}
\label{fig:alzheimer_prf}
\end{figure}

Figure~\ref{fig:alzheimer_prf} provides a clearer view of the
classification behavior, particularly for the ZZ-based feature maps.

For Angle-RY, H+RY, and Pearson-based RY+CRY, Precision, Recall, and F1
were 0.732, 0.525, and 0.612, respectively. Although MI produced a higher
Precision of 0.752, its Recall decreased to 0.480, resulting in a lower F1
score of 0.586.

For Z feature Map, Precision was 0.683, whereas Recall was only 0.316 and F1
was 0.432. Thus, despite its high training accuracy, Z feature Map was not
able to identify a sufficient proportion of positive-class samples in the
test set.

The most pronounced behavior was observed for ZZ linear and ZZ circular.
Both methods produced test Precision, Recall, and F1 values of zero,
indicating that they failed to correctly identify any positive-class
samples in the test set.

Although ZZ full achieved a Precision of 1.000, its Recall was only 0.011
and its F1 score was 0.022. Therefore, high Precision alone does not
indicate successful classification. The model produced very few positive
predictions and missed nearly all actual positive samples.


\begin{figure}[!htbp]
\centering
\includegraphics[width=0.90\columnwidth]
{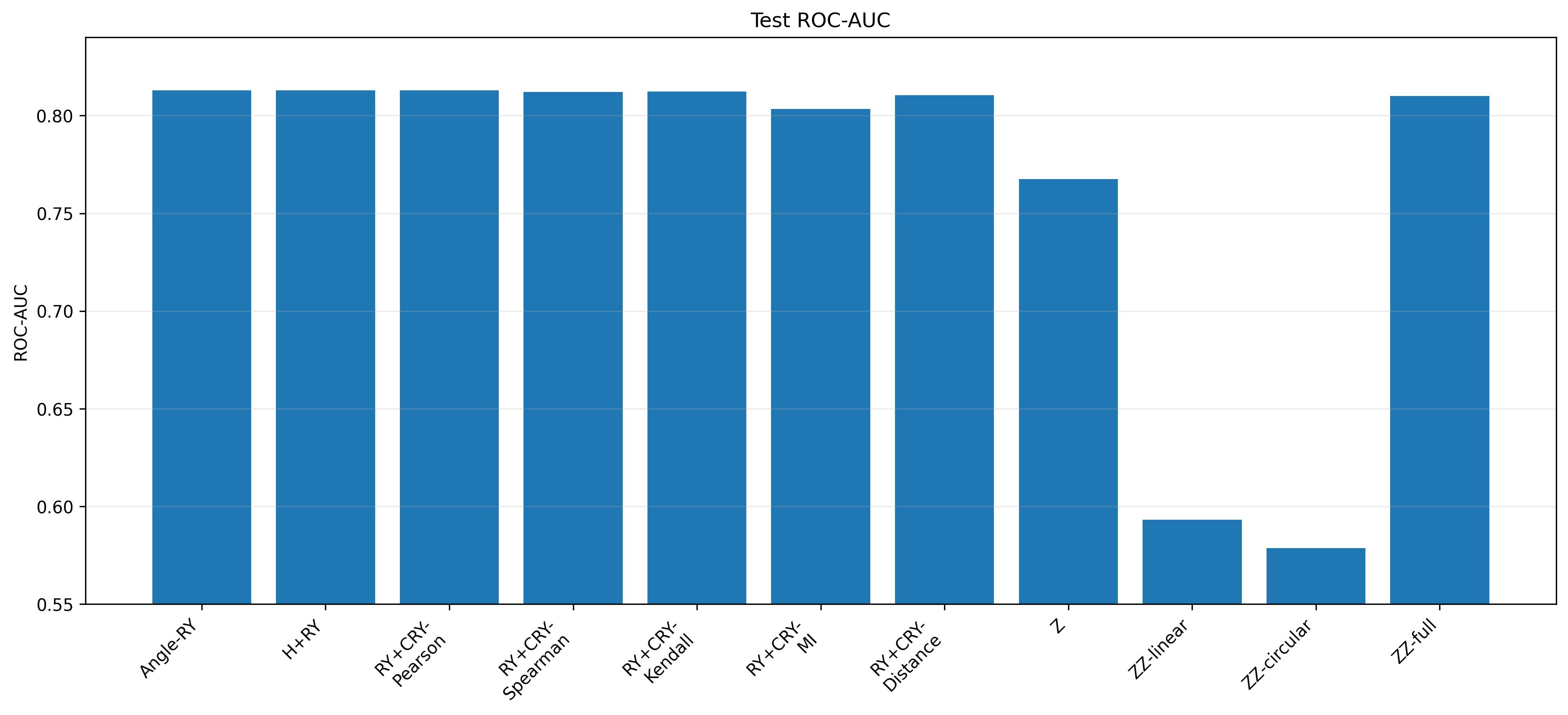}
\caption{Comparison of test ROC-AUC values for the Alzheimer's Disease
Dataset.}
\label{fig:alzheimer_auc}
\end{figure}

The ROC-AUC results provide additional information beyond the
threshold-dependent classification metrics. Angle-RY, H+RY, and
Pearson-based RY+CRY achieved the highest values at approximately 0.813.

Spearman and Kendall achieved approximately 0.812, while Distance
Correlation and MI produced 0.810 and 0.803, respectively. Therefore,
changing the statistical dependency measure within the RY+CRY group
produced only limited differences in class-separation capability.

Although ZZ full achieved an Accuracy of only 0.650 and an F1 score of only
0.022, its ROC-AUC was 0.810. This result suggests that the model retained
some ability to rank samples according to its decision scores, while
producing very poor positive-class decisions at the selected classification
threshold.

Therefore, ROC-AUC should be interpreted together with F1 and Recall for
the Alzheimer's Disease Dataset.

\subsubsection{Generalization Behavior and Log-Loss}

\begin{table*}[!htbp]
\centering
\caption{Training--test performance gaps and Log-Loss values for the
Alzheimer's Disease Dataset.}
\label{tab:alzheimer_generalization}

\footnotesize
\setlength{\tabcolsep}{3.8pt}
\renewcommand{\arraystretch}{1.10}

\begin{tabular}{lccccccc}
\toprule
\textbf{Encoding} &
\makecell{\textbf{Acc.}\\\textbf{Gap}} &
\makecell{\textbf{Prec.}\\\textbf{Gap}} &
\makecell{\textbf{Recall}\\\textbf{Gap}} &
\makecell{\textbf{F1}\\\textbf{Gap}} &
\makecell{\textbf{AUC}\\\textbf{Gap}} &
\makecell{\textbf{Train}\\\textbf{Loss}} &
\makecell{\textbf{Test}\\\textbf{Loss}} \\
\midrule
Angle-RY        & 0.049 & 0.055 & 0.118 & 0.096 & 0.068 & 0.427 & 0.496 \\
H+RY            & 0.049 & 0.055 & 0.118 & 0.096 & 0.068 & 0.427 & 0.496 \\
RY+CRY-Pearson  & 0.049 & 0.055 & 0.118 & 0.096 & 0.068 & 0.427 & 0.496 \\
RY+CRY-Spearman & 0.052 & 0.063 & 0.120 & 0.100 & 0.070 & 0.425 & 0.496 \\
RY+CRY-Kendall  & 0.051 & 0.061 & 0.118 & 0.098 & 0.070 & 0.426 & 0.496 \\
RY+CRY-MI       & 0.111 & 0.131 & 0.250 & 0.213 & 0.129 & 0.374 & 0.515 \\
RY+CRY-Distance & 0.084 & 0.111 & 0.182 & 0.159 & 0.083 & 0.412 & 0.499 \\
Z               & 0.259 & 0.291 & 0.608 & 0.516 & 0.227 & 0.220 & 0.551 \\
ZZ linear       & 0.354 & 1.000 & 1.000 & 1.000 & 0.407 & 0.001 & 0.636 \\
ZZ circular     & 0.354 & 1.000 & 1.000 & 1.000 & 0.421 & 0.006 & 0.642 \\
ZZ full         & 0.350 & 0.000 & 0.989 & 0.978 & 0.190 & 0.001 & 0.507 \\
\bottomrule
\end{tabular}
\end{table*}

\begin{figure}[!htbp]
\centering
\includegraphics[width=0.90\columnwidth]
{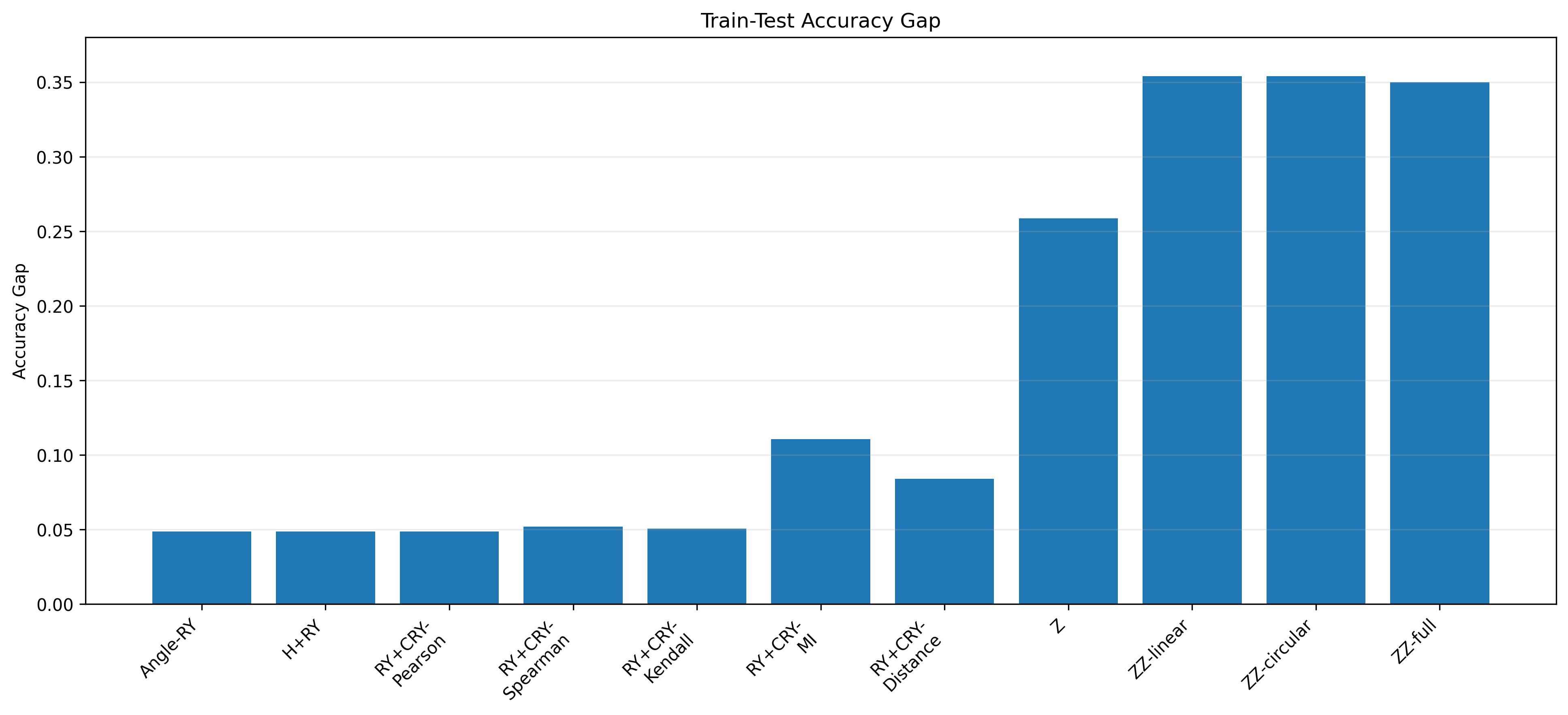}
\caption{Training--test Accuracy Gap across encoding strategies for the
Alzheimer's Disease Dataset.}
\label{fig:alzheimer_accuracy_gap}
\end{figure}

The lowest Accuracy Gap, 0.049, was obtained with Angle-RY, H+RY, and
Pearson-based RY+CRY. Spearman and Kendall also maintained gaps of
approximately 0.05.

The gap increased to 0.111 for MI and to 0.259 for Z feature Map. The highest
values were obtained with ZZ linear and ZZ circular at 0.354, followed by
ZZ full at 0.350.

Thus, as the complexity of the quantum feature map increased, training
performance improved while test performance did not improve to the same
extent and, in several cases, deteriorated substantially.

In particular, although ZZ linear and ZZ circular classified the training
data with an accuracy of 1.000, their test accuracy decreased to 0.646,
indicating pronounced overfitting.

\begin{figure}[!htbp]
\centering
\includegraphics[width=0.90\columnwidth]
{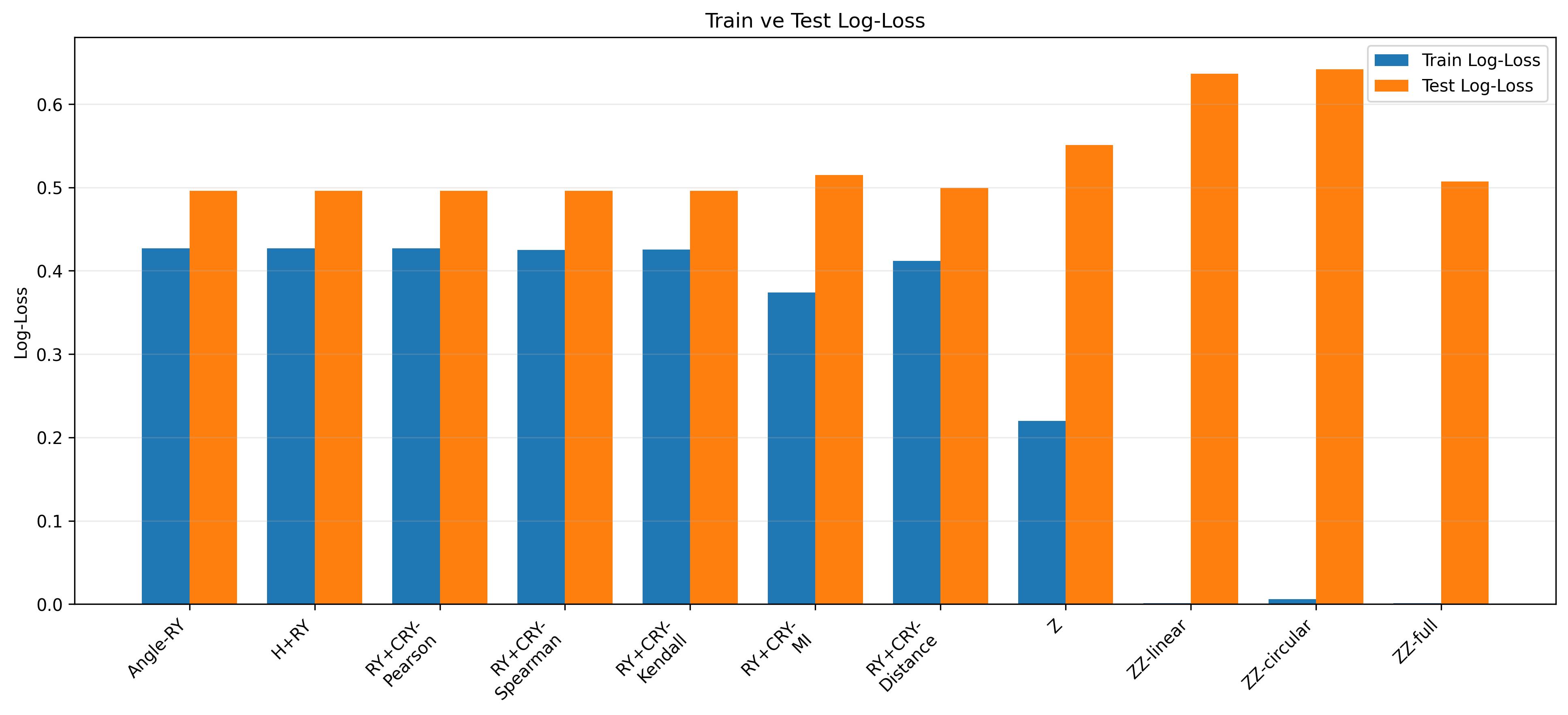}
\caption{Comparison of training and test Log-Loss values for the
Alzheimer's Disease Dataset.}
\label{fig:alzheimer_logloss}
\end{figure}

The Log-Loss results support the same trend. Angle-RY and Pearson-based
RY+CRY produced a Train Log-Loss of approximately 0.427 and a Test Log-Loss
of 0.496.

Although Z feature Map reduced the training loss to 0.220, its test loss
increased to 0.551. This indicates that the model produced more confident
predictions for the training samples, while the quality of its predictions
on unseen samples deteriorated.

For ZZ linear and ZZ circular, the training loss approached zero, whereas
the Test Log-Loss increased to 0.636 and 0.642, respectively. These were
the highest test losses among all methods and further support the evidence
of strong overfitting.

The Test Log-Loss of ZZ full was lower than those of the linear and
circular variants at 0.507. Nevertheless, since its training loss was only
0.001, the separation between training and test behavior remained
substantial.

\subsubsection{Quantum Circuit Complexity}

\begin{table*}[!htbp]
\centering
\caption{Circuit and gate complexities of quantum feature encoding
strategies for the Alzheimer's Disease Dataset.}
\label{tab:alzheimer_circuit}

\footnotesize
\setlength{\tabcolsep}{3.5pt}
\renewcommand{\arraystretch}{1.10}

\begin{tabular}{lrrrrrrrrr}
\toprule
\textbf{Encoding} &
\textbf{H} &
\textbf{RY} &
\textbf{CRY} &
\textbf{CX} &
\makecell{\textbf{Total}\\\textbf{Gates}} &
\textbf{Depth} &
\makecell{\textbf{Dec.}\\\textbf{CX}} &
\makecell{\textbf{Dec.}\\\textbf{Gates}} &
\makecell{\textbf{Dec.}\\\textbf{Depth}} \\
\midrule
Angle-RY        & 0  & 14 & 0  & 0   & 14  & 1  & 0   & 14  & 1 \\
H+RY            & 14 & 14 & 0  & 0   & 28  & 2  & 0   & 28  & 2 \\
RY+CRY-Pearson  & 0  & 14 & 91 & 0   & 105 & 26 & 182 & 378 & 89 \\
RY+CRY-Spearman & 0  & 14 & 91 & 0   & 105 & 26 & 182 & 378 & 89 \\
RY+CRY-Kendall  & 0  & 14 & 91 & 0   & 105 & 26 & 182 & 378 & 89 \\
RY+CRY-MI       & 0  & 14 & 91 & 0   & 105 & 26 & 182 & 378 & 89 \\
RY+CRY-Distance & 0  & 14 & 91 & 0   & 105 & 26 & 182 & 378 & 89 \\
Z               & 14 & 0  & 0  & 0   & 28  & 2  & 0   & 28  & 2 \\
ZZ linear       & 14 & 0  & 0  & 26  & 67  & 41 & 26  & 67  & 41 \\
ZZ circular     & 14 & 0  & 0  & 28  & 70  & 44 & 28  & 70  & 44 \\
ZZ full         & 14 & 0  & 0  & 182 & 301 & 77 & 182 & 301 & 77 \\
\bottomrule
\end{tabular}
\end{table*}

\begin{figure*}[!htbp]
\centering
\begin{subfigure}{0.48\textwidth}
\centering
\includegraphics[width=\linewidth]{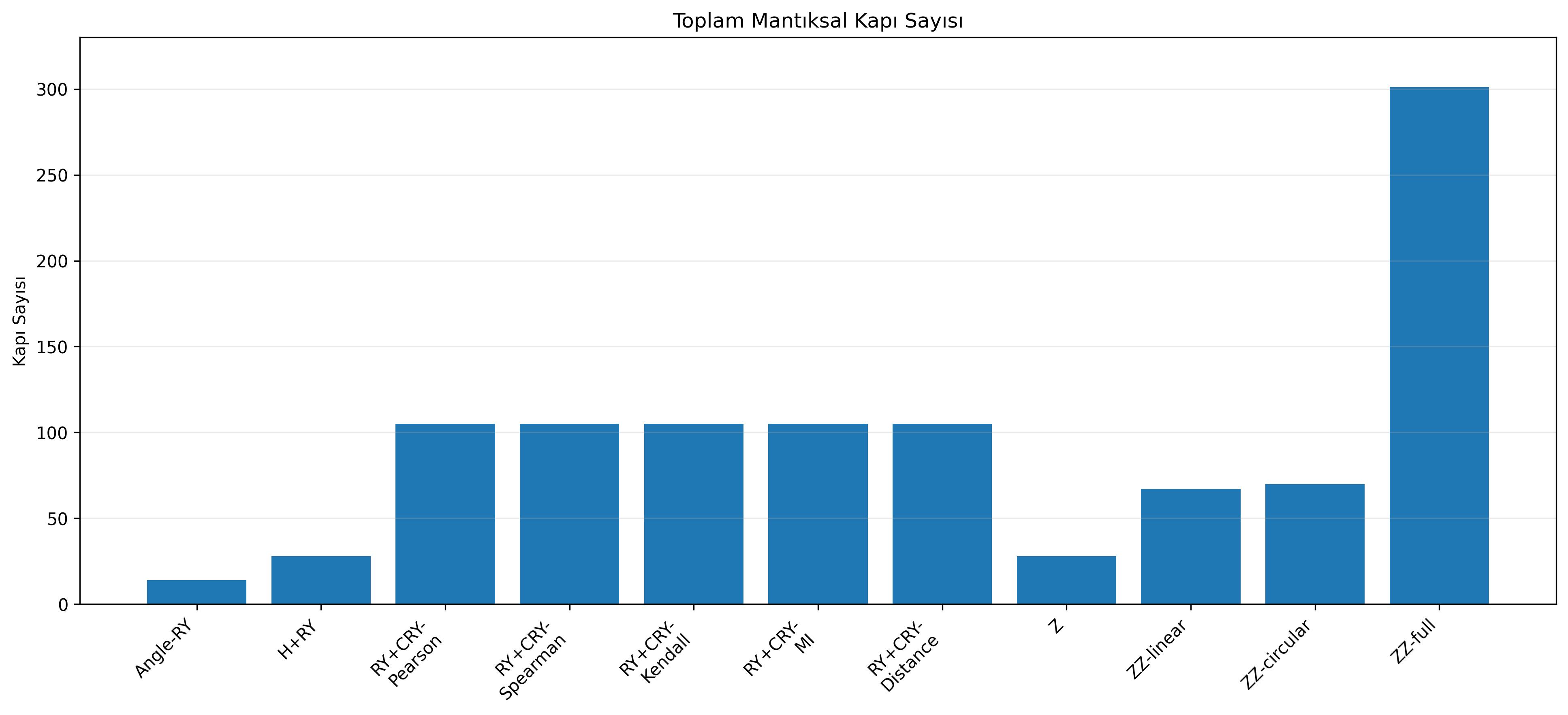}
\caption{Total number of logical gates.}
\end{subfigure}
\hfill
\begin{subfigure}{0.48\textwidth}
\centering
\includegraphics[width=\linewidth]{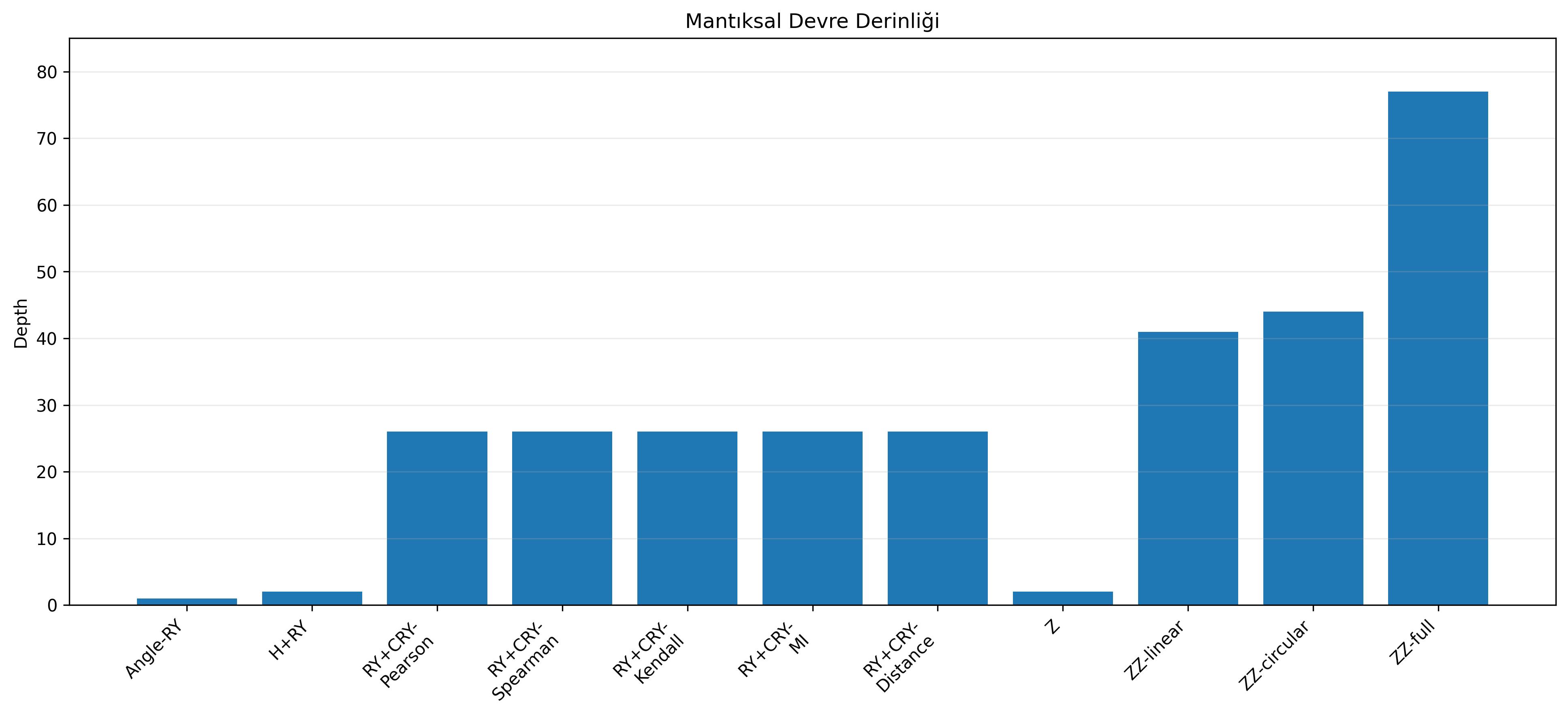}
\caption{Logical circuit depth.}
\end{subfigure}
\caption{Comparison of total gate count and circuit depth for the
Alzheimer's Disease Dataset.}
\label{fig:alzheimer_gate_depth}
\end{figure*}

Since 14 PCs were retained after PCA, the Angle-RY circuit contains 14 RY
gates. The number of all unique pairwise relationships among the 14
components is

\begin{equation}
\binom{14}{2}=91.
\end{equation}

Therefore, each RY+CRY method contains 91 CRY gates.

At the logical level, these circuits contain 105 gates and have a depth of
26. After decomposition, the circuits increase to 378 total gates,
182 CX gates, and a circuit depth of 89.

However, Pearson-based RY+CRY produced exactly the same test performance as
Angle-RY. Therefore, within the 14-dimensional PCA representation, encoding
all 91 pairwise relationships did not provide an additional test-performance
benefit over basic Angle-RY for the Pearson measure.

\begin{figure}[!htbp]
\centering
\includegraphics[width=0.95\columnwidth]
{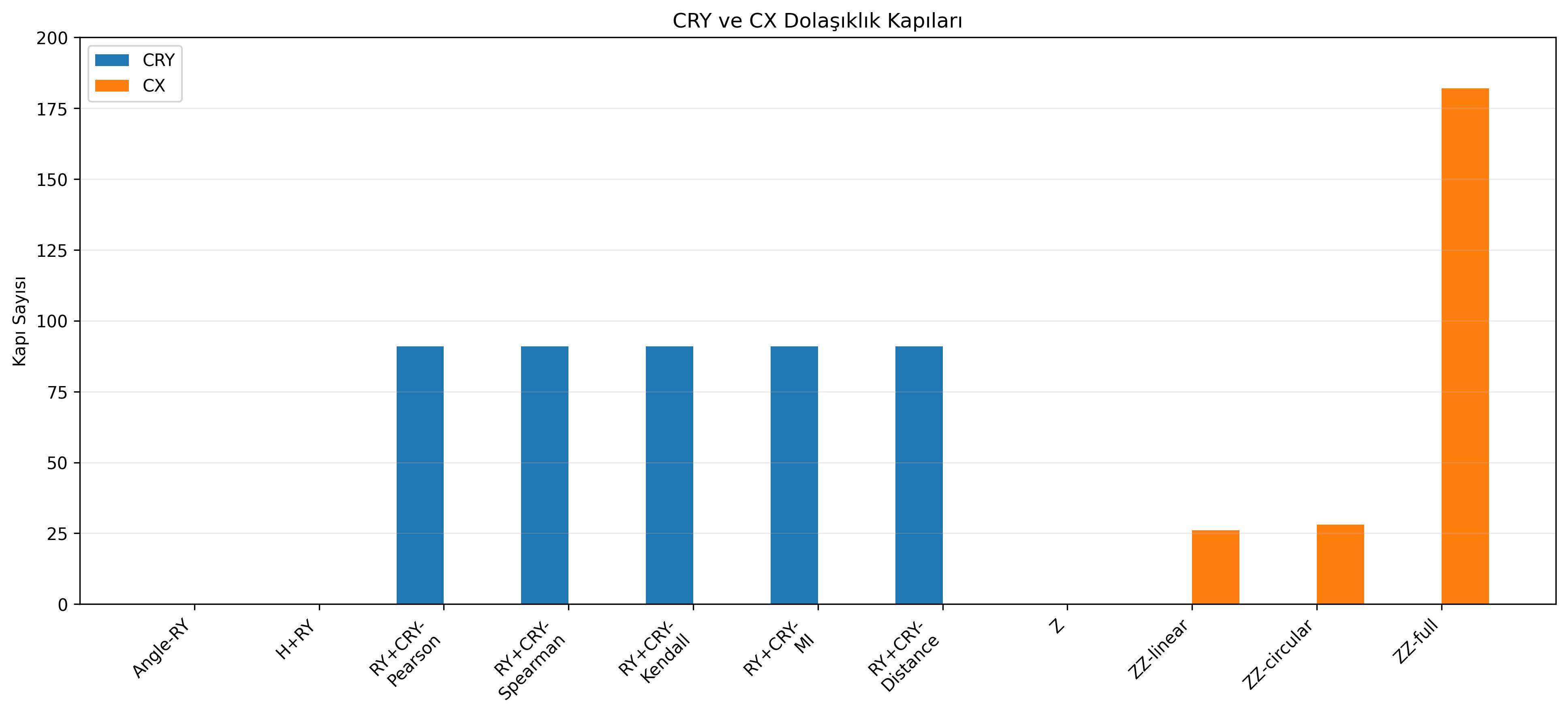}
\caption{Comparison of CRY and CX entangling gates for the Alzheimer's
Disease Dataset.}
\label{fig:alzheimer_entangling}
\end{figure}

ZZ linear, ZZ circular, and ZZ full contain 26, 28, and 182 CX gates,
respectively. Despite the increase in entanglement density, Test Accuracy,
Recall, and F1 did not improve and, particularly for the ZZ variants,
decreased substantially.

This result demonstrates that a higher level of entanglement does not
necessarily imply better classification performance for the Alzheimer's
Disease Dataset.

\begin{figure}[!htbp]
\centering
\includegraphics[width=0.95\columnwidth]
{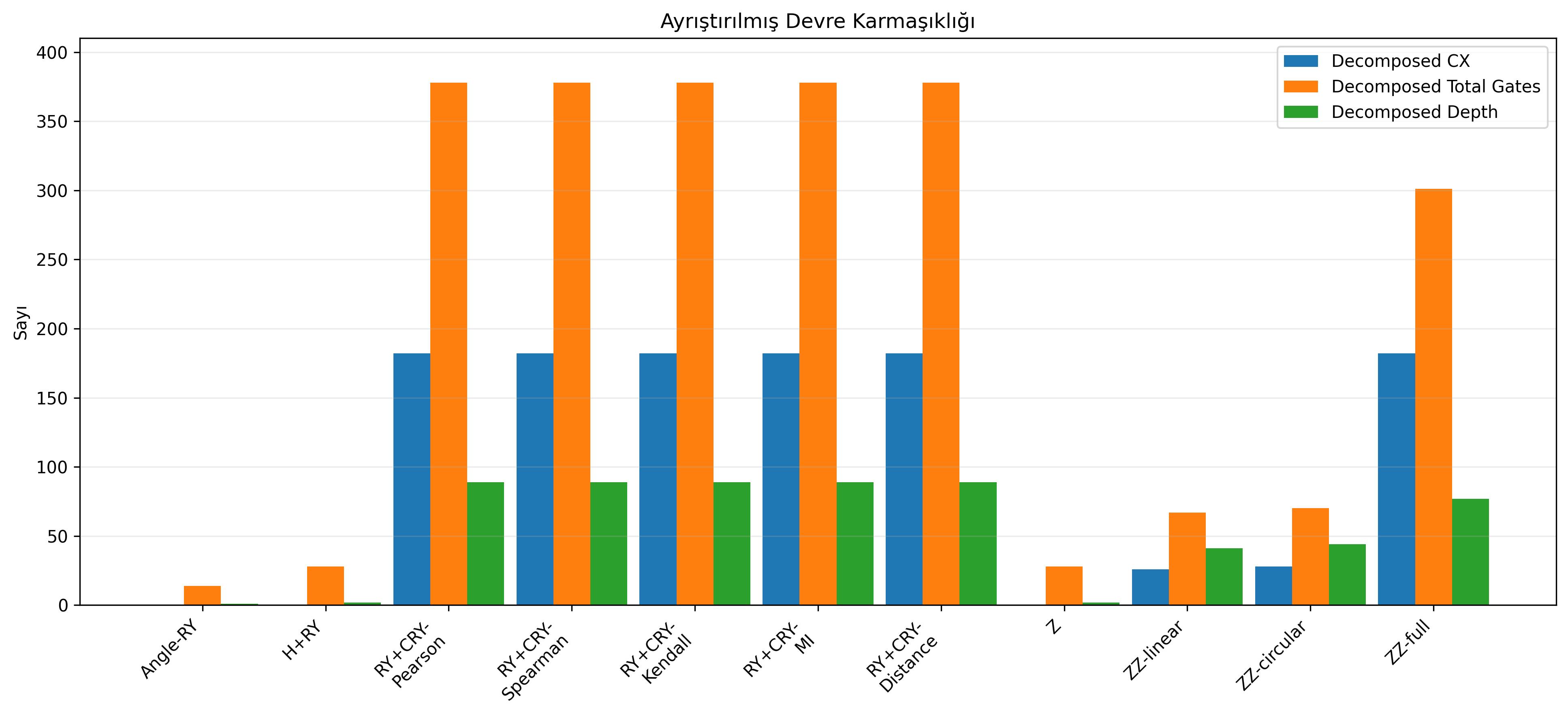}
\caption{Comparison of decomposed circuit complexity for the Alzheimer's
Disease Dataset.}
\label{fig:alzheimer_decomposed}
\end{figure}

In terms of decomposed circuit complexity, the RY+CRY structures are among
the most resource-intensive approaches. They require 378 total gates,
182 CX gates, and a circuit depth of 89. By comparison, Angle-RY requires
only 14 gates and a depth of 1.

The absence of any test-performance difference between Angle-RY and
Pearson-based RY+CRY indicates that encoding all pairwise relationships
with controlled rotations is not resource-efficient in the PCA-transformed
Alzheimer feature space.

Although ZZ full is substantially more complex, with 301 gates and a depth
of 77, its test accuracy decreases from 0.764 for Angle-RY to 0.650.
Therefore, no positive relationship between increasing circuit complexity
and classification performance was observed for this dataset.

\subsubsection{Computational Cost}

\begin{table}[!htbp]
\centering
\caption{Computation times for the Alzheimer's Disease Dataset (seconds).}
\label{tab:alzheimer_time}

\footnotesize
\setlength{\tabcolsep}{4pt}
\renewcommand{\arraystretch}{1.10}

\begin{tabular}{lrrr}
\toprule
\textbf{Encoding} &
\makecell{\textbf{state vector}\\\textbf{Time}} &
\makecell{\textbf{Kernel}\\\textbf{Time}} &
\makecell{\textbf{Training}\\\textbf{Time}} \\
\midrule
Angle-RY        & 5.038   & 4.985 & 0.093 \\
H+RY            & 9.268   & 4.871 & 0.094 \\
RY+CRY-Pearson  & 53.200  & 4.892 & 0.098 \\
RY+CRY-Spearman & 53.790  & 4.917 & 0.092 \\
RY+CRY-Kendall  & 54.285  & 4.965 & 0.091 \\
RY+CRY-MI       & 53.897  & 4.969 & 0.125 \\
RY+CRY-Distance & 54.379  & 4.863 & 0.088 \\
Z               & 9.380   & 4.905 & 0.130 \\
ZZ linear       & 24.458  & 4.915 & 0.151 \\
ZZ circular     & 25.650  & 4.906 & 0.160 \\
ZZ full         & 119.889 & 5.034 & 0.148 \\
\bottomrule
\end{tabular}
\end{table}

\begin{figure}[!htbp]
\centering
\includegraphics[width=0.90\columnwidth]
{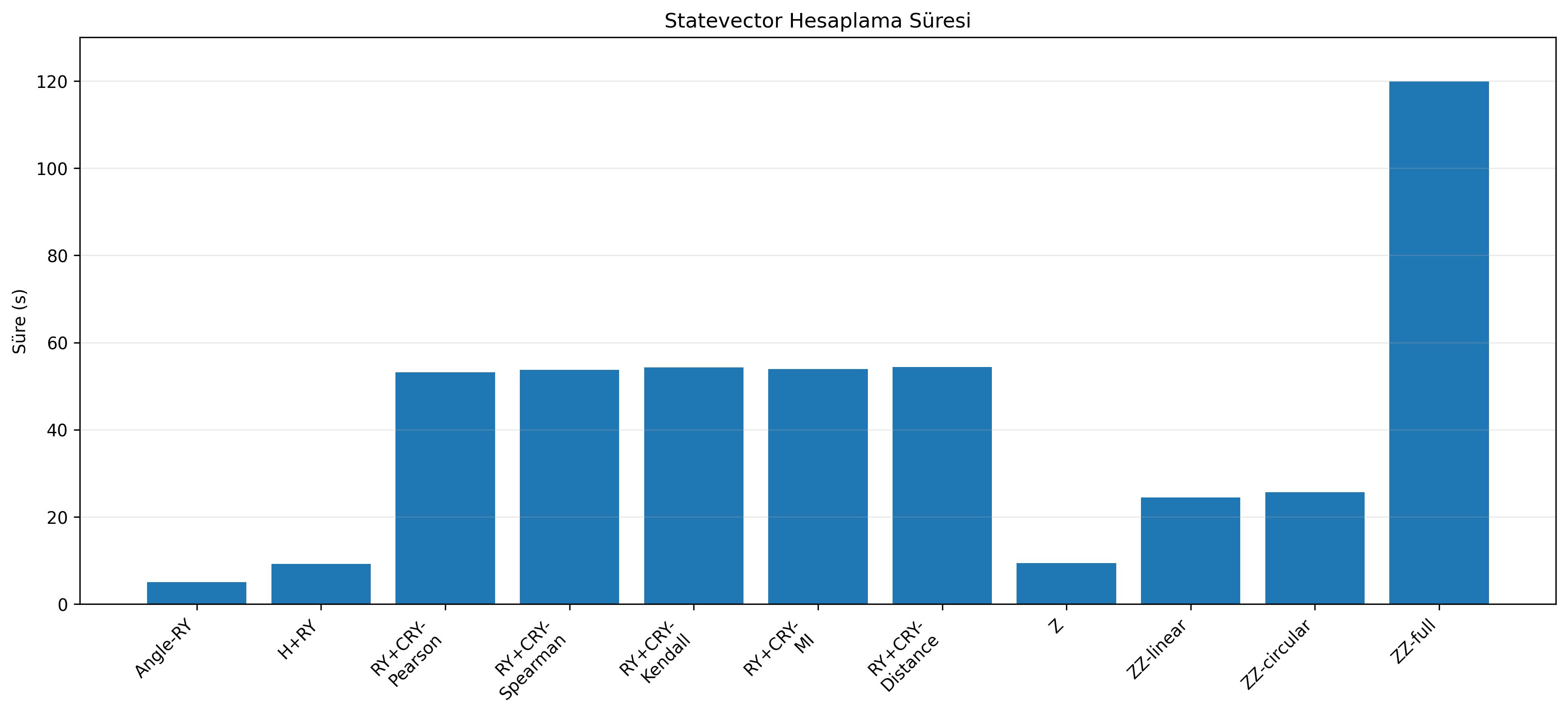}
\caption{state vector computation times for the Alzheimer's Disease Dataset.}
\label{fig:alzheimer_state vector}
\end{figure}

The state vector computation times follow the differences in circuit
complexity. Angle-RY had the lowest state vector cost at approximately
5.04 seconds. Z feature Map required approximately 9.38 seconds, whereas the
RY+CRY methods required approximately 53--54 seconds.

The highest computation time, \textbf{119.89 seconds}, was obtained with
ZZ full.

The state vector cost of ZZ full relative to Angle-RY was approximately

\begin{equation}
\frac{119.89}{5.04}\approx23.8.
\end{equation}

Thus, ZZ full required approximately 23.8 times more state vector computation
time, despite producing lower test accuracy and a dramatically lower F1
score.

This observation strongly demonstrates that increased quantum circuit cost
did not translate into improved classification performance for the
Alzheimer's Disease Dataset.

\begin{figure}[!htbp]
\centering
\includegraphics[width=0.90\columnwidth]
{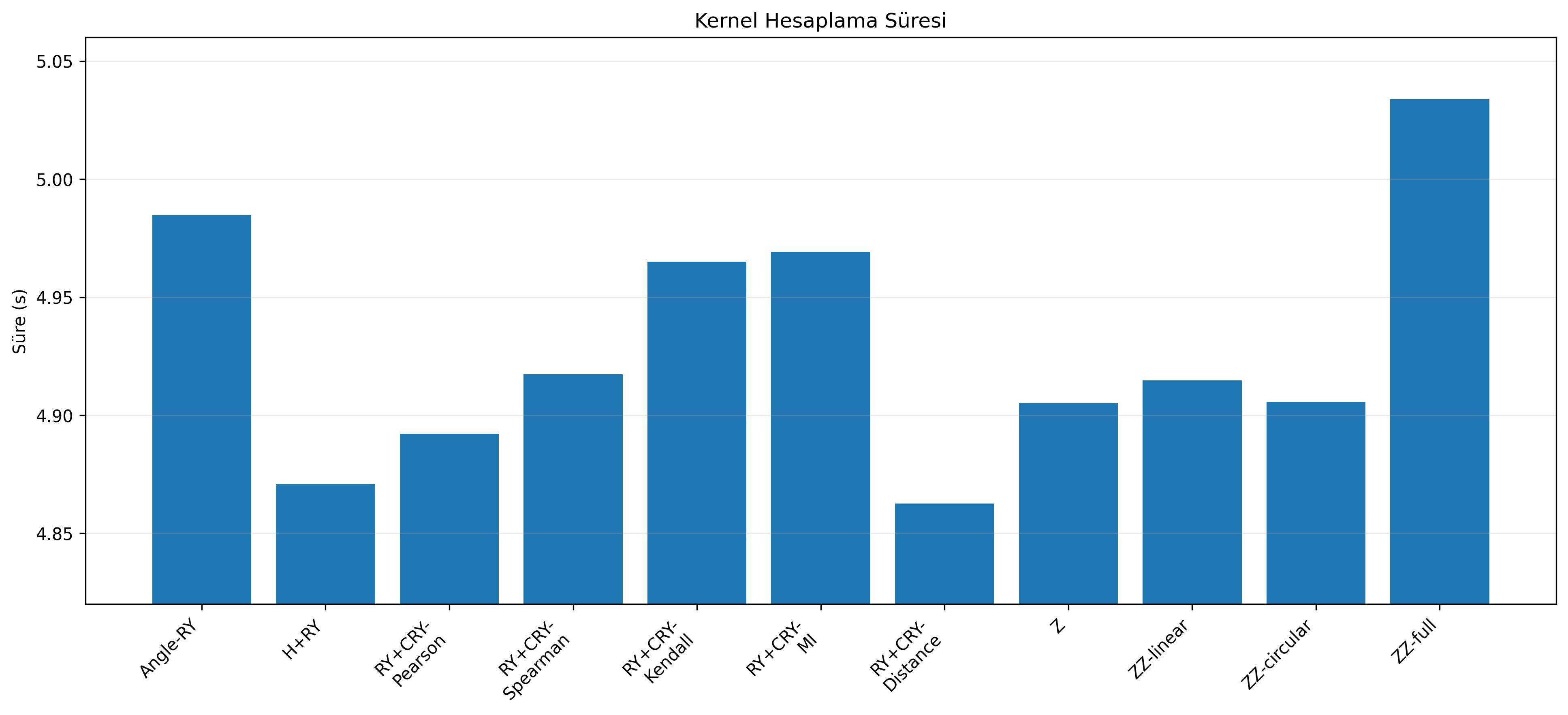}
\caption{Kernel computation times for the Alzheimer's Disease Dataset.}
\label{fig:alzheimer_kernel}
\end{figure}

Kernel Time remained between approximately 4.86 and 5.03 seconds for all
methods. Therefore, the substantial differences in computational cost among
the circuit structures originated primarily from state vector generation
rather than the kernel computation stage.

\begin{figure}[!htbp]
\centering
\includegraphics[width=0.90\columnwidth]
{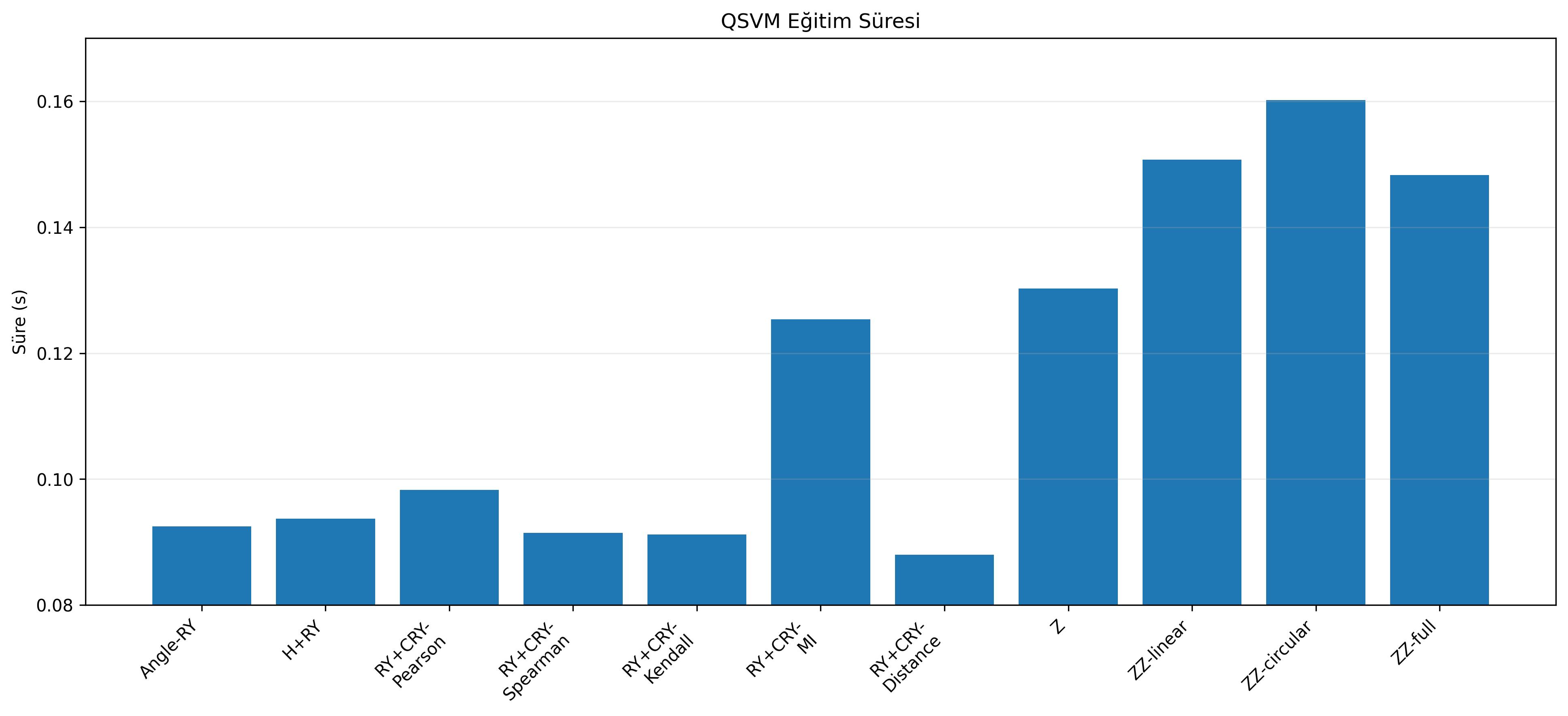}
\caption{QSVM training times for the Alzheimer's Disease Dataset.}
\label{fig:alzheimer_training}
\end{figure}

Training Time ranged from approximately 0.088 to 0.160 seconds across all
methods. These values are very small compared with the state vector
computation times.

Therefore, the primary determinant of the total computational cost in the
Alzheimer experiments was the preparation of quantum feature states rather
than classical SVM optimization.

\subsubsection{Overall Evaluation of the Alzheimer's Disease Dataset}

The results obtained for the Alzheimer's Disease Dataset differ
substantially from those observed for the other datasets in this study.

First, the feature space reduced to 14 PCs through PCA retained
approximately 49.61\% of the total variance. Although this dimensionality
reduction was necessary to make the quantum simulation computationally
feasible, some information potentially relevant to classification may have
been lost. Therefore, the effect of PCA cannot be considered independently
of the quantum encoding methods when interpreting the Alzheimer results.

Second, the highest test performance was obtained with the simpler methods.
Angle-RY, H+RY, and Pearson-based RY+CRY achieved a Test Accuracy of 0.764,
an F1 score of 0.612, and a ROC-AUC of 0.813. Although Pearson-based
RY+CRY used a substantially more complex circuit, it provided no test
performance advantage over Angle-RY.

Third, training performance increased sharply as circuit complexity
increased, while test performance decreased. This behavior was particularly
pronounced for the ZZ-based feature maps. All three ZZ methods achieved a
training Accuracy of 1.000, whereas their test Accuracy decreased to
approximately 0.65.

The Recall and F1 values of ZZ linear and ZZ circular decreased to zero,
indicating that these methods were unable to correctly identify the positive
class in the test set. Although ZZ full achieved a Precision of 1.000, its
Recall was only 0.011 and its F1 score remained at 0.022.

In terms of generalization, the lowest Accuracy Gap, approximately 0.049,
was obtained with Angle-RY, H+RY, and Pearson-based RY+CRY, whereas the
corresponding values for the ZZ methods increased to approximately 0.35.

A similar pattern is observed in circuit cost. Angle-RY achieved one of the
highest test performances with only 14 gates and a circuit depth of 1,
whereas ZZ full required 301 logical gates, a circuit depth of 77, and
approximately 120 seconds of state vector computation while producing lower
test performance.

Overall, the Alzheimer experiments clearly demonstrate that more complex
quantum feature maps do not necessarily lead to stronger classification
performance. For this dataset, the simple Angle-RY encoding provides a
highly competitive solution in terms of performance, generalization, and
computational cost.

However, the use of PCA should be explicitly considered when directly
comparing the Alzheimer results with those of the other datasets. Since the
first 14 PCs explain only approximately half of the total variance, the
observed performance limitations may reflect not only the choice of quantum
feature map but also the representational feature space produced by
dimensionality reduction.

Thus, the Alzheimer's Disease Dataset demonstrates the importance of jointly
considering feature representation, dimensionality reduction,
generalization, and circuit complexity when evaluating quantum feature
encoding strategies.

\subsection{Cross-Dataset Discussion}

This study comparatively investigated the effects of different quantum
feature encoding strategies on QSVM classification performance,
generalization behavior, and quantum circuit complexity across the EEG Eye
State Classification, Heart Failure Prediction, Credit Risk, Student
Performance, and Alzheimer's Disease datasets. In addition to relatively
simple angle encoding strategies such as Angle-RY and H+RY, RY+CRY
structures based on different statistical dependency measures, Z feature Map,
and ZZ feature Map variants with different connectivity topologies were
evaluated. The results demonstrate that the effect of the quantum feature
encoding strategy is neither dataset-independent nor unidirectional;
instead, it depends on the interaction among data structure, feature
representation, entanglement topology, and model complexity.

When the five datasets are considered together, the first important finding
is that more complex quantum feature maps do not consistently provide higher
test performance. On the EEG dataset, ZZ full achieved the highest test
accuracy of 0.850, but it was also associated with an Accuracy Gap of
approximately 0.150 and a high circuit cost. ZZ circular achieved a very
similar test accuracy of 0.846 while requiring substantially fewer gates
and a much shorter state vector computation time. This observation indicates
that maximum test accuracy alone is insufficient to determine the optimal
encoding strategy.

A different pattern emerged for the Heart Failure Prediction Dataset. The
highest test accuracy, approximately 0.883, was obtained with the Pearson-
and Distance Correlation-based RY+CRY structures, whereas Z feature Map
achieved a very similar test accuracy of 0.878. In contrast, ZZ linear,
ZZ circular, and ZZ full achieved very high training accuracies but
decreased in test performance. In particular, ZZ circular achieved a
training accuracy of 0.994 but only a test accuracy of 0.757, demonstrating
that high circuit complexity did not provide a strong generalization
advantage.

For the Credit Risk Dataset, no differences were observed among the quantum
feature maps in terms of Test Accuracy, Precision, Recall, or F1; all
methods achieved a test accuracy of 0.908. However, ROC-AUC values differed
substantially. The Distance Correlation-based RY+CRY method achieved the
highest ROC-AUC of 0.663, whereas Z feature Map produced only 0.464. Thus,
feature maps that appear equivalent when only Accuracy is considered can
behave differently in terms of decision scores and class separability.
Accordingly, complementary measures such as ROC-AUC and Log-Loss should be
considered together with threshold-dependent metrics when evaluating
quantum kernel methods.

For the Student Performance Dataset, Z feature Map produced the strongest and
most balanced overall result. It achieved a test accuracy of 0.944, an F1
score of 0.900, and a ROC-AUC of 0.975, while its Accuracy Gap remained
only 0.024. In contrast, although ZZ full achieved a training accuracy of
1.000, its test accuracy decreased to 0.712, Recall to 0.020, and F1 to
0.040. This finding clearly demonstrates that a highly expressive quantum
feature map can represent the training data very strongly while losing its
ability to generalize to unseen data.

The Alzheimer's Disease Dataset differs methodologically from the other
datasets because PCA was applied to manage the high feature dimensionality,
and the first 14 PCs were retained. These PCs explained approximately
49.61\% of the total variance. Therefore, the Alzheimer results reflect not
only the effect of the quantum feature map but also the amount of
information retained after dimensionality reduction. Angle-RY, H+RY, and
Pearson-based RY+CRY achieved the highest test accuracy of 0.764. In
contrast, the Z- and ZZ-based structures substantially increased training
performance while test performance decreased. For ZZ linear and
ZZ circular, the test F1 score decreased to zero, whereas ZZ full achieved
a Precision of 1.000 but only a Recall of 0.011 and an F1 score of 0.022.
Therefore, the Alzheimer results should not be attributed solely to the
entanglement structure; the amount of classification-relevant information
preserved in the PCA-reduced feature space must also be considered.

These differences among the datasets demonstrate that no single quantum
feature map is universally superior for all classification problems.
ZZ full achieved the highest test accuracy for EEG; Pearson- and Distance
Correlation-based RY+CRY were strongest for HEART; Z feature Map was superior
for Student Performance; and the simpler Angle-RY and Pearson-based RY+CRY
structures performed best for Alzheimer. For Credit Risk, all methods
produced the same Accuracy. Thus, the effectiveness of a feature map appears
to depend not only on circuit complexity or entanglement density but also on
the compatibility between the data distribution and the structure of the
resulting quantum feature space.

When the RY+CRY results are considered separately, incorporating statistical
relationships between features into controlled rotations was beneficial for
some datasets, although the magnitude of the effect was dataset-dependent.
For EEG, Pearson-based RY+CRY increased test accuracy from 0.748 with
Angle-RY to 0.806, representing a clear improvement. For HEART, the same
approach provided a smaller but consistent improvement. For Credit Risk,
classification accuracy remained unchanged, but an advantage was observed
in ROC-AUC. For Alzheimer, Pearson-based RY+CRY produced exactly the same
test performance as Angle-RY. These results indicate that statistical
dependency information can be useful for quantum encoding, but its
effectiveness depends on the structure of the data.

The cost of RY+CRY circuits must also be considered. For an input dimension
$n$, encoding all pairwise relationships requires

\begin{equation}
N_{\mathrm{CRY}} = \binom{n}{2}
= \frac{n(n-1)}{2}
\end{equation}

controlled connections. For EEG and Alzheimer, 91 CRY connections were
used for 14 features or PCs; Student Performance required 78 connections
for 13 features; HEART required 55 connections for 11 features; and Credit
Risk required 36 connections for 9 features. Thus, an RY+CRY structure in
which all pairwise relationships are encoded grows approximately
quadratically with the number of input features. In the EEG and Alzheimer
experiments, the decomposed RY+CRY circuits reached 182 CX gates and a
circuit depth of 89. This indicates that limited performance improvements
may sometimes be obtained at the cost of substantial circuit complexity.

An important design implication follows from this observation. It may not
be necessary to transfer all pairwise feature relationships into the
quantum circuit. Sparse RY+CRY structures in which only the strongest,
most stable, or most informative relationships are selected may reduce
circuit depth and the number of entangling gates while preserving the
performance contribution of dependency information. Such an approach may be
particularly important for reducing error accumulation and implementation
cost on NISQ devices.

The Z feature Map results also represent an important finding of this study.
Z feature Map achieved the strongest overall test performance on the Student
Performance Dataset and produced competitive results on both EEG and HEART
despite its low circuit complexity. In particular, for Student Performance,
it achieved a test accuracy of 0.944 and ROC-AUC of 0.975 using only 26
logical gates and a circuit depth of 2. For EEG, it achieved a test accuracy
of 0.834 with 28 logical gates and a depth of 2, while maintaining an
Accuracy Gap of 0.060. Therefore, Z feature Map provides a strong balance
among performance, generalization, and circuit cost for some datasets.

The ZZ feature Map results clearly demonstrate that the relationship between
circuit complexity and classification performance is not linear. Although
ZZ full achieved the highest test accuracy for EEG, its improvement over
ZZ circular was only 0.004, while the logical gate count increased from 70
to 301. In contrast, denser ZZ connectivity generally increased training
accuracy but reduced test performance for HEART, Student Performance, and
Alzheimer. Thus, increasing entanglement density does not automatically
produce a more effective quantum feature space.

Generalization behavior represents another important common dimension
across the five datasets. Particularly for the more complex ZZ-based
feature maps, training accuracies frequently approached 1.000 while test
performance declined. This behavior was especially pronounced for the
Student Performance and Alzheimer's Disease datasets. When Accuracy Gap,
F1 Gap, and Log-Loss are considered together, increasing circuit complexity
appears to cause some models to represent the training samples excessively
strongly. Therefore, quantum feature encoding strategies should be evaluated
not only according to test accuracy but also according to the differences
between training and test performance.

A common trend was also observed in computation times. The largest
differences among methods occurred during state vector generation, whereas
Kernel Time and classical SVM Training Time generally exhibited much smaller
variations. For example, ZZ full required approximately 126.6 seconds for
EEG and 119.9 seconds for Alzheimer, whereas Angle-RY remained near
5 seconds. In contrast, kernel computation times within the same dataset
remained largely stable. This indicates that the primary computational cost
in the investigated QSVM framework arises from the generation of quantum
feature states rather than classical SVM optimization.

The PCA application to the Alzheimer's Disease Dataset further emphasizes
the relationship between classical preprocessing and quantum feature
encoding. The first 14 PCs explained only approximately 49.6\% of the total
variance, indicating that a substantial amount of variance was excluded
during dimensionality reduction. Therefore, particularly for
high-dimensional datasets, QML performance should be understood as depending
not only on feature-map selection but also on the classical dimensionality
reduction strategy. Changing the number of PCA components or employing
alternative dimensionality reduction methods may alter the results obtained
for the Alzheimer's Disease Dataset.

\subsection{Quantum Feature Encoding Score (QFES)}

Evaluating different quantum feature encoding strategies solely in terms of
classification performance may overlook differences in generalization
ability and computational cost. More complex quantum feature maps may
provide high classification performance while also requiring greater
circuit depth, larger gate counts, and longer computation times. Therefore,
the Quantum Feature Encoding Score (QFES) was defined in this study to
jointly evaluate classification performance, generalization behavior, and
computational cost.

The performance component of QFES is constructed from Accuracy ($A$),
Precision ($P$), Recall ($R$), F1-score ($F1$), and ROC-AUC ($AUC$)
obtained on the test set. Since all these metrics naturally lie within the
$[0,1]$ interval, they are used directly without additional normalization.
The classification performance component is defined as

\begin{equation}
S = A + P + R + F1 + AUC.
\label{eq:qfes_performance}
\end{equation}

A higher $S$ value therefore indicates stronger overall classification
performance for the corresponding quantum feature encoding strategy.

To evaluate generalization behavior, the absolute difference between
training and test accuracy is used. The Accuracy Gap ($G$) is defined as

\begin{equation}
G = |A_{\rm train} - A_{\rm test}|.
\label{eq:qfes_gap}
\end{equation}

A low $G$ value indicates that training and test performance are close to
each other and that the model exhibits more consistent generalization,
whereas a larger $G$ indicates greater separation between training and test
performance. Since Accuracy Gap naturally lies within $[0,1]$, no
additional normalization is applied.

Circuit depth ($D$), total gate count ($Q$), and state vector computation
time ($T$) are used to represent quantum circuit complexity and
computational cost. Since these variables have different numerical scales,
Min--Max normalization is applied before they are incorporated into QFES.

The normalized circuit depth is calculated as

\begin{equation}
D_N =
\frac{D-D_{\min}}
{D_{\max}-D_{\min}}.
\label{eq:qfes_depth}
\end{equation}

The normalized total gate count is calculated as

\begin{equation}
Q_N =
\frac{Q-Q_{\min}}
{Q_{\max}-Q_{\min}}.
\label{eq:qfes_gate}
\end{equation}

Similarly, the normalized state vector computation time is defined as

\begin{equation}
T_N =
\frac{T-T_{\min}}
{T_{\max}-T_{\min}}.
\label{eq:qfes_time}
\end{equation}

Thus, $D_N$, $Q_N$, and $T_N$ are all transformed into the $[0,1]$
interval. Min--Max normalization is performed separately across the quantum
feature encoding strategies compared within each dataset.

To account for the effect of different input dimensions on circuit and
computational cost, the generalization and cost terms are scaled by the
number of inputs $n$. Here, $n$ denotes the number of input features
transferred to the quantum feature map. When a direct feature--qubit mapping
is used, $n$ also corresponds to the number of qubits.

The generalization and computational cost per input are defined as

\begin{equation}
C =
\frac{G + D_N + Q_N + T_N}{n}.
\label{eq:qfes_cost}
\end{equation}

By combining performance, generalization, and computational cost, QFES is
defined as

\begin{equation}
QFES =
\frac
{A + P + R + F1 + AUC}
{1 + (G + D_N + Q_N + T_N)/n}.
\label{eq:qfes}
\end{equation}

In Eq.~\ref{eq:qfes}, the numerator represents the classification
performance of the quantum feature encoding strategy. Increasing Accuracy,
Precision, Recall, F1-score, and ROC-AUC therefore increases both the
numerator and QFES.

In contrast, increasing Accuracy Gap, normalized circuit depth, gate count,
or state vector computation time increases the denominator and consequently
reduces QFES.

The constant $1$ in the denominator prevents the denominator from
approaching zero when the generalization and computational cost terms
approach their minimum values. This prevents methods with minimal circuit
complexity from obtaining artificially inflated QFES values.

No weighting coefficients are used in the QFES formulation. Accuracy,
Precision, Recall, F1-score, and ROC-AUC contribute equally to the
classification performance component, whereas Accuracy Gap, circuit depth,
gate count, and state vector computation time represent generalization and
computational cost. This design avoids introducing subjective predefined
weights into the final ranking of the methods.

Consequently, a higher QFES indicates a quantum feature encoding strategy
that achieves a more favorable balance among high classification
performance, a small training--test accuracy gap, low circuit complexity,
and low computation time. QFES is therefore used as a composite evaluation
criterion to identify not only the method with the highest predictive
performance, but also the method that provides a more balanced
performance--generalization--computational cost trade-off.

\subsubsection{EEG Eye State Classification Dataset}

\begin{table*}[!htbp]
\centering
\caption{QFES results for the EEG Eye State Classification Dataset.}
\label{tab:qfes_eeg}

\footnotesize
\setlength{\tabcolsep}{3.7pt}
\renewcommand{\arraystretch}{1.10}

\begin{tabular}{clrrrrrr}
\toprule
\textbf{Rank} &
\textbf{Encoding} &
\textbf{$S$} &
\textbf{Gap} &
\textbf{$D_N$} &
\textbf{$Q_N$} &
\textbf{$T_N$} &
\textbf{QFES} \\
\midrule
1  & Z feature Map         & 4.1589 & 0.0600 & 0.0132 & 0.0488 & 0.0362 & 4.1124 \\
2  & ZZ circular         & 4.2255 & 0.1493 & 0.5658 & 0.1951 & 0.1850 & 3.9189 \\
3  & ZZ linear           & 4.1993 & 0.1520 & 0.5263 & 0.1847 & 0.1671 & 3.9115 \\
4  & RY+CRY-Pearson      & 4.0224 & 0.0767 & 0.3289 & 0.3171 & 0.4172 & 3.7196 \\
5  & RY+CRY-Spearman     & 3.9891 & 0.0627 & 0.3289 & 0.3171 & 0.4167 & 3.6923 \\
6  & RY+CRY-Distance     & 3.9689 & 0.0800 & 0.3289 & 0.3171 & 0.4229 & 3.6679 \\
7  & Angle-RY            & 3.6212 & 0.0007 & 0.0000 & 0.0000 & 0.0000 & 3.6210 \\
8  & RY+CRY-Kendall      & 3.8937 & 0.0560 & 0.3289 & 0.3171 & 0.4236 & 3.6039 \\
9  & H+RY                & 3.6212 & 0.0007 & 0.0132 & 0.0488 & 0.0394 & 3.5950 \\
10 & RY+CRY-MI           & 3.8539 & 0.0513 & 0.3289 & 0.3171 & 0.4282 & 3.5671 \\
11 & ZZ full             & 4.2441 & 0.1500 & 1.0000 & 1.0000 & 1.0000 & 3.4646 \\
\bottomrule
\end{tabular}
\end{table*}

For the EEG dataset, Z feature Map achieved the highest QFES value of 4.1124.
Although ZZ full had the highest overall classification performance
component, $S=4.2441$, it ranked last in QFES because its normalized circuit
depth, gate count, and state vector computation time reached their maximum
values. In contrast, Z feature Map combined strong classification performance
with low circuit and computational cost, resulting in the highest composite
score. ZZ circular and ZZ linear ranked second and third, respectively,
because they provided strong predictive performance at substantially lower
resource cost than ZZ full.

\subsubsection{Heart Failure Prediction Dataset}

\begin{table*}[!htbp]
\centering
\caption{QFES results for the Heart Failure Prediction Dataset.}
\label{tab:qfes_heart}

\footnotesize
\setlength{\tabcolsep}{3.7pt}
\renewcommand{\arraystretch}{1.10}

\begin{tabular}{clrrrrrr}
\toprule
\textbf{Rank} &
\textbf{Encoding} &
\textbf{$S$} &
\textbf{Gap} &
\textbf{$D_N$} &
\textbf{$Q_N$} &
\textbf{$T_N$} &
\textbf{QFES} \\
\midrule
1  & Angle-RY            & 4.4610 & 0.0287 & 0.0000 & 0.0000 & 0.0000 & 4.4493 \\
2  & Z feature Map         & 4.4883 & 0.0156 & 0.0172 & 0.0625 & 0.0688 & 4.4223 \\
3  & H+RY                & 4.4610 & 0.0287 & 0.0172 & 0.0625 & 0.0566 & 4.3950 \\
4  & RY+CRY-Pearson      & 4.5008 & 0.0316 & 0.3276 & 0.3125 & 0.4409 & 4.0874 \\
5  & RY+CRY-Distance     & 4.5028 & 0.0331 & 0.3276 & 0.3125 & 0.4682 & 4.0795 \\
6  & RY+CRY-Spearman     & 4.4872 & 0.0360 & 0.3276 & 0.3125 & 0.4373 & 4.0748 \\
7  & RY+CRY-Kendall      & 4.4842 & 0.0316 & 0.3276 & 0.3125 & 0.4495 & 4.0695 \\
8  & RY+CRY-MI           & 4.4718 & 0.0447 & 0.3276 & 0.3125 & 0.4468 & 4.0547 \\
9  & ZZ linear           & 4.2052 & 0.1869 & 0.5345 & 0.2330 & 0.2218 & 3.7990 \\
10 & ZZ circular         & 4.0394 & 0.2377 & 0.5862 & 0.2500 & 0.2486 & 3.6059 \\
11 & ZZ full             & 4.3696 & 0.1565 & 1.0000 & 1.0000 & 1.0000 & 3.3953 \\
\bottomrule
\end{tabular}
\end{table*}

For the Heart Failure Prediction Dataset, the highest QFES value, 4.4493,
was obtained with Angle-RY. Z feature Map produced a very similar second-best
score of 4.4223. Although RY+CRY-Distance and RY+CRY-Pearson produced
higher classification performance components than Angle-RY, their greater
circuit and state vector costs reduced their overall QFES rankings. In
particular, the maximum normalized resource costs of ZZ full reduced its
QFES to 3.3953. These results indicate that the simple Angle-RY structure
provided the most efficient performance--cost balance for this dataset.

\subsubsection{Credit Risk Dataset}

\begin{table*}[!htbp]
\centering
\caption{QFES results for the Credit Risk Dataset.}
\label{tab:qfes_credit}

\footnotesize
\setlength{\tabcolsep}{3.7pt}
\renewcommand{\arraystretch}{1.10}

\begin{tabular}{clrrrrrr}
\toprule
\textbf{Rank} &
\textbf{Encoding} &
\textbf{$S$} &
\textbf{Gap} &
\textbf{$D_N$} &
\textbf{$Q_N$} &
\textbf{$T_N$} &
\textbf{QFES} \\
\midrule
1  & Angle-RY            & 4.1585 & 0.0013 & 0.0000 & 0.0000 & 0.0000 & 4.1579 \\
2  & H+RY                & 4.1585 & 0.0013 & 0.0217 & 0.0769 & 0.0725 & 4.0803 \\
3  & Z feature Map         & 3.9686 & 0.0067 & 0.0217 & 0.0769 & 0.0949 & 3.8822 \\
4  & RY+CRY-Distance     & 4.1681 & 0.0000 & 0.3261 & 0.3077 & 0.4346 & 3.7259 \\
5  & RY+CRY-Pearson      & 4.1624 & 0.0013 & 0.3261 & 0.3077 & 0.4511 & 3.7142 \\
6  & RY+CRY-Spearman     & 4.1636 & 0.0013 & 0.3261 & 0.3077 & 0.4773 & 3.7056 \\
7  & RY+CRY-MI           & 4.1587 & 0.0013 & 0.3261 & 0.3077 & 0.4921 & 3.6958 \\
8  & RY+CRY-Kendall      & 4.1586 & 0.0013 & 0.3261 & 0.3077 & 0.4981 & 3.6936 \\
9  & ZZ linear           & 4.1236 & 0.0253 & 0.5435 & 0.2821 & 0.3011 & 3.6557 \\
10 & ZZ circular         & 4.1103 & 0.0253 & 0.6087 & 0.3077 & 0.3209 & 3.6047 \\
11 & ZZ full             & 4.1179 & 0.0373 & 1.0000 & 1.0000 & 1.0000 & 3.0788 \\
\bottomrule
\end{tabular}
\end{table*}

For the Credit Risk Dataset, Angle-RY achieved the highest QFES value of
4.1579. Because Test Accuracy, Precision, Recall, and F1 were largely
identical across the methods, resource costs became more influential in the
QFES ranking. Although RY+CRY-Distance produced one of the highest
performance components, its greater circuit depth, gate count, and
state vector cost resulted in fourth place. ZZ full produced the lowest QFES,
3.0788, because it achieved similar test performance at substantially
higher resource cost.

\subsubsection{Student Performance Dataset}

\begin{table*}[!htbp]
\centering
\caption{QFES results for the Student Performance Dataset.}
\label{tab:qfes_student}

\footnotesize
\setlength{\tabcolsep}{3.7pt}
\renewcommand{\arraystretch}{1.10}

\begin{tabular}{clrrrrrr}
\toprule
\textbf{Rank} &
\textbf{Encoding} &
\textbf{$S$} &
\textbf{Gap} &
\textbf{$D_N$} &
\textbf{$Q_N$} &
\textbf{$T_N$} &
\textbf{QFES} \\
\midrule
1  & Z feature Map         & 4.6238 & 0.0240 & 0.0143 & 0.0526 & 0.0474 & 4.5752 \\
2  & Angle-RY            & 4.1791 & 0.1133 & 0.0000 & 0.0000 & 0.0000 & 4.1430 \\
3  & H+RY                & 4.1791 & 0.1133 & 0.0143 & 0.0526 & 0.0446 & 4.1081 \\
4  & ZZ linear           & 4.3672 & 0.0887 & 0.5286 & 0.1984 & 0.1795 & 4.0567 \\
5  & RY+CRY-Pearson      & 4.2431 & 0.1080 & 0.3286 & 0.3158 & 0.4030 & 3.8968 \\
6  & RY+CRY-Spearman     & 4.2429 & 0.1080 & 0.3286 & 0.3158 & 0.4041 & 3.8963 \\
7  & RY+CRY-Distance     & 4.2441 & 0.1080 & 0.3286 & 0.3158 & 0.4133 & 3.8948 \\
8  & RY+CRY-Kendall      & 4.2421 & 0.1067 & 0.3286 & 0.3158 & 0.4115 & 3.8939 \\
9  & ZZ circular         & 4.2139 & 0.1080 & 0.5714 & 0.2105 & 0.2003 & 3.8879 \\
10 & RY+CRY-MI           & 4.2150 & 0.1107 & 0.3286 & 0.3158 & 0.4177 & 3.8663 \\
11 & ZZ full             & 2.6811 & 0.2880 & 1.0000 & 1.0000 & 1.0000 & 2.1399 \\
\bottomrule
\end{tabular}
\end{table*}

For the Student Performance Dataset, Z feature Map clearly achieved the
highest QFES value, 4.5752. It also produced the highest total performance
component, $S=4.6238$, while maintaining a low Accuracy Gap and very low
normalized circuit costs. Angle-RY ranked second because of its minimum
circuit cost despite its lower performance component. ZZ full produced a
substantially lower QFES of 2.1399 because of its poor test Recall and F1,
large Accuracy Gap, and maximum resource costs.

\subsubsection{Alzheimer's Disease Dataset}

\begin{table*}[!htbp]
\centering
\caption{QFES results for the Alzheimer's Disease Dataset.}
\label{tab:qfes_alzheimer}

\footnotesize
\setlength{\tabcolsep}{3.7pt}
\renewcommand{\arraystretch}{1.10}

\begin{tabular}{clrrrrrr}
\toprule
\textbf{Rank} &
\textbf{Encoding} &
\textbf{$S$} &
\textbf{Gap} &
\textbf{$D_N$} &
\textbf{$Q_N$} &
\textbf{$T_N$} &
\textbf{QFES} \\
\midrule
1  & Angle-RY            & 3.4464 & 0.0487 & 0.0000 & 0.0000 & 0.0000 & 3.4345 \\
2  & H+RY                & 3.4464 & 0.0487 & 0.0132 & 0.0488 & 0.0368 & 3.4105 \\
3  & RY+CRY-Pearson      & 3.4464 & 0.0487 & 0.3289 & 0.3171 & 0.4193 & 3.1924 \\
4  & RY+CRY-Spearman     & 3.4360 & 0.0520 & 0.3289 & 0.3171 & 0.4245 & 3.1809 \\
5  & RY+CRY-Kendall      & 3.4362 & 0.0507 & 0.3289 & 0.3171 & 0.4288 & 3.1805 \\
6  & RY+CRY-MI           & 3.3820 & 0.1107 & 0.3289 & 0.3171 & 0.4254 & 3.1186 \\
7  & RY+CRY-Distance     & 3.3373 & 0.0840 & 0.3289 & 0.3171 & 0.4296 & 3.0820 \\
8  & Z feature Map         & 2.9053 & 0.2587 & 0.0132 & 0.0488 & 0.0378 & 2.8327 \\
9  & ZZ full             & 2.4938 & 0.3500 & 1.0000 & 1.0000 & 1.0000 & 2.0123 \\
10 & ZZ linear           & 1.2391 & 0.3540 & 0.5263 & 0.1847 & 0.1691 & 1.1387 \\
11 & ZZ circular         & 1.2246 & 0.3540 & 0.5658 & 0.1951 & 0.1795 & 1.1210 \\
\bottomrule
\end{tabular}
\end{table*}

For the Alzheimer's Disease Dataset, Angle-RY achieved the highest QFES
value of 3.4345. Although H+RY and Pearson-based RY+CRY produced the same
overall classification performance component, their additional circuit
complexity and computation time caused them to rank below Angle-RY.
Z feature Map exhibited a lower QFES because its higher training performance
was accompanied by a large Accuracy Gap. The ZZ methods ranked near the
bottom because of both low test Recall and F1 values and higher resource
costs. Since the dataset was represented using 14 PCs after PCA, these
results should also be interpreted in the context of the information
retained in the reduced feature space.


When the QFES results across all five datasets are considered together, no
single quantum feature encoding method is optimal for every dataset.
Z feature Map achieved the highest QFES for the EEG Eye State Classification
and Student Performance datasets, whereas Angle-RY ranked first for the
Heart Failure Prediction, Credit Risk, and Alzheimer's Disease datasets.

This finding demonstrates that the method with the highest raw
classification performance does not necessarily produce the highest QFES.
In particular, methods with high circuit complexity, such as ZZ full, can
produce strong Accuracy or ROC-AUC values on some datasets but receive lower
QFES rankings because of increased Accuracy Gap, circuit depth, gate count,
and state vector computation time.

In contrast, simpler structures such as Angle-RY and Z feature Map frequently
provided sufficient classification performance at substantially lower
computational cost, resulting in a more favorable
performance--generalization--resource cost balance.

\section{Conclusion}

This study systematically evaluated different quantum feature encoding
strategies for QSVM-based binary classification across multiple datasets.
The results demonstrate that the choice of quantum feature encoding affects
not only classification performance but also model generalization, quantum
circuit complexity, and computational cost.

When the results obtained from the different datasets are considered
together, no single quantum feature map was found to be universally superior
across all classification problems. While more complex and highly entangled
feature maps achieved higher test performance for some datasets, simpler
encoding strategies provided better generalization and a more favorable
balance between predictive performance and circuit cost for others. These
findings indicate that the effectiveness of a quantum feature map depends
not only on its circuit structure but also on the statistical characteristics
of the dataset and the structure of the corresponding feature space.

Incorporating statistical dependencies between features into quantum
encoding through RY and CRY gates improved classification performance over
basic angle encoding for some datasets. However, encoding all pairwise
feature relationships through controlled rotations substantially increased
circuit depth and the number of entangling gates. Therefore, the benefit of
transferring statistical dependency information into quantum circuits should
be evaluated by considering the resulting improvement in classification
performance together with the additional circuit cost.

Similarly, the results obtained with ZZ feature Map demonstrate that increasing
entanglement density does not necessarily produce a more effective quantum
feature space. In several experiments, denser connectivity structures
substantially increased training performance without providing a comparable
improvement in test performance, resulting in larger training--test gaps.
This observation suggests that the expressive capacity of a quantum feature
map should be considered together with its ability to generalize to unseen
data.

To provide a unified assessment of classification performance,
generalization, and quantum computational cost, the Quantum Feature Encoding
Score (QFES) was also employed in this study. The QFES results showed that
the encoding strategy with the highest raw classification performance was
not necessarily the most favorable overall approach. Methods that achieved
competitive test performance with lower circuit depth, fewer gates, and
shorter state vector computation times could provide a better balance among
predictive performance, generalization, and resource requirements.

Overall, the selection of a feature encoding strategy for QSVM and
quantum-kernel-based classification should not be based solely on individual
performance measures such as Accuracy or ROC-AUC. Classification
performance, training--test differences, circuit depth, gate count, and
computational time should instead be considered jointly. The findings of
this study suggest that data-aware and resource-conscious approaches to
quantum feature encoding may provide a more appropriate framework for
designing efficient quantum machine learning models.

Future studies may investigate sparse RY+CRY structures in which only the
strongest or most informative statistical relationships are encoded, rather
than introducing controlled rotations between all feature pairs. Such an
approach may preserve the benefits of dependency-aware encoding while
reducing circuit depth and entangling-gate requirements. Furthermore,
experiments involving different input dimensions, alternative dimensionality
reduction techniques, and real NISQ hardware may provide further insight
into the scalability, robustness, and practical applicability of quantum
feature encoding strategies.

\bibliographystyle{unsrt}
\bibliography{references}

\end{document}